\documentclass[a4paper,11pt]{article}
\pdfoutput=1 
\usepackage{jcappub}
\usepackage{graphicx}
\pdfoutput=1 
\usepackage{color}
\usepackage[dvipsnames]{xcolor}
\usepackage{xspace}
\usepackage{multirow}
\usepackage{cleveref}		
\usepackage[normalem]{ulem}
\usepackage{makecell}

\newcommand{\rr}{\mathrm}
\newcommand{\fulleqref}[1]{Equation~\eqref{#1}}
\newcommand{\fullmanyeqref}[1]{Equations~\eqref{#1}}
\newcommand{\tquote}[1]{``#1''}
\newcommand{\ssec}[1]{\subsection*{#1}}
\newcommand{\dnew}{\par\vspace{0.5\baselineskip}\mbox{}}
\newcommand{\dnewnoindent}{\par\vspace*{0.5\baselineskip}\noindent\mbox{}}
\providecommand{\abs}[1]{\ensuremath{\lvert#1\rvert}}
\newcommand{\parder}[2]{\ensuremath{\frac{\partial #1}{\partial #2}}}
\newcommand{\parderconst}[3]{\ensuremath{\left(\frac{\partial #1}{\partial #2}}\right)_#3}
\newcommand{\der}[2]{\ensuremath{\frac{\mathrm{d} #1}{\mathrm{d} #2}}}

\providecommand{\CS}{CS\xspace}
\providecommand{\DC}{DC\xspace}
\providecommand{\BS}{BR\xspace}
\providecommand{\BB}{BB\xspace}
\providecommand{\SD}{SD\xspace}
\providecommand{\ppsd}{PPSD\xspace}
\providecommand{\lcdm}{$\Lambda$CDM\xspace}
\DeclareMathAlphabet\mathbfcal{OMS}{cmsy}{b}{n}

\providecommand{\thus}{\ensuremath{\qquad \quad \Longrightarrow \qquad \quad}}
\title{The synergy between CMB spectral distortions and anisotropies}

\newcommand{\final}[1]{#1}
\newcommand{\finalr}[2]{#2}

\newcommand\Tstrut[1]{\rule{0pt}{#1ex}}       

\newcommand\id{{\rm d}}       

\author[1,2]{Matteo Lucca,}
\author[1]{Nils Sch\"{o}neberg,}
\author[1,2]{Deanna C. Hooper,}
\author[1]{Julien Lesgourgues,}
\author[3]{Jens Chluba} 

\affiliation[1]{Institute for Theoretical Particle Physics and Cosmology (TTK), \\ RWTH Aachen University, D-52056 Aachen, Germany.}
\affiliation[2]{Service de Physique Th\'{e}orique, \\Universit\'{e} Libre de Bruxelles, C.P. 225, B-1050 Brussels, Belgium}
\affiliation[3]{Jodrell Bank Centre for Astrophysics, \\ University of Manchester, Manchester M13 9PL, UK.}

\emailAdd{matteo.lucca@ulb.be}
\emailAdd{schoeneberg@physik.rwth-aachen.de}
\emailAdd{hooper@physik.rwth-aachen.de}

\abstract{Spectral distortions and anisotropies of the CMB provide independent and complementary probes to study energy injection processes in the early universe. Here we discuss the synergy between these observables, and show the promising future of spectral distortion missions to constrain both exotic and non-exotic energy injections. We show that conventional probes such as Big Bang Nucleosynthesis and CMB anisotropies can benefit from and even be surpassed by future spectral distortion experiments. For this, we have implemented a unified framework within the Boltzmann code {\sc class} to consistently treat the thermal evolution of photons and baryons. Furthermore, we give an extensive and pedagogical introduction into the topic of spectral distortions and energy injections throughout the thermal history of the universe, highlighting some of their unique features and potential as a novel probe for cosmology and particle physics.}

\begin{document}

\hfill{\small TTK-19-39}\\
\vspace{-0.5 cm}
\hfill{\small ULB-TH/19-08}

\maketitle

\newpage
\section{Introduction \label{intro}}
With the recent results of the Planck collaboration \cite{Aghanim:2018eyx}, the wealth of information gained from cosmic microwave background (CMB) temperature and polarization anisotropies has increased dramatically. Upcoming observations, like for instance the ground-based Simons Observatory \cite{SOWP2018} and CMB-S4 \cite{Abazajian:2016yjj, Abitbol2017CMB, Abazajian:2019eic} or the space mission LiteBIRD \cite{Matsumura:2013aja, Suzuki2018LiteBIRD}, will reach an unprecedented degree of precision.
\dnew 
However, despite their incredible precision, these observations still face limitations. These include cosmic variance on large scales, the diffusion damping for scales ${k\gg 1}$~Mpc$^{-1}$, and the difficulty of accurate astrophysical foreground subtraction \cite{Adam:2015wua}. As a consequence, some parameter degeneracies remain, such as between the reionization optical depth $\tau_{\rm reio}$ and the amplitude of the primordial power spectrum $A_s$. Nonetheless, there is still some information in the CMB that has not yet been exploited to its full potential, and which could help overcome some of the aforementioned limitations. Important examples are the observation of primordial polarization B-modes (e.g., \cite{Ade:2014xna, Ade:2014afa, Creminelli:2015oda}), the Bispectrum (e.g.,  \cite{Wang2000, Fergusson2010}) and higher N-point correlators (e.g., \cite{Hindmarsh2010}), or observations of CMB secondary anisotropies (e.g., \cite{Aghanim_2008}) and lensing effects (e.g., \cite{Seljak:1995ve, Zaldarriaga:1998ar,Challinor:2005jy})
\dnew
One particularly interesting opportunity to extract more information is given by CMB spectral distortions (\SD{s}) \cite{zeldovich1969interaction, Sunyaev1970, Zeldovich1972Influence, Illarionov1975, danese1982double, burigana1991formation, hub1993thermalization, chluba2011evolution}. These distortions are created whenever the energy or number density of the CMB photons is modified. Many physical effects that cause deviations from a perfect blackbody (\BB) are predicted even within the standard $\Lambda$CDM model, and are linked to a variety of processes spanning from cosmological effects, such as the adiabatic cooling of electrons and baryons, to more particle physics based reactions, as in the case of photon emission and absorption during recombination \cite{sunyaev2009signals, chluba2011evolution, Sunyaev2013, Tashiro2014, DeZotti:2015awh, chluba2016spectral}. Even astrophysical models for galaxy and star formation can produce detectable SDs {\cite{hill2015taking} through the Sunyaev-Zeldovich (SZ) effect \cite{Sunyaev1972CoASP, BIRKINSHAW199997, Carlstrom2002Cosmology, Mroczkowski2019}.  
\dnew	
Furthermore, several non-minimal cosmological models intrinsically predict some level of energy injection. For instance, this is the case in models with dark matter (DM) annihilating or decaying into standard model particles \cite{hub1993thermalization, McDonald2001, chluba2010could, chluba2013distinguishing}, DM interacting with baryons or photons \cite{ali2015constraints, slatyer2018early}, Primordial Black Hole (PBH) evaporation \cite{macgibbon1990quark, macgibbon1991quark, tashiro2008constraints, carr2010new, Nakama2017xvq}, and different inflationary scenarios \cite{Sunyaev1970diss, daly1991spectral, barrow1991jd, Hu1994, chluba2012cmb, Chluba2012inflaton}. All these models can be constrained with future observations of \SD{s}. Additionally, it has been shown that \SD{s} can help us distinguish between different proposed solutions to the so-called \emph{small scale crisis} of cosmology \cite{Nakama2017, diacoumis2017using}. 
\dnew
\final{Moreover, SDs are expected  to be anisotropic -- analogously to the CMB blackbody temperature and polarization spectra. This has already been measured in the case of Sunyaev-Zeldovich distortions induced by clusters of galaxies \cite{Aghanim:2015eva}. For other sources of SDs, anisotropies are expected to be very difficult to detect. However, a measurement of the power spectrum of SD anisotropies would offer a unique way of investigating the reionization epoch \cite{Pitrou:2009bc, Renaux-Petel:2013zwa, Pitrou:2014ota} as well as the thermal history prior to recombination \cite{Ota:2018iso}, although the latter signal is even lower.}
\dnew
As such, the amount of information that could be gained from \SD{s} is very rich and would cover times and scales yet unexplored by any other experiment. For a recent review, we refer the reader to \cite{Chluba2019BAAS} and references therein.
\newpage
Since the pioneering works of the early '70s \cite{zeldovich1969interaction, Sunyaev1970, Zeldovich1972Influence, Illarionov1975, Illarionov1975b}, the theoretical framework surrounding \SD{s} has been developed considerably, with significant progress over the last decade. In particular, with the development of {\sc CosmoTherm} \cite{chluba2011evolution} it became possible to precisely compute \SD{} shapes for several physical mechanisms by directly following the full time dependence of the processes involved, which had been approximated in previous numerical studies (e.g., \cite{burigana1991formation, hub1993thermalization}). It then became possible to build approximate solutions based on the Green's function method, which greatly speeds up  calculations \cite{chluba2011evolution, Khatri2012mix, chluba2013green}. A few years later, several efficient schemes have been developed to precisely compute other contributions to \SD{s} like those from non-thermal photon-injection processes \cite{Chluba:2015hma}, the cosmological recombination radiation (CRR) \cite{Ali2013RecSpec, chluba2016cosmospec}, and late-time contributions from reionization and structure formation~\cite{chluba2012fast, chluba2013sunyaev}.
\dnew
Thus, today SD theory relies on a remarkably solid analytical and numerical base. However, the experimental counterpart has unfortunately stayed behind. In fact, the only observation of the energy power spectrum of CMB photons was conducted in the '90s by the COBE/FIRAS satellite at a level of precision such that no \SD{s} were observed \cite{Mather1994, fixsen1996cosmic}. Nevertheless, two important results emerged. First, COBE/FIRAS accurately determined the average CMB temperature \cite{Mather1994, fixsen1996cosmic, fixsen2009temperature}, which fixes the energy scale for understanding the evolution of the pre-recombination, radiation-dominated universe. Second, it set upper bounds on the $y$ and $\mu$ parameters describing the final shape of the \SD{s} at approximately $|y|< 1.5\times10^{-5}$ and $|\mu|<9\times10^{-5}$ (95\% confidence level (CL)), which constrains cosmological models with exotic energy release at the level $\Delta \rho_\gamma/\rho_\gamma<6\times 10^{-5}$ (95\% CL). Despite their wide-ranging implications, these values are still too loose to touch on the SDs predicted by the \lcdm model (e.g., \cite{chluba2016spectral}). With current technology, significant improvements over the long-standing COBE/FIRAS bounds could be expected, and even the detection of SDs from \lcdm should be possible~\cite{kogut2011primordial, Kogut2019, Chluba2019Voyage}.

\dnew
In this work we investigate the synergy between CMB anisotropies and \SD{s}, and show the surprising wealth of information to be gained from futuristic experimental setups, covering large ranges of parameter space otherwise unconstrained. To achieve this goal, we first present the implementation of \SD{s} in the Boltzmann code {\sc class} \cite{blas2011cosmic}, thus incorporating the already well developed \SD formalism in a general cosmological code, in a fully consistent way and without redundant steps. This generalizes and improves on similar studies carried out previously by \cite{chluba2013distinguishing, chluba2014teasing}.
We subsequently select a few interesting cosmological scenarios and perform parameter sensitivity forecasts, to illustrate the synergy between future \SD{} missions and other cosmological probes. Our results clearly demonstrate that CMB \SD{s} are an independent and exciting new probe of physics.

\dnew
This paper is organized as follows. In Section~\ref{Theory} we review the formalism used to describe \SD{s}, paying special attention to the parameter dependency of the \SD{} shape and amplitude, before discussing in Section~\ref{SD_causes} several different mechanisms that can generate \SD{s}. In Sections~\ref{class} and \ref{Greens} we present the ingredients used for our numerical implementation of \SD{s} in {\sc class}, while in Section~\ref{likelihoods} we describe the mock likelihoods that we build to account for future experiments. In Section~\ref{results} we show how this framework can be used to forecast the sensitivity of parameter reconstruction for different cosmological models, and we illustrate the advantage of combining \SD{s} with other cosmological observables. Our conclusions are presented in Section~\ref{conclusions}, while the Appendices provide more in-depth details on the \SD{s} formalism.

\dnew
\textit{Remarks on notation:} Throughout this paper we use Greek indices for four-dimensional quantities (e.g., $q^\mu$) and adopt the Einstein convention for the summation over repeated indices. \finalr{Massless particles will have $E=p$, with the}{The} spacetime coordinates and 4-momentum of a particle \final{are} denoted by $q^\mu$ and $p^\mu$ respectively. Three-dimensional forms will be written in bold. \final{We define the proper momentum measured by comoving observers\footnote{\final{This definition of the proper momentum can be extended in presence of metric fluctuations, but we will only be concerned with homogeneous cosmology within this work.}} as $p=a \, \abs{\textbf{p}}$, which for massless particles in a flat homogeneous FLRW metric implies ${p = p^0 = E}$, where $E$ is the particle energy measured by the same comoving observers.}
Furthermore, we use an overdot to indicate derivatives with respect to \textit{physical} time, and -- unless stated otherwise -- we work in natural units with $\hbar=c=k_B=1$. \final{All these quantities are defined in the context of the homogeneous flat FLRW metric
\begin{align}
ds^2=-dt^2+a^2(t)[dr^2+r^2d\theta^2+r^2\sin^2\theta d\phi^2]\,.
\end{align}
}Moreover, a small $n$ will correspond to number density, and $\rho$ to energy density. The symbols $e, \gamma, b, \nu, \mathrm{cdm}$, H, and He denote respectively electrons, photons, baryons, neutrinos, cold dark matter, Hydrogen, and Helium. Additionally, we will refer to $x_{e}=n_e/n_H$ as the fraction of free electrons.
\section{Theory\label{Theory}}
Here we aim to unify and streamline the theory of \SD{s}, building on the theory reviews and lecture notes of \cite{chluba2014science, Chluba2015IJMPD, Chluba2018}. In Section \ref{SD_what} we introduce the photon Boltzmann equation governing the evolution of the photon phase-space distribution (\ppsd). In Section \ref{SD_shapes} we infer what kind of distortions of the \BB spectrum of the CMB are allowed. In Section \ref{SD_calc} we show how the amplitude of the \SD{s} can be calculated for a given thermal history of the universe. Finally, in Section \ref{SD_causes} we list the most significant heating processes within the standard cosmological model and review several exotic heating mechanisms as well.

\subsection{Photon Boltzmann equation \label{SD_what}}
The goal of this section is to describe the Boltzmann equation of the \ppsd in the presence of Compton scattering (\CS), double Compton scattering (\DC), and Bremsstrahlung (\BS). The study of the evolution of \ppsd directly provides a description of \SD{s}, as the observable intensity spectrum is just given by the \ppsd multiplied by a factor of $2 h \nu^3/c^2$.
\dnewnoindent
In the homogeneous FLRW metric we have
\begin{equation}\label{eq:p prop to a}
p^0 \parder{p^\mu}{t} + \Gamma^{\mu}_{\alpha\beta} \, p^\alpha p^\beta = 0  \qquad \quad \Rightarrow  \qquad \quad 
\der{p}{t} = - H p \qquad \quad \Rightarrow \qquad \quad p \propto a^{-1}~,
\end{equation}
where $\Gamma^{\mu}_{\alpha\beta}$ are the Christoffel symbols. 
\final{For convenience, we define the time-invariant and} dimensionless frequency $x=x(t,p)$,
\begin{equation}\label{eq:definition x}
x \equiv \frac{p}{T_z} \propto a \cdot p \qquad \qquad \Rightarrow \qquad \qquad \der{x}{t} = 0~,
\end{equation}
where \final{$T_0$ is a reference temperature, and $T_z \equiv T_{0} (1+z)$ scales} exactly as if photons were a decoupled species\footnote{This quantity should not be confused with the actual photon temperature $T_\gamma$\,, which may have a more complicated evolution. The precise normalization of $T_z$ is arbitrary, but $T_0\equiv T_z(0)$ will be chosen close to the actual temperature today, $T_\gamma(0) = (2.7255 \pm 0.0005) \mathrm{K}$ \cite{fixsen1996cosmic,fixsen2009temperature}, in order to have $T_\gamma \simeq T_z$ at least in the late universe.}.
\final{The dimensionless frequency is related to the current observed frequency $\nu$ through $x=h\nu/T_0$.}
This definition of $x$ absorbs the momentum redshifting and simplifies the frequency dependence of the \BB spectrum, as can be seen in \fullmanyeqref{eq:Boltzmann 1} and \eqref{eq:distortion}, respectively. 
\dnew
Furthermore, we are going to assume a homogeneous background-distribution for the \ppsd $f(q^\alpha,p^\alpha)$ such that $\partial f/\partial \textbf{q} = 0$ and $\partial f/\partial \textbf{n}=0$ with $\textbf{p} = p\,\textbf{n}$. Substituting then ${p(t)\to x(t,p)}$, we obtain that $f(q^\alpha,p^\alpha)=f(t,x)$. Thus, the \ppsd obeys the general homogeneous Boltzmann equation 
\begin{equation}\label{eq:Boltzmann 1}
C[f] = \der{f(t,x)}{t} = \parderconst{f}{t}{x} + \der{x}{t}\parderconst{f}{x}{t} =  \parderconst{f}{t}{x}~,
\end{equation}
denoting the indices of the brackets as the variables we hold constant when evaluating the derivatives (see Appendix~\ref{ap:parder} for more details on our treatment of partial derivatives and a more general discussion regarding \fulleqref{eq:Boltzmann 1}). This result clearly shows that in absence of collisions, i.e., setting $C[f]=0$, the homogeneous \ppsd\footnote{Note that with perturbations in the \ppsd this is not true anymore as shown in e.g., \cite{Sunyaev1970diss, daly1991spectral, barrow1991jd, Hu1994, chluba2012cmb}.} is constant in time. In other words, only the collision term $C[f]$ can change the \ppsd, as it adds or removes photons or changes the momenta of existing photons.
\dnew
The main effect capable of modifying the momentum distribution of the photon bath is \CS. The solution of the collision term for this process has been found by \cite{kompaneets1957establishment} assuming a Maxwellian electron phase-space distribution. The result is the famous Kompaneets equation
\begin{equation}\label{eq:Kompaneets eq}
C[f]\rvert_\mathrm{CS} = \dot{\tau} \frac{T_e}{m_e} \frac{1}{x^2} \parder{}{x} \left( x^4 \left[\parder{f}{x} + \frac{T_z}{T_e} f(1+f)\right] \right)~,
\end{equation}
where $\sigma_T$ is the Thomson cross section and $\dot{\tau} = n_e \sigma_T$. If \CS is very efficient, the system will tend towards an equilibrium solution where $C[f]\rvert_\mathrm{CS}$ is \textit{functionally} identical to zero. This can be fulfilled as long as $f$ is a solution to the differential equation
\begin{equation}
0 = \left[\parder{f}{x} + \frac{T_z}{T_e} f(1+f)\right]~,
\end{equation}
which has a physically relevant solution
\begin{equation}\label{eq:kompaneetssolution}
f(x) = \frac{1}{\exp(\tilde{x}+C)-1}~,
\end{equation}
where $\tilde{x} = x \, T_z/T_e = p/T_e$ and $C$ is an integration constant. As expected, this solution coincides with the Bose-Einstein distribution for photons in kinetic equilibrium with electrons with a chemical potential $\mu=C$.
\dnew
{\CS} conserves the number of photons and is compatible with a non-zero chemical potential, but this is not the case for additional processes like \DC scattering and \BS emission. When those processes are also in equilibrium, the chemical potential must vanish, since reactions like $n \gamma \longleftrightarrow m \gamma$ with $n\neq m$ are permitted and efficient (more details are provided in Appendix~\ref{ap:mu}).}
As such, the efficiency of the \DC and \BS processes is crucial to minimize the effective chemical potential of the photon bath, and their inefficiency will subsequently cause a distortion. 
\dnew
When including these additional processes to the collision term expressed in \fulleqref{eq:kompaneetssolution}, one finds the complete evolution equation for the \ppsd, including the most important processes (with $\mathrm{d}\tau = \sigma_T n_e \mathrm{d}t$):
\begin{equation}\label{eq:kompaneetsgeneralsolution}
\parder{f}{\tau} = \frac{T_e}{m_e} \frac{1}{x^2} \parder{}{x} \left( x^4 \left[\parder{f}{x} + \frac{T_z}{T_e} f(1+f)\right] \right) + \frac{K_\mathrm{BR} e^{-\widetilde{x}}}{\widetilde{x}^3} \mathcal F+\frac{K_\mathrm{DC} e^{-2x}}{x^3} \mathcal F~,
\end{equation}
with $\mathcal{F}(x) = 1 - f(x)\cdot (e^{\tilde{x}}-1)$. Here $K_\mathrm{BR}$ and $K_\mathrm{DC}$ are both temperature and frequency dependent factors describing the efficiency of \BS and \DC, respectively. Note that there are several conventions for the definition of these factors. In particular, the differences between $\tilde{x}$ and $x$ can be included in the definitions of $K_\mathrm{BR}$ and $K_\mathrm{DC}$ or not. Here we follow \cite{chluba2014science}, where, together with \cite{lightman1981double, chluba2005spectral, chluba2011evolution}, the interested reader can find the full derivations for these factors and more in-depth discussions.

\subsection{Shapes of the distortions}\label{SD_shapes}
The thermalization of the CMB takes place through various processes, the most prominent of which are \CS, \DC, and \BS. As long as all of these processes are efficient, the CMB spectrum will locally remain a \BB. Their gradual inefficiency causes the \SD{s} of the BB spectrum to be generated. To model the \ppsd $f(t,x)$ we will thus always decompose it as
\begin{equation}\label{eq:distortion}
f(t,x) = B(x) + \Delta f(t,x)~,
\end{equation}
where $B(x) \equiv 1/(e^x-1)$ is the phase-space distribution of a \BB at the temperature $T_z$\,. We will treat any contribution to $\Delta f(t,x)$ as a distortion of the spectrum. 
\dnew
Note that this also includes a deviation of the radiation temperature $T_\gamma$ from the simple $T_z \propto (1+z)$ law.
\final{In this case, there are no actual distortions with respect to a blackbody spectrum, but only a departure from the arbitrarily chosen reference one. For this reason, we will refer to such deviations as temperature shifts instead of distortions}.
As argued in the following sections, temperature shift\final{s} \finalr{distortions}{} are \finalr{in principle}{very} difficult to observe\footnote{\final{Note that the temperature history $T_\gamma(z)$ at different times could still be constrained through different probes, such as recombination constraints from CMB anisotropies \cite{Hamann:2007sk,Chluba:2014wda,ade2016planck} or entropy constraints from BBN \cite{Steigman2007, Chluba:2015hma, Schoneberg:2019wmt}. This would, in principle, allow for a measurement of temperature shifts between the corresponding epochs. However, experimental uncertainties are usually much larger than predicted shifts.}}. Consequently, keeping them cleanly separated from the other \final{true} distortions will be crucial. As a possible distinguishing criterion to isolate the components of the other distortions, which should not be confused with a shift in temperature, one can use the shift in photon number $\Delta N$ caused by thermalization. Indeed, this precisely separates the temperature shift\final{s} \finalr{distortions}{}, involving \DC, \BS and $\Delta N \neq 0$, form the other distortions, involving only \CS and $\Delta N = 0$.
\dnew
\final{When the PPSD does not follow exactly a thermal shape, several definitions of temperature can be introduced. By choosing the above criterion, we are implicitly introducting a definition of temperature based on number density: $T_\gamma$ is the temperature of a blackbody that would share the same number density as the distorted PPSD. Other authors occasionally refer to alternative definitions, such as the energy density temperature, or the Rayleigh-Jeans temperature \cite{chluba2011evolution,chluba2012cmb}. In any case, our final results will be expressed in terms of the full observable photon energy spectrum, and will thus be independent of the temperature definition.}
\dnew
\newpage
All necessary equations and definitions to describe the three major types of distortions expected throughout the thermal history are now assembled. These include, in chronological order of importance, the temperature shift $g$ \finalr{distortions}{}, the chemical potential $\mu$ distortion, and the Compton $y$ distortion.

\ssec{Temperature shift $g$ \finalr{distortion}{}}
The solution for the real photon temperature $T_\gamma$ will deviate from $T_z$ whenever energy is injected, and from the electron temperature $T_e$ when their thermal coupling becomes inefficient. Solutions such as \fulleqref{eq:kompaneetssolution} \finalr{or~\eqref{eq:kompaneetsgeneralsolution}}{} will then not be applicable. The temperature of the spectrum will be shifted, even if it can still be described as a \BB spectrum.
\dnewnoindent
According to \fulleqref{eq:distortion}, this can be written at first order as
\begin{equation}
	f(x)=B\left(\frac{p}{T_\gamma}\right) = B\left(\frac{x}{1+\Delta T/T_z}\right)\approx B(x) - x \parder{B(x)}{x} \frac{\Delta T}{T_z} \equiv B(x) + G(x) \frac{\Delta T}{T_z}~,
\end{equation}
with $\Delta T = T_\gamma - T_z \ll T_z$\,. Thus, the shift of the phase space distribution reads 
\begin{equation}
\Delta f(x) = G(x) \frac{\Delta T}{T_z}~,
\end{equation} 
where we defined the shape of the temperature shift \finalr{$g$ distortion}{}
\begin{equation}
G(x) = - x \parder{B(x)}{x} = \frac{x e^{x}}{(e^{x}-1)^2}~.
\end{equation}
The amplitude of the temperature shift \finalr{$g$ distortion}{} is determined by the true \BB temperature today $T_\gamma(z=0)$ and the chosen reference temperature \mbox{$T_0 \equiv T_z(z=0)$}. Consequently, it can only be constrained up to the experimental uncertainty on $T_\gamma(z=0)$. \final{In practice, however, it is always possible to readjust the reference temperature to coincide with the observed one.}
\finalr{Note that the temperature history $T_\gamma(z)$ at different times can still be constrained through other probes, e.g.,~through the entropy constraints from BBN \cite{}.}{}

\ssec{Chemical potential $\mu$ distortion}
We have seen above that the general solution to the Kompaneets equation in full equilibrium is \fulleqref{eq:kompaneetssolution}, which involves a chemical potential. This chemical potential vanishes only as long as processes changing the number of photons are efficient. Otherwise, one finds\footnote{To be more rigorous, we should write this solution in terms of $\tilde{x}$ instead of $x$. However, the difference between $x$ and $\tilde{x}$ is equivalent to a simple temperature shift distortion, not relevant for this section.}
\begin{equation}
f(x)=B(x+\mu) = \frac{1}{e^{x+\mu} - 1} \approx \frac{1}{e^x -1} - \mu\frac{G(x)}{x} = B(x) - \mu \frac{G(x)}{x}~.
\end{equation}
We find that the shift in the total photon phase-space distribution reads
\begin{equation}
\Delta f(x) = -\mu \frac{G(x)}{x}~,
\end{equation} 
suggesting a \textit{possible} definition of the $\mu$ distortion shape as  
\begin{equation}\label{eq:mu shape 1}
\widetilde{M}(x) = - \frac{G(x)}{x}~.
\end{equation}
Note, however, that the above \ppsd shift does not respect the number count changing criterion employed here to separate the distortions. In fact, the definition expressed in \fulleqref{eq:mu shape 1} can be seen as a superposition of a BB temperature shift and pure $\mu$ distortion. To correct this, we can subtract the temperature shift away and obtain
\begin{equation}\label{eq:mu shape 2}
M(x) = - G(x)\left( \frac{1}{x}- \alpha_\mu\right)~,
\end{equation}
where the coefficient $\alpha_\mu$ is found by imposing that the remaining $\mu$ distortion conserves the photon number density\footnote{We recall that the number density is given as $n(t) = \int f(p,t) \mathrm{d}^3p = 4 \pi T_z^3 \int f(x) x^2 \mathrm{d}x$}, 
\begin{equation}\label{eq:number conservation mu-dist}
\int x^2 M(x) \mathrm{d}x \stackrel{!}{=} 0 \thus \int (-x + \alpha_\mu x^2)G(x) \mathrm{d}x = (-G_1+\alpha_\mu G_2) \stackrel{!}{=} 0~.
\end{equation}
Here we have defined the useful quantity \mbox{$G_k = \int x^k G(x) \mathrm{d}x = (k+1)! \zeta(k+1)$}, and one subsequently obtains \mbox{$\alpha_\mu = G_1/G_2 \approx 0.4561$}. Finally, the $\mu$ distortion reads
\begin{equation}\label{eq:mu shape 3}
\Delta f(x) = \mu \, M(x) ~.
\end{equation}
Note that one could have defined $\mu$ distortions in such way to conserve energy rather number density \cite{chluba2012cmb}, but the current definition leads to simpler and more consistent formulas.

\ssec{Compton $y$ distortion}
The $y$ distortion occurs when the Kompaneets equation \eqref{eq:Kompaneets eq} applies without reaching its equilibrium solution. This occurs when \CS still takes place, but is not very efficient. Following any departure from equilibrium, and starting from an initial \BB spectrum, the photons will be redistributed on some timescale $\Delta \tau$ according to\footnote{Note that $-\partial{B(x)}/\partial x = B(x)(1+B(x)) = G(x)/x$.}
\begin{equation}
\frac{\Delta f}{\Delta \tau} \approx \frac{T_e}{m_e} \frac{1}{x^2} \parder{}{x} \left( x^4 \left[\parder{B(x)}{x} + \frac{T_z}{T_e} B(x)(1+B(x))\right] \right) = \frac{T_z-T_e}{m_e} \frac{\parder{}{x} (x^3 G(x))}{x^2}~.
\end{equation}
Therefore, the shift in the total photon phase-space distribution reads
\begin{equation}\label{eq:yfromtau}
\Delta f(x) \approx \Delta \tau \frac{T_e-T_z}{m_e} Y(x)~,
\end{equation}
which then defines the $y$ distortion shape as $\Delta f(x) = y \, Y(x)$ with
\begin{equation}
Y(x) \equiv -\frac{\parder{}{x} (x^3 G(x))}{x^2} = G(x) \left[x \frac{e^x+1}{e^x-1} - 4\right]~.
\end{equation}
We can immediately see that the photon number density is conserved by such a distortion, since
\begin{equation}
\int x^2 Y(x) \mathrm{d}x = - \int \parder{}{x} (x^3 G(x)) \mathrm{d}x = 0~,
\end{equation}
and thus there is no need to subtract any additional temperature shift.
\dnew
\newpage
Note that inefficient \CS is one possible example of generating a $y$ distortion, but other processes can cause the same distortion shape of the \ppsd. This implies that the general distortion amplitude $y$ can include additional contributions on top of the historical Compton parameter $y_C$, which is commonly defined as
\begin{equation}
	y_C = \int \frac{T_e-T_z}{m_e} \mathrm{d}\tau = \int \frac{T_e-T_z}{m_e} \sigma_T n_e \mathrm{d}t ~.
\end{equation}
We will come back to the definition of the full amplitude $y$ and its relation to $y_C$ in Section~\ref{SD_calc}.

\ssec{Normalization}
For simplicity and consistency, we can normalize the shape distortion functions such that a distortion amplitude equal to one induces a relative variation of the photon energy density of one. For that purpose, we first calculate the factors $C_x$ such that a distortion amplitude $g=C_g$, $\mu=C_\mu$ or $y=C_y$ gives $\Delta \rho_\gamma/\rho_\gamma = 1$. For the \finalr{$g$ distortion}{temperature shift} we obtain\footnote{Remember that $\rho(t) = \int f(p,t) E d^3p = 4 \pi T_z^4 \int f(x)x^3 dx$, using $E=p$ for photons.}
\begin{equation}\label{eq:normalization 1}
\frac{\Delta\rho_\gamma}{\rho_\gamma} = C_g \frac{\int x^3 G(x)\mathrm{d}x}{\int x^3 B(x)\mathrm{d}x} = C_g \frac{G_3}{1/4\,\,G_3} = 4 C_g \stackrel{!}{=} 1 \thus C_g = 1/4~.
\end{equation}
For the $\mu$ distortion we obtain
\begin{equation}\label{eq:normalization 3}
\frac{\Delta\rho_\gamma}{\rho_\gamma} = C_\mu \frac{\int x^3 M(x)\mathrm{d}x}{\int x^3 B(x)\mathrm{d}x} = C_\mu \frac{ - G_2 + \alpha_\mu G_3}{1/4\,\,G_3} = C_\mu \frac{\kappa_\mu}{3} \stackrel{!}{=} 1 \hspace*{-1em}\thus\hspace*{-1em}C_\mu = \frac{3}{\kappa_\mu}~,
\end{equation}
with the constants $\kappa_\mu = 12 ( G_1/G_2 -  G_2/G_3 ) \approx 2.1419$ and $C_\mu = 3/\kappa_\mu \approx 1.401$.
Moreover, for the $y$ distortion we obtain
\begin{equation}\label{eq:normalization 2}
\frac{\Delta\rho_\gamma}{\rho_\gamma} = C_y \frac{\int x^3 Y(x)\mathrm{d}x}{\int x^3 B(x)\mathrm{d}x} = C_y \frac{G_3}{1/4\,\,G_3} = 4 C_y \stackrel{!}{=} 1 \hspace*{-1em}\thus\hspace*{-1em}C_y = 1/4~.
\end{equation}
We can then define the renormalized amplitudes as
\begin{align}
\widetilde{y} \equiv y/C_y = 4y~,  \qquad  \quad \widetilde{\mu} \equiv \mu/C_\mu \approx \mu/1.401~, \qquad \quad \widetilde{g} \equiv g/C_g = 4g~.
\end{align}
These three renormalized contributions to the shift in the \ppsd now have the desired property that $\tilde{g}=1$, $\tilde{\mu} =1$, or $\tilde{y} =1$ result in $\Delta \rho_\gamma/\rho_\gamma = 1$.
\dnew
Finally, the distortions of the intensity spectrum are given by those of the \ppsd multiplied by $2 h \nu^3/c^2$. Using $x = p/T_z = h\nu/(k_B T_0)$, we can write this factor as $2 h \nu^3/c^2 = \mathcal{N} x^3$ with $\mathcal{N} \equiv 2 (k_B T_0)^3/(h c)^2$. Then the intensity spectrum in presence of the three types of distortions reads
\begin{align}
I(x) = \mathcal{B}(x) + \tilde{g} \, \mathcal{G}(x) + \tilde{\mu} \, \mathcal{M}(x) + \tilde{y} \, \mathcal{Y}(x) ~,
\end{align}
where we have defined the normalized shapes
\begin{equation}
\begin{split}
\begin{aligned}
\mathcal{B}(x) = & \quad \, \mathcal{N} x^3 & \hspace*{-0.8em}B(x) &~, \\
\mathcal{G}(x) = & C_g \, \mathcal{N} x^3 & \hspace*{-0.8em}G(x) & = &\hspace*{-0.5em} 1/4 \,\mathcal{N}x^3 & G(x)~, \\
\mathcal{M}(x) = & C_\mu \, \mathcal{N} x^3 &\hspace*{-0.8em} M(x) & \approx &\hspace*{-0.5em} 1.401\,\mathcal{N}x^3 & G(x) \left[-1/x + \alpha_\mu \right] ~,\\
\mathcal{Y}(x) = & C_y \, \mathcal{N} x^3 & \hspace*{-0.8em}Y(x) & = &\hspace*{-0.5em}  1/4\, \mathcal{N}x^3 & G(x) \left[x \frac{e^x+1}{e^x-1} - 4\right] ~.
\end{aligned}
\end{split}
\end{equation}

\ssec{Other distortions}
There are several ways to produce distortions that fall in none of the $g$, $\mu$, or $y$ categories (e.g., \cite{chluba2011evolution, Khatri2012mix, chluba2013green, Chluba:2015hma}). 
\dnew
In the epochs when the redistribution of the $y$ distortion towards a chemical potential is neither fully inefficient nor fully efficient, an intermediate (or hybrid) distortion will be obtained. 
Other types of distortions can also result from highly energetic exotic energy injections when the \CS term is very inefficient. The Compton redistribution term is proportional to $\Delta \tau = \sigma_T n_e \Delta t$, which can be very small after recombination when the free electron fraction, and correspondingly $n_e$, drop towards zero, allowing for the injected photon spectrum to remain \tquote{frozen} at the initial injection frequencies \cite{Chluba:2015hma} (see e.g., \cite{pospelov2018new} for a recent proposal of injections in the Rayleigh-Jeans tail). Other non-thermal distortions can be created by atomic transitions in the pre-recombination era \cite{Liubarskii83, chluba2009pre} or non-thermal particle distributions \cite{Ensslin2000, Chluba:2015hma, Slatyer2015, Acharya2018, Acharya:2019owx, Acharya:2019uba}. All of these particular distortions provide, in principle, additional opportunities for testing the standard cosmological model.
\dnew
Therefore, general distortions are usually modeled as a sum of $g$, $\mu$, and $y$ distortions plus a residual distortion $R(x)$, which has to be calculated knowing the full thermal history. This can be accomplished using, for example, the Green's function method, as described in Section~\ref{Greens}. For convenience, we shall also assume the residual distortion to be normalized to $\Delta \rho_{\gamma}/\rho_{\gamma} = 1$.
\subsection{Amplitudes of the distortions \label{SD_calc}}
\begin{figure}[t]
	\centering
	\includegraphics[height= 6 cm, width=7.5 cm]{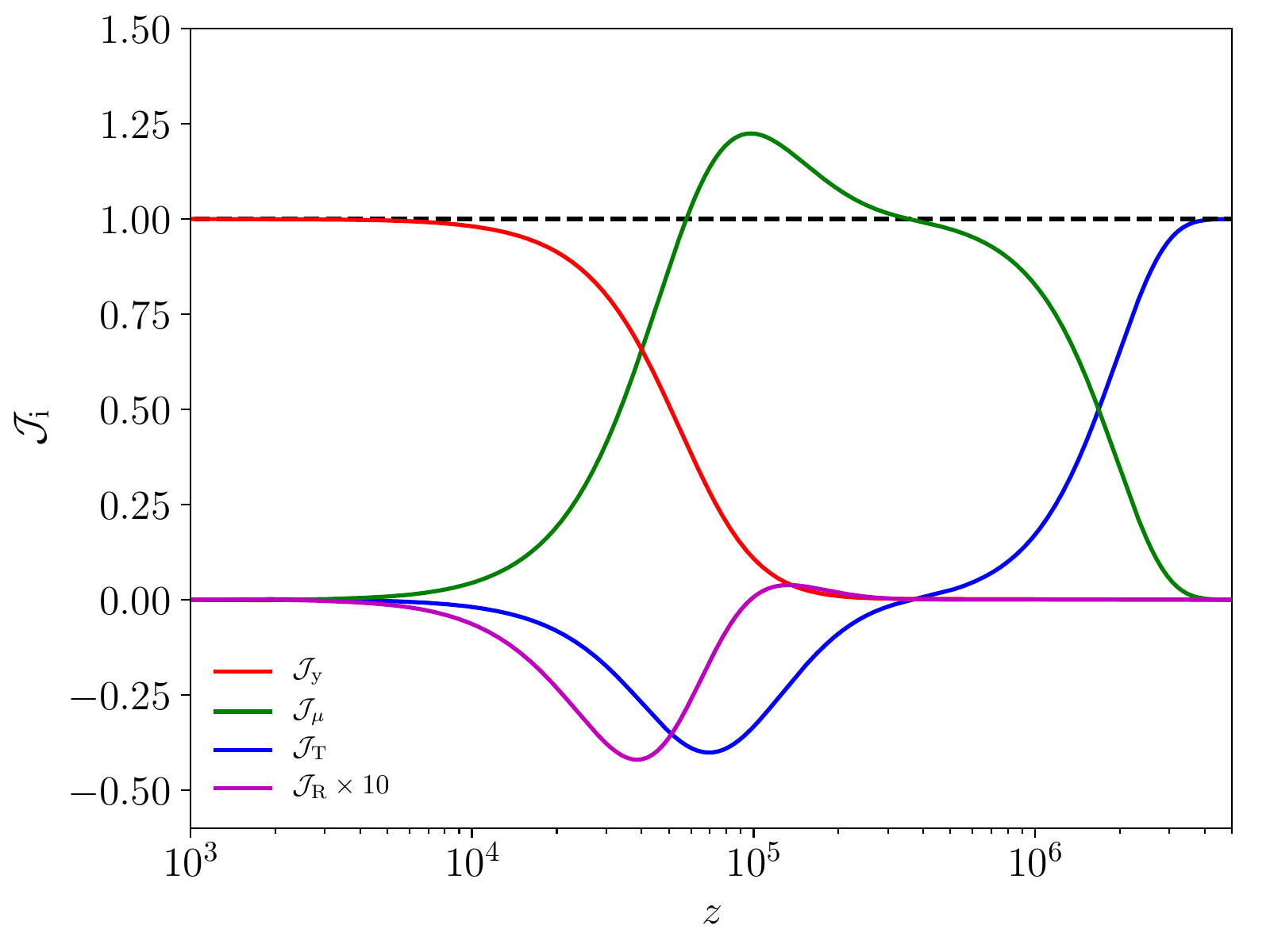}
	\includegraphics[height= 6 cm, width=7.5 cm]{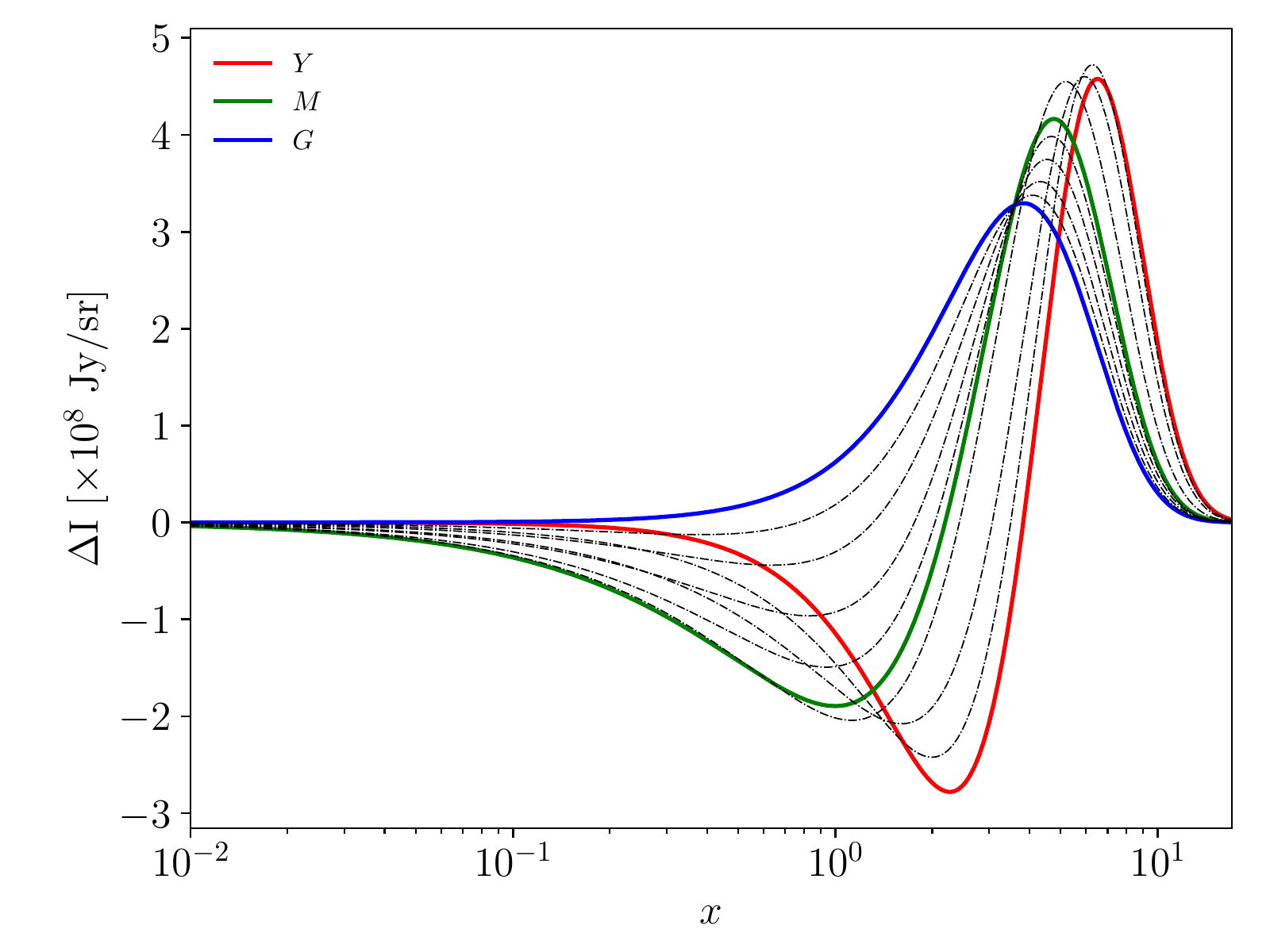}
	\caption{\textbf{Left panel}: Branching ratios of the different \SD types calculated according to Section~\ref{Greens} and defined in \fulleqref{eq: SDs amplitudes definition 1}. The red, green, and blue lines correspond respectively to a pure $y$ distortion, $\mu$ distortion, and \finalr{$g$ distortion}{temperature shift}.The magenta line refers to the contribution from the residuals. 
	\textbf{Right panel}: The corresponding changes in intensity of the photon spectrum given a single \textit{instantaneous} energy injection, plotted for different times of injection.  Since the injection is modelled by a $\delta$-function \final{in redshift}, the Green\finalr{s}{}'s functions of Section~\ref{Greens} correspond to the total \SD. The dashed black lines represent the shapes obtained at redshifts where a mixture of two types of distortions are present. The time of injection is progressing from the blue curve (\finalr{$g$ distortion}{temperature shift}) to the red curve ($y$ distortion).}
	\label{fig: plot_branching_ratios}
\end{figure}

According to previous definitions, the total distortion of the photon intensity spectrum is given at first order by 
\begin{align}\label{eq: DI definition 1}
\Delta I_{\rm tot} = \Delta I_{y}+ \Delta I_{\mu}+ \Delta I_{\rm T}+ \Delta I_{\rm R}~,
\end{align}
where $\Delta I_{y}=\widetilde{y} \mathcal{Y}(x)$ determines the contribution from $y$ distortions, $\Delta I_{\rm \mu}=\widetilde{\mu} \mathcal{M}(x)$ the contribution from $\mu$ distortions, $\Delta I_{\rm T}=\widetilde{g} \mathcal{G}(x)$ the contribution from temperature shift $g$ distortions, and $\Delta I_{\rm R}=R(x) \epsilon$ the contribution from residuals, with $\epsilon$ denoting the energy stored within the residual distortion. For higher precision, the \finalr{$g$ distortion}{temperature shift} can easily be written at second order in \finalr{the temperature shift}{$\widetilde{g}$}, $\Delta I_{\rm T}=\widetilde{g}(1+\widetilde{g}/4)\mathcal{G}(x)+\widetilde{g}^2/8 \,\mathcal{Y}(x)$ (see Appendix \ref{ap:secondorder_g} for more details). By means of the decomposition of the $\Delta I_\mathrm{tot}$ into shapes and amplitudes as in \fulleqref{eq: DI definition 1}, the full knowledge of the distortion is given by a set of four amplitudes ($y$, $\mu$, $g$, and $\epsilon$) and one normalized shape $R(x)$. 
\dnew
Our definitions for the normalization factors match those in Ref.~\cite{Chluba2015IJMPD} and allow to reduce the number of parameters in several equations. In particular, they are such that the relative shift in photon density $\Delta \rho_\gamma/\rho_\gamma$ is given at first order by the amplitudes  $\tilde{y}$, $\tilde{\mu}$, $\tilde{g}$, and $\epsilon$,
\begin{align}\label{eq: delta rho definition 2}
\frac{\Delta\rho_\gamma}{\rho_\gamma}\biggr\rvert_\mathrm{tot}=\left.\frac{\Delta\rho_\gamma}{\rho_\gamma}\right|_{y}+\left.\frac{\Delta\rho_\gamma}{\rho_\gamma}\right|_{\mu}+\left.\frac{\Delta\rho_\gamma}{\rho_\gamma}\right|_{g}+\left.\frac{\Delta\rho_\gamma}{\rho_\gamma}\right|_{R}=\widetilde{y}+\widetilde{\mu}+\widetilde{g}+\epsilon~,
\end{align}
where the indices $y$, $\mu$, $g$, and $R$ refer to the corresponding fractions of the total injected energy that generate the given distortion.
\dnew
\final{To calculate each of these amplitudes, we need to know how the \ppsd changes throughout the thermal history. The collision operator $C[f] = C[f(t,x)]$ that appears in the photon Boltzmann equation of Appendix \ref{eq:photonboltzmann} accounts precisely for  this. To calculate the final spectrum of photons after all injections, one would in principle need to know the full spectral dependence of the injection/collision term \cite{Chluba:2015hma}. Then the Boltzmann equation would tell us how photon momenta get redistributed or new photons are added, and how the \ppsd evolves as a function of the injected spectrum $C[f]$.
\dnew
However, in many interesting cases, the redistribution of photons is quick and efficient enough to erase any dependency of the \ppsd on the spectral shape of the injected energy spectrum $C[f]$. In this case, what matters is just how much energy is injected in total, and when it is injected \cite{Chluba:2015hma}. This applies to any injection occurring at dimensionless frequencies $x<10^{-4}$, because the \BS and \DC processes introduced in Section \ref{Theory} are very efficient at such frequencies at any point in the history of the universe. The same is true also at high redshifts ($z>10^3$) and high frequencies ($x>10^{-1}$), where the redistribution processes of Section \ref{Theory} are efficient enough to impose a precise shape for the spectrum $f(t,x)$, known up to a normalization factor.
At each redshift, this precise shape is a combination of the three basic shapes introduced in Section~\ref{SD_shapes}. We show it for different redshifts on the right panel of Figure~\ref{fig: plot_branching_ratios}. Details on this calculations can be found in~\cite{Chluba:2015hma}. In this regime, all that is required to calculate the normalisation factor and infer the \ppsd is the frequency-integrated heating rate $\dot{Q}(z)$. It is defined as the energy-weighted integral over the collision operator
\begin{equation}
\dot{Q} \equiv \int C[f] E \, \mathrm{d}^3p~.
\end{equation}
One can then show (Appendix~\ref{ap:drho_rho}) that the total energy density injected into photons is related to the heating rate through
\begin{align}\label{eq: rhotot from heating}
\frac{\Delta \rho_{\gamma}}{\rho_{\gamma}}\biggr\rvert_\mathrm{tot} = \int\limits_z^\infty \frac{\dot{Q}}{(1+z)H\rho_{\gamma}} \mathrm{d}z~.
\end{align}
\dnew
The exact approach based on the knowledge of the full photon injection spectra $C[f]$ could give results slightly different from the simplified approach based on $\dot{Q}(z)$ at intermediate frequencies and after recombination, because in this regime \CS can re-scatter a fraction of the total \ppsd (see Equation (30) of \cite{Chluba:2015hma}). However, the actual frequency dependence of the energy injection term (that should account for the full spectrum of primary and secondary particles which inject energy into the thermal plasma) is currently not well understood, except in a few particular cases. For instance, the case of injections in the Rayleigh-Jeans tail of the CMB spectrum has been investigated in \cite{pospelov2018new}.
\dnew
Luckily, the injection models considered in this work involve frequencies much higher than that of typical CMB photons, $x\gg1$. Thus, energy gets efficiently redistributed at early times. At late times, the effects of reionization are anyway much more significant, and happen in hot clusters where Comptonization is once again efficient. This is why, in this paper as well as in most of the literature, one uses the simplified approach involving only the frequency-integrated heating rate $\dot{Q}(z)$.  For alternative approaches including the full frequency dependence see \cite{Chluba:2015hma}. A more exhaustive study including the evolution of the full photon energy spectrum for given injection spectra is left for future work.
\dnew
Additionally, the assumption of $\Delta \rho_{\gamma}/\rho_{\gamma} \ll 1$ implies that the problem can be linearized \cite{chluba2013green}, and treated with a Green's function approach (see Appendix \ref{ap:model} for further details).}
\dnew
During different eras of the thermal evolution of the universe any energy injection will be differently redistributed depending on the availability of number count changing processes and the efficiency of \CS. To quantify which part of the injected energy generates each of the distortions, we define for each distortion type $a$ the branching ratio of deposited energy into the distortion, such that
\begin{align}\label{eq: SDs amplitudes definition 1}
a = \left.\frac{\Delta \rho_\gamma}{\rho_\gamma}\right|_a \equiv \int \frac{\id Q/\id z}{\rho_{\gamma}} \,\cdot\, \mathcal{J}_{a}(z)\mathrm{d}z~,
\end{align}
where the branching ratio $\mathcal{J}_{a}(z)$ determines the fractional energy release into a given distortion $a$ as a function of redshift. Here we have used the relation (see Appendix~\ref{ap:drho_rho} for additional clarification)
\begin{equation}
\frac{\dot{Q}}{(1+z)H\rho_{\gamma}}=-\frac{\mathrm{d}Q/\mathrm{d}z}{\rho_{\gamma}}~
\end{equation}
to recover an expression for the heating rate similar to the one employed in \fulleqref{eq: rhotot from heating}. 
\dnew
In this way, we have effectively split the problem into the model-dependent heating function $\mathrm{d}Q/\mathrm{d}z$ and the model independent branching ratios $\mathcal{J}_a(z)$. To find the precise values of the branching ratios, multiple approaches can be taken. Appendix~\ref{ap:br} deals with different common approximations to the branching ratios, and the left panel of Figure~\ref{fig: plot_branching_ratios} displays the results of a quasi-exact calculation based on the Green's function method (see Section~\ref{Greens}).
\newpage 
The three main eras visible in the left panel of Figure~\ref{fig: plot_branching_ratios} are the $y$, $\mu$, and $g$ eras. For redshifts higher than
\begin{equation}\label{eq:z_th}
z_{\rm th} \equiv 1.98\times10^6\left(\frac{1-Y_{\rm He}/2}{0.8767}\right)^{-2/5}\left(\frac{\Omega_bh^2}{0.02225}\right)^{-2/5}\left(\frac{T_0}{2.726\text{ K}}\right)^{1/5} \approx 2 \cdot 10^6~,
\end{equation}
most of the injected energy tends to fully thermalize as the number count changing processes of \DC and \BS are very efficient \cite{burigana1991formation,hu1993thermalization}. Hence, mostly temperature shift\final{s} \finalr{$g$ distortions}{} will be caused, and the era is named the $g$ or thermal era. For redshifts between $z = z_{\mu y} \approx 5 \times 10^4$ and $z = z_\mathrm{th}$, the number count changing processes become inefficient, while \CS is still efficient. This is the so-called $\mu$ era, during which the dominant contribution will be a $\mu$ distortion. The final era is the $y$ era, where \CS is inefficient and the injected energy is only partially redistributed, so that a $y$ distortion is created. This era lasts between $z = z_{\mu y}$ and today. Finally, the residual distortions $R(x)$ account for deviations from this simplified picture. The corresponding shapes of the distortions at different times can be seen in the right panel of Figure~\ref{fig: plot_branching_ratios}.
\dnew
\finalr{
As a final note, we want to mention that one can go beyond a frequency independent heating function $\dot{Q}(z)$ and instead consider a general photon injection $S(\nu,z)$ which can have any kind of frequency dependence \cite{Chluba:2015hma}. However, the frequency dependence can be ignored for $x<10^{-4}$ as the \BS and \DC efficiently convert photons into heat at any point in the history of the universe. The same is true also at high redshifts ($z>10^3$) and high frequencies ($x>10^{-1}$), where the efficient redistribution processes happen on much faster time-scales than the expansion of the universe, as \cite{Chluba:2015hma} pointed out.
\dnew
On the other hand, the exact approach based on a general photon injection $S(\nu,z)$ could give different results from the simplified approach based on $\dot{Q}(z)$ at intermediate frequencies and after recombination, where the \CS can re-scatter a fraction of the total \ppsd (see Equation (30) of \cite{Chluba:2015hma}). As an example, the case of injections in the Rayleigh-Jeans tail of the CMB spectrum has been further investigated in \cite{pospelov2018new}.
\dnew
However, the actual frequency dependence of the energy injection term (that should account for the full spectrum of primary and secondary particles that inject energy into the thermal plasma) is currently not well understood. This is why, in this paper as well as in most of the literature, one uses the simplified approach involving only the frequency-integrated heating contribution $\dot{Q}(z)$. Luckily, the energy injection models considered in this work happen at frequencies much higher than that of the CMB and are thus efficiently redistributed at early times. Moreover, at late times the effects of reionization are anyway much more significant, and happen in hot clusters, where Comptonization is once again efficient. For alternative approaches including the full frequency dependence see \cite{Chluba:2015hma}.}{}

\subsection{Causes of the distortions \label{SD_causes}}
As shown in the previous section, the magnitude of the final observed \SD{s} has a complete and unique dependence on the heating history of the universe, which can be parameterized using the heating rate $\dot{Q}$. To better understand \final{how to calculate }this heating rate, we start with a general discussion regarding the difference between injected and deposited energy in Section~\ref{ssec:injdep}, and then focus on energy deposition into heating in Section~\ref{ssec:dep_heat}.
Furthermore, in Section~\ref{ssec:heatingLCDM} we discuss the different injection mechanisms predicted by the standard $\Lambda$CDM model. This catalogue relies on the work of many recent publications like \cite{chluba2011evolution, chluba2013distinguishing, chluba2016spectral}. Finally, in Section~\ref{ssec:heatingExotic} we additionally discuss a few of the most common non-standard injection mechanisms. 

\subsubsection{Injection and deposition} \label{ssec:injdep}
The energy injection into the intergalactic medium (IGM) through various processes does not necessarily immediately heat the IGM and the photon bath. As such, we differentiate energy injection, energy deposition, and various deposition channels. The injected energy is the energy released by a given process. The deposited energy is the fraction of this energy that eventually affects the medium after the radiative transfer and electron cooling. The deposition channels (labelled by an index $c$) describe the final impacts on the IGM.
\dnew
The deposition \textit{function} $f_c(z)$ represents the fraction of injected energy that is deposited in channel $c$ at redshift $z$. It can be decomposed into an injection efficiency function $f_{\rm eff}(z)$ and a deposition \textit{fraction} $\chi_{c}(z)$, with all deposition fractions across all channels summing up to one, $\sum_c \chi_{c}(z)=1$. The deposition fraction usually depends only on the free electron fraction $x_{e}$ at a given redshift, and can thus be written as $\chi_{c}(x_e(z))$. In summary, the injection and deposition rates are related through
\begin{equation}\label{eq:definition_E_dep}
\left.\frac{\id E}{\id t\id V}\right|_{\mathrm{dep, c}} = \left.\frac{\id E}{\id t\id V}\right|_{\mathrm{inj}} f_\mathrm{c}= \left.\frac{\id E}{\id t\id V}\right|_{\mathrm{inj}} f_\mathrm{eff}\,\, \chi_c\, \equiv \mathcal{\dot{Q}} \,\chi_c~,
\end{equation}
where we have defined the effective rate of energy injection $\mathcal{\dot{Q}}$ as a useful shorthand. It should not be confused with $\dot{Q}$, which is the effective heating term (see also \fulleqref{eq: qdef}).
\dnew
The so-called injection efficiency $f_{\rm eff}$ determines how much of the heating is deposited at all, regardless of the form. In general, this function depends on the emitting process and on the characteristics of the universe, such as transparency and energy densities at the time of emission  (see for example \cite{slatyer2009cmb, slatyer2013energy} for recent overviews). For instance, a given process may emit not only particles interacting electromagnetically with the medium, but also neutrinos that do not affect the surrounding environment at all. 
\dnew
Furthermore, the injection efficiency $f_{\rm eff}$ does not necessarily coincide with the full fraction of electromagnetically released energy $f_{\rm em}$\,, as particles emitted at one moment in the history of the universe might lose energy through redshifting or secondary interactions before effectively depositing their energy into the medium. For this reason, we 
define the fraction of electromagnetic energy lost before the deposition, $f_{\rm loss}$\,, such that $f_{\rm eff}=f_{\rm em}(1-f_{\rm loss})$. Both $f_{\rm em}$  and $f_{\rm loss}$ vary in the range from 0 to 1. Note that, although $f_{\rm loss}$ might be relevant in the so-called dark ages when the particle density is very low, it is not relevant in high density environments, such as in the pre-recombination plasma, when scatterings are so frequent that particles do not have enough time to lose a significant amount of energy between injection and deposition. Therefore, a common approximation, called \textit{on-the-spot}, assumes the deposition to be instantaneous, and thus sets $f_{\rm loss}=0$. A detailed calculation of these quantities can be performed with tools like {\sc DarkAges} \cite{stocker2018exotic} or {\sc DarkHistory} \cite{liu2019darkhistory}.
\dnew
Next, the deposition fraction $\chi_{c}$ partitions the deposited energy into different channels~$c$ depending on their main impact on the thermal bath. 
For the calculation of SDs we are only interested in the channel corresponding to the heating of the photon bath and intergalactic medium, but for other purposes, like the study of recombination, many other channels play a role. In general, the deposited energy may also ionize hydrogen and helium atoms, or play a role for the excitation of the different transitions of hydrogen (the one with the biggest impact being the Lyman-$\alpha$ transition).
Furthermore, some energy could even be lost into photons with too low energies to initiate atomic reactions. Several models with different levels of approximations have been proposed during the last few decades to define how much of the injected energy affects each scenario, and more details on the representative cases \cite{chen2004particle, padmanabhan2005detecting, galli2013systematic} are given in Appendix~\ref{ap:f_eff_chi}. In this paper in all the curves of Section \ref{SD_causes} and all the analyses of Section \ref{results}, we always use the $\chi_c$ from Table V of \cite{galli2013systematic} (from now on labeled GSVI2013) described further in Appendix~\ref{ap:f_eff_chi}.

\subsubsection{Energy deposition into heat} \label{ssec:dep_heat}
When investigating the impact of an energy injection on \SD{s}, the fraction of energy deposited in the form of heat plays a particularly important role. In fact, as already shown in Section~\ref{SD_calc}, this quantity is intrinsically linked with the amplitude of the final distortion. In the next few paragraphs we will underline some particular aspects that are important for later discussions.
\dnew
First of all, it is useful to differentiate between two kinds of heating: the heating of the baryons and the heating of the photons. In both cases, the most general approach to account for the presence of energy injections would be to evaluate their effects on the evolution of the photon/matter temperature $T_{\gamma/m}$\,. 
\dnew
\final{For photons, one can treat any deviation from a \BB spectrum with a temperature scaling like the reference temperature $T_z$ as a distortion. These distortions always remain small, since the injected energy is always much smaller than the total energy of the photon bath, i.e. $\Delta \rho_{\gamma}/\rho_{\gamma} \ll 1$. Even when no energy is injected into the IGM and photon bath, distortions can be generated by an internal redistribution among photon momenta, or by energy and momentum exchange between photons and baryons. Examples are provided by the adiabatic cooling of electrons and baryons, and by the dissipation of acoustic waves (see Section~\ref{ssec:heatingLCDM} for more details).  In that case, equation (\ref{eq: rhotot from heating}) features a contribution to the rate $\dot{Q}$ despite the fact that there is no actual energy injection: we will call it the non-injected heating rate $\dot{Q}_\mathrm{non-inj}$\,. Summing up  this contribution with the actual energy injection rate in the form of heat defined in  Section~\ref{SD_calc}, we can express the effective deposition rate in the form of heat as
	\begin{equation} \label{eq: qdef}
	\dot{Q}=\left.\frac{\id E}{\id t \id V}\right|_{\mathrm{dep},h}+\dot{Q}_{\rm non-inj}= \mathcal{\dot{Q}} \, \chi_{h}+\dot{Q}_{\rm non-inj}~.
	\end{equation}
}
\dnew
Note that at early times the \textit{on-the-spot} approximation is valid and the entire injected energy is deposited in the form of heat, as clear from Appendix~\ref{ap:f_eff_chi} and particularly \fulleqref{eq:dep_func}. For this reason, and since we are primarily interested in the pre-recombination generation of \SD{s}, in the following discussions it will often be possible at early times to employ the approximation $\dot{Q}\approx\dot{\mathcal{Q}}+\dot{Q}_{\rm non-inj}$\,. 
\dnew
For the baryons, on the other hand, the full temperature evolution is calculated assuming a Maxwellian phase space distribution (see Section \ref{class}). Due to the very strong and poorly constrained galactic influences, however, the calculation is still very uncertain. Explicitly, we do not even attempt to define or calculate the \SD{s} of the baryon phase space distribution. An improved treatment of the baryon thermal evolution is left for future work.

\subsubsection{Heating mechanisms in $\Lambda$CDM \label{ssec:heatingLCDM}}

\begin{figure}[t]
\centering
\begin{minipage}{\textwidth}
\includegraphics[width=7.5 cm, height=5.5 cm]{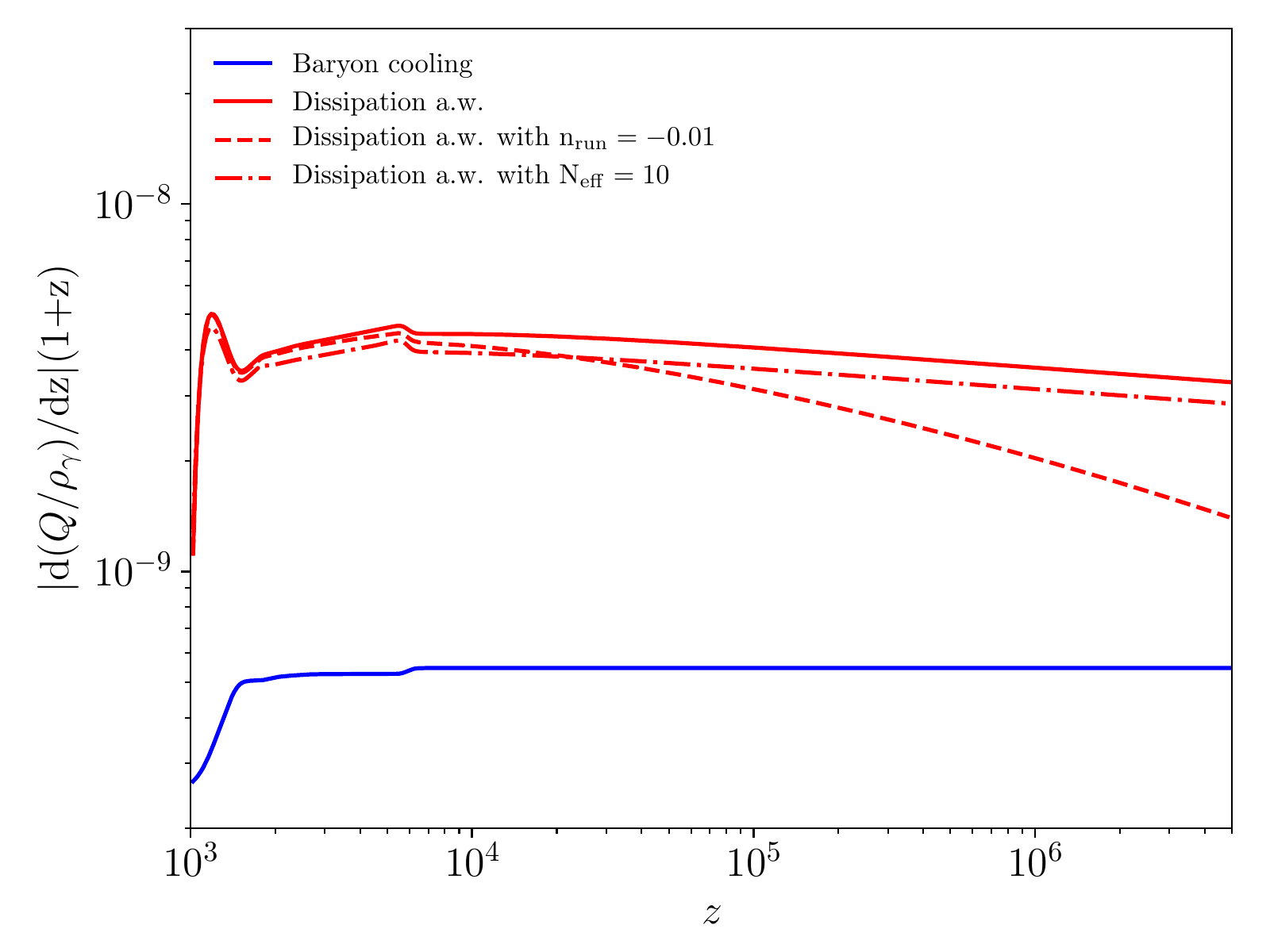}
\includegraphics[width=7.5 cm, height=5.5 cm]{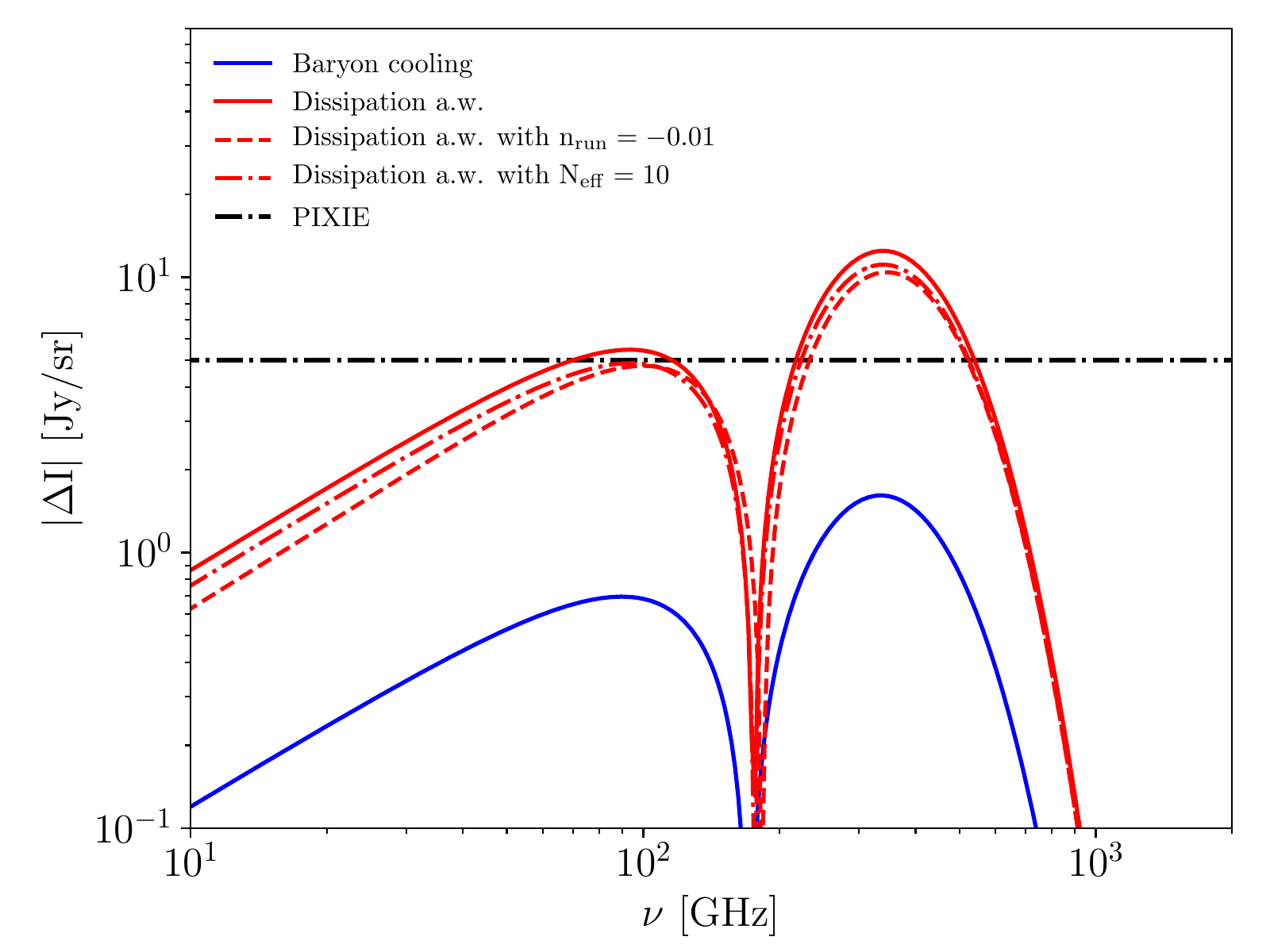}
\end{minipage}
\caption{Heating rate (left panel) and SDs (right panel) caused by the adiabatic cooling of electrons and baryons (blue line) and by the dissipation of acoustic waves (red line). The difference between the shape of the solid and dashed red lines reflects the influence of the primordial power spectrum on the heating rate. We changed the value of $n_{\rm run}$ from $0$ to $-0.01$ as a representative example. Similarly, the difference between the shape of the solid and dashed-dotted red lines reflects the impact of $N_{\rm eff}$ on the heating rate, where we increased the value of $N_{\rm eff}$ from 3.046 to the extreme value 10. In the right panel, the black dot-dashed line represents the predicted PIXIE sensitivity. Note that the absolute values of each quantity are plotted so that the curve describing the baryon cooling also appears positive, even though the actual contribution is negative.}
\label{fig: heating_LCMD}
\end{figure}

\ssec{Adiabatic cooling of electrons and baryons}
If the interaction with the CMB photons can be neglected, the temperature of non-relativistic matter\footnote{Comoving number density conservation gives $\mathrm{d}(a^3n) = 0$, and from the first law of thermodynamics one derives $\mathrm{d}(\rho a^3) = - p \mathrm{d}V$. Inserting the expression of $\rho$ and $p$ for a non-relativistic species at first order in $T/m$ gives $\mathrm{d}(na^3 \cdot( m + 3/2\, T)) = - nT \mathrm{d}(a^3)$. We then find $na^3 \cdot 3/2 \,\mathrm{d}T = - 3 n a^3 T \mathrm{d}a/a$, and thus $\mathrm{d}T/T = -1/2\, \mathrm{d}a/a$, and finally $T \propto a^{-2} \propto (1+z)^2$ \cite{roos2015introduction}.} scales as $T_{m} \propto (1+z)^2$, while the photon temperature scales roughly as $T_{\gamma} \propto (1+z)$. At very low redshifts ($z < 200$), when \CS becomes inefficient, this difference in the adiabatic index of baryonic matter and radiation leads to a significant difference in the CMB and matter temperatures, with $T_{m} < T_\gamma$ \cite{zeldovich1969interaction}. However, at higher redshifts the CMB photons are tightly coupled to baryons. This implies that the baryonic matter in the Universe must continuously extract energy from the CMB in order to establish $T_m \approx T_\gamma$\,. As a consequence of this energy extraction, photons shift towards lower energies \cite{chluba2005spectral, chluba2011evolution}. 
\dnew
In the steady state approximation \cite{scott2009matter} the cooling rate associated to this process can be determined as
\begin{align}\label{eq: heating rate adiabatic cooling 1}
\dot{Q}_{\rm non-inj}=-H\alpha_{h}T_{\gamma}~,
\end{align}
where we define the heat capacity of the intergalactic medium \cite{seager2000exactly,chluba2005spectral,chluba2011evolution} as
\begin{align}\label{eq:heat_capacity}
\alpha_{h}  = \frac{3}{2} n_\mathrm{bar} = \frac{3}{2} (n_\mathrm{H}+n_e+n_\mathrm{He}) =\frac{3}{2} n_\mathrm{H} (1+x_{e}+f_\mathrm{He})~,
\end{align}
where $n_\mathrm{bar}$ is the number density of all baryonic constituents of the IGM and $f_{He}=n_\mathrm{He}/n_H$ is the relative abundance of He to H. \enlargethispage{1\baselineskip}
\dnew
The evolution of the heating rate expressed in \fulleqref{eq: heating rate adiabatic cooling 1} can be seen in the left panel of Figure~\ref{fig: heating_LCMD} as a blue line. Note that the process described here \textit{extracts} energy from the system, so that the net heating is negative, while in Figure~\ref{fig: heating_LCMD} the absolute value is plotted. The same is true for the \SD{s} parameters $y$ and $\mu$ resulting from this process\footnote{Although $\tilde{\mu}$ and $\tilde{y}$ are used within this work to simplify the equations, for a more direct comparison to previous literature, we quote here the $\mu$ and $y$ values.}, which take the approximate values of $-5\times10^{-10}$ and $-3\times10^{-9}$. The shape of the corresponding \SD{s} is displayed in the right panel of Figure~\ref{fig: heating_LCMD}. In fact, as opposed to the case of positive contributions, for adiabatic cooling the low frequency peak is the positive one, whereas the high frequency peak is negative.

\ssec{Dissipation of acoustic waves}
In the early universe, the presence of primordial density fluctuations causes some regions of space to be hotter and denser than others. At the approach of decoupling, the photon mean free path increases, such that photons diffuse from overdense to underdense regions and vice-versa. This random process leads to an isotropization of the \ppsd, and thus to an erasure of density perturbations that is called diffusion damping or Silk damping~\cite{silk1968cosmic}. This diffusion leads to a superposition of BB spectra with slightly different temperatures which causes \SD{s} \cite{Sunyaev1970diss, daly1991spectral, barrow1991jd}.
\dnew
The first comprehensive calculation of the consequent heating rate was performed in~\cite{chluba2012cmb}, where the photon Boltzmann equation was calculated at second order in cosmological perturbation theory and second order in the energy transfer by \CS. The results \final{in a flat FLRW universe} can be summarized in the following heating rate \cite{khatri2012mixing,chluba2012cmb}
\begin{align}\label{eq: heating rate acoustic waves 1}
\dot{Q}_{\rm non-inj}=4\dot{\tau}\rho_{\gamma}\int \frac{\mathrm{d}k k^2}{2\pi^2} P_{\mathcal{R}}(k)\left[\frac{(v_\gamma-v_{\rm b})^2}{3}+\frac{9}{2}\Theta_{2}^{2}-\frac{1}{2}\Theta_{2} (\Theta_{0}^{\rm P}+\Theta_{2}^{\rm P})+\sum_{\ell\geq 3}(2\ell+1)\Theta_{\ell}^{2}\right]~.
\end{align}
Here $P_{\mathcal{R}}(k)$ refers to the primordial power spectrum, $v_\gamma$ and $v_{\rm b}$ are respectively the electron and photon longitudinal velocity. \final{$\Theta_\ell(k, z)$ is the transfer functions of the $\ell^{\rm th}$ photon temperature Legendre multipole moment, related to the Fourier-expanded temperature anisotropy $\Theta(\textbf{k}, \hat{n})$ through
\begin{align}
\Theta(\textbf{k}, \hat{n})=\sum_{\ell=0}(-\mathrm{i})^{\ell}(2 \ell+1)\Theta_\ell(k) P_{\ell}(\hat{k} \cdot \hat{n})\,.
\end{align}
$\Theta_{\ell}^{\rm P}(k, z)$ is the transfer functions of the polarization multipole moments, related to the Stokes parameter $Q(\textbf{k}, \hat{n})$ in the same way. These transfer functions relate to those of Ma \& Bertschinger \cite{Ma:1995ey} through $\Theta_{\ell}=F_\ell/4$ and $\Theta_{\ell}^{\rm P} = G_\ell/4$.
Additional polarization corrections were originally introduced in \cite{Chluba2015} and later generalized in \cite{Pitrou:2019hqg}, but will be omitted here.}
\dnew
To simplify \fulleqref{eq: heating rate acoustic waves 1}, as done in \cite{chluba2013cmb}, we can employ the tight-coupling approximation, i.e. $v_{\rm b} \approx v_\gamma$, $\Theta_{\ell\geq 2}\approx0$, $\Theta^{\rm P}_{\ell\geq 0}\approx0$, and on subhorizon scales $\Theta_1$ can be inferred from the the approximate WKB solution
\begin{align}
v_\gamma/3 = \Theta_1\approx A\frac{c_{\rm s}^{2}}{(1+R)^{1/4}}\sin(kr_{\rm s})e^{-(k/k_{\rm D})^2}~.
\end{align}
The normalization of the transfer function $v_\gamma$ to adiabatic initial conditions with ${\cal R}=1$ gives
\begin{align}
A\approx\left(1+\frac{4}{15}f_\nu\right)^{-1} ~,
\end{align}
where $f_\nu = \rho_\nu/\rho_r$ is the ratio between the energy density of neutrinos and of the total relativistic species (see \cite{diacoumis2017using} for a more detailed discussion), $R=4\rho_{\gamma}/(3\rho_b)$ is the ratio between baryon and photon energy density (equal to zero when neglecting baryon loading), $c_{\rm s}$ is the sound speed of the fluid, $r_{\rm s}(z)$ is the comoving sound horizon, and $k_{\rm D}(z)$ is the comoving damping scale. As a result, we can reduce \fulleqref{eq: heating rate acoustic waves 1} to
\begin{align}
\dot{Q}_{\rm non-inj}=8A^2\rho_{\gamma}\int \frac{\mathrm{d}k k^2}{2\pi^2} P_{\mathcal{R}}(k)\sin^2(kr_{\rm s})k^2(\partial_t k_{\rm D}^{-2})e^{-2(k/k_{\rm D})^2}~.
\label{eq: heating rate acoustic waves 2}
\end{align}
For a definition for the damping scale we follow \cite{hu1995wandering}, i.e.
\begin{align}
k_{\rm D}=\frac{2\pi}{r_{\rm D}}=2\pi\left[\int \mathrm{d}z \frac{c_{\rm s}^{2}}{2\dot{\tau}H} \left(\frac{R^2}{1+R}+\frac{16}{15}\right)\right]^{-1/2}~.
\end{align}
Note that since the shape of the primordial power spectrum is assumed to be very smooth, one can make use of the average of the quickly oscillating sine over many periods $\langle \sin^2 (kr_{\rm s} )\rangle~=~1/2$, and hence
\begin{align}
\dot{Q}_{\rm non-inj}=4A^2\rho_{\gamma}\int \frac{\mathrm{d}k k^2}{2\pi^2} P_{\mathcal{R}}(k)k^2(\partial_t k_{\rm D}^{-2})e^{-2(k/k_{\rm D})^2}~.
\label{eq: heating rate acoustic waves 3}
\end{align}
The evolution of this heating rate is displayed in the left panel of Figure~\ref{fig: heating_LCMD} as a red line. 
\dnew
By inserting \fulleqref{eq: heating rate acoustic waves 3} in the definition~\eqref{eq: SDs amplitudes definition 1}, it possible to show that the $y$ and $\mu$ parameters have a value of approximately $4\times10^{-9}$ and $2\times10^{-8}$, respectively. The corresponding \SD{s} calculated with \fulleqref{eq: DI definition 1} are shown in the right panel of Figure~\ref{fig: heating_LCMD} as a red line. We also show there as a horizontal line the expected PIXIE sensitivity, to stress the fact that with the current technological status it would in principle already be possible to observe \SD{s} generated before recombination.
\dnew
The dependence on the primordial power spectrum allows us to use \SD{s} caused by the dissipation of acoustic waves to constrain the running of the spectral index $n_\mathrm{run}$, as schematically shown in Figure~\ref{fig: heating_LCMD} (dashed red line). Additionally, the dependence of the amplitude~$A$ on the neutrino energy density fraction $f_\nu$ allows a weak constraint on $N_\mathrm{eff}$ as well, since we can write
\begin{equation}
	f_\nu = \frac{\rho_\nu}{\rho_r} = \left(1+\frac{1}{\alpha(N_\mathrm{eff})}\right)^{-1}~,
\end{equation}
where $\alpha(N_\mathrm{eff}) = \rho_\nu/\rho_{\gamma} = (7/8) \left(4/11\right)^{1/3}N_\mathrm{eff}$. However, changing $N_\mathrm{eff}$ simply results in an overall change of the heating and \SD{} amplitude, which is degenerate with the amplitude $A_s$ of the primordial power spectrum $P_{\mathcal{R}}(k)$. 
\dnew By combining with CMB anisotropy constraints, this degeneracy can in principle be broken. However, even with very futuristic \SD{s} measurements, it would be difficult to supersede current $N_\mathrm{eff}$ constraints from Planck \cite{Aghanim:2018eyx} or future CMB anisotropy surveys.
\dnew
Primordial scalar perturbations are not the only source of \SD{s} through dissipation of acoustic waves. Since the effect is at second order in perturbation theory, tensor modes can additionally source it \cite{Ota2014, Chluba2015}. However, due to the tight constraint of the tensor-to-scalar ratio $r_{0.002}\lesssim0.1$ \cite{Aghanim:2018eyx,Akrami:2018odb}, we expect such a contribution to be subdominant, unless the tensor fluctuations are significantly enhanced on scales $k \ll 0.002\, h$Mpc$^{-1}$ and do not follow a simple nearly scale-invariant power spectrum. In principle, \SD{s} could be sensitive to tensor fluctuations with wavenumber as large as $k\simeq 10^{7}-10^{8}\,{\rm Mpc^{-1}}$ \cite{Chluba2015}. We thus leave a more detailed analysis for future work.

\ssec{Cosmological recombination radiation}
Another source of \SD{s} is given by the so-called cosmological recombination radiation (CRR) \cite{zeldovich1969interaction, peebles1968recombination, Dubrovich1975} (see \cite{sunyaev2009signals, Glover2014} for pedagogical reviews, and \cite{chluba2016cosmospec} for the most recent calculations). As the name suggests, this effect is driven by the emission and absorption of photons due to the recombination of H and He, relevant in the redshift intervals $500<z<2000$ for HII$\to$HI, $1600<z<3500$ for the transition HeII$\to$HeI, and $5000<z<8000$ for the transition HeIII$\to$HeII. These processes occur when \CS is gradually becoming inefficient, thus resulting in residual distortions on top of $y$ distortions. These residual distortions are dubbed CRR peaks (see e.g., Figure~1 of \cite{chluba2006free} for a representative example).
\dnew
It is interesting to note that, although He makes up only a small fraction of the total matter content, its contribution cannot be overlooked (see e.g., Figure~1 of \cite{rubino2008lines} an quantitative example). Indeed the role of He is enhanced compared to naive expectations for three reasons.
Firstly, there are two epochs of helium recombination. Secondly, HeII recombination occurred when the photon-baryon plasma was still in thermal equilibrium, so that the recombination process followed the Saha solution more closely. Thirdly, the number of photons related to helium atoms is enhanced by detailed radiative transfer effects and feedback processes \cite{Chluba2009c, Chluba2012HeRec}, overall resulting in several emitted photons per helium nucleus from recombination \cite{Chluba2009c}. As a consequence, the emission lines are more sharply peaked and can thus reach higher amplitudes and change the shape of the broader H lines considerably.
\dnew
Of course the emission processes strongly depend on the characteristics of the plasma they are taking place in, such as the photon temperature and baryon fraction. Therefore, as discussed for instance in \cite{chluba2008there}, the eventual observation of the CRR peaks would provide an additional and independent test of the parameters describing the standard cosmological model. Furthermore, \cite{chluba2009pre} argues that the influence of possible exotic energy injections might be analyzed through the shape of the CRR spectrum (see, e.g., Figure~5 therein). In contrast to the standard $y$ distortions, CRR-induced distortions have the additional advantage that their characteristic shape remains unchanged between the end of Hydrogen recombination and today, simplifying thus the extraction of information from an eventual observation \cite{chluba2009pre}.
\dnew
Today it is possible to predict the amplitude of the \SD{s} caused by the CRR extremely precisely, using tools such as {\sc CosmoRec} \cite{chluba2011towards} and {\sc CosmoSpec} \cite{chluba2016cosmospec}. The total contribution to the final distortion is then calculated to be roughly $\Delta I_\mathrm{tot}\approx 0.01-1$~Jy/sr. Although CRR is the smallest of the \lcdm contributions to the $\Delta I_\mathrm{tot}$\,, the precision required to observe it might be within reach with futuristic detectors \cite{Vince2015, Mayuri2015}.
\dnew
As a final remark, note that, among all the effects mentioned in Sections~\ref{ssec:heatingLCDM} and \ref{ssec:heatingExotic}, this is the only guaranteed contribution to the final distortion shape that is not currently implemented in our code and left as future work.

\ssec{CMB multipoles}
It is well known \cite{Zeldovich1972Sup, chluba2004superposition, Stebbins2007} that the sum of \BB{}s of different temperature is not in itself a \BB. One particularly important example of varying temperature is given by the CMB multipoles. For instance, {\sc COBE/FIRAS} measured a difference of $3.381\pm0.007$ mK between the all-sky average and the dipole temperature. For the dipole, this temperature difference arises from the earth's movement relative to the CMB rest frame. Under different angles, through the relativistic Doppler effect one observes CMB photons blueshifted (in the direction of motion) or redshifted (opposite to the direction of motion), and thus with different temperatures. Furthermore, even the definition of the all-sky averaged temperature $T_\mathrm{ref}$ will no longer directly correspond to the intrinsic temperature $T_\gamma(z=0) \approx T_0$\,, and induce a temperature shift at second order in $\Delta T/T_0$ \citep{chluba2004superposition}.
\dnew
The angle-dependent temperature of incident photons for an observer moving through the CMB can be calculated through the relativistic Doppler effect to be
\begin{align}\label{eq: definition temperature BB 1}
T(\cos\vartheta)=\frac{T_0}{\gamma[1-\beta\cos\vartheta]}~,
\end{align}
with the observer's relativistic velocity $\beta$, corresponding Lorentz factor $\gamma$,  and angle $\vartheta$ between the Earth's velocity vector and the line of sight direction.
\dnewnoindent
The corresponding full-sky average temperature $T_\mathrm{ref}$ can be computed as \cite{Chluba2011ab}
\begin{equation}
\begin{aligned}
T_\mathrm{ref} = \frac{1}{4\pi} \int_{0}^{2\pi} \mathrm{d}\phi \int_{0}^{\pi} \mathrm{d}\vartheta \sin \vartheta \,\, T(\cos \vartheta)
=\frac{1}{2}\int^{1}_{-1}T(\cos\vartheta)\mathrm{d}\cos\vartheta =\frac{T_0}{2\gamma\beta}\ln \left(\frac{1+\beta}{1-\beta}\right)~.
\end{aligned}
\end{equation}
As such, for every angle $\vartheta$ we obtain some deviation of the temperature from the reference temperature, which according to Appendix~\ref{ap:secondorder_g}, gives rise to distortions of the size 
\begin{equation}
	I(\vartheta)-I_\mathrm{ref} = \epsilon(\vartheta) (1+\epsilon(\vartheta)) \mathcal{G}(x) + \epsilon(\vartheta)^2/2 \mathcal{Y}(x)~,
\end{equation}
where the relative temperature $\epsilon = \Delta T/T$ is simply given by
\begin{align}\label{eq: definition delta BB}
\epsilon(\vartheta)=\frac{T(\cos\vartheta)- T_{\rm ref}}{T_{\rm ref}}~.
\end{align}
At any angle $\vartheta$ we will thus find a distortion $I(\vartheta)-I_\mathrm{ref}$ due to the peculiar motion of the observer within the CMB rest frame. 
\dnew
However, we might also be interested in the sky-average of the distortion, which can be calculated to be
\begin{align}
\langle\epsilon\rangle & =\frac{1}{4\pi} \int_{0}^{2\pi} \mathrm{d}\phi \int_{0}^{\pi} \mathrm{d}\vartheta \sin \vartheta \,\,\epsilon\,\, \mathrm{d}\cos\vartheta=0~, \\
\langle\epsilon^{2}\rangle & =\frac{1}{4\pi} \int_{0}^{2\pi} \mathrm{d}\phi \int_{0}^{\pi} \mathrm{d}\vartheta \sin \vartheta \,\,\epsilon^2 \mathrm{d}\cos\vartheta= 4\beta^2 \frac{1}{(1-\beta^2) \ln^2 \left(\frac{1+\beta}{1-\beta}\right)} \approx \beta^2/3 + \mathcal{O}(\beta^4)~,
\end{align}
and thus find that the average distortion is $\beta^2/3$ for the \finalr{$g$ distortion}{temperature shift} and $\beta^2/6$ for the $y$ distortion. Note that the presence of a relative velocity between photons and observer is mainly given by the proper motion of the Solar System with a speed $\beta=(1.231\pm0.003)\times10^{-3}$ \cite{hinshaw2009five}. Therefore, we are left with the sky-averaged $y_{\rm dipole}\approx \beta^2/6\approx(2.525\pm0.012)\times10^{-7}$. Higher order terms would be of the order of $10^{-9}$ and are thus ignored for the current analysis.
\dnew
In a similar way, one can also account for higher CMB multipoles as they also introduce an angle-dependence of the temperature, but detailed studies \cite{chluba2004superposition} have shown the effects to be negligible, as the higher multipoles are $\sim 100$ times smaller than the CMB dipole.

\ssec{Reionization and structure formation}
Note that the $y$ parameter determining $\Delta I_{y}$ in \fulleqref{eq: DI definition 1} can be decomposed in an early-time and late-time component. Here we discuss the late-time component generated by reionization and structure formation. At these times, due to the inefficiency of \CS, the induced distortions are mainly $y$ distortions.

\begin{figure}[t]
\centering
\includegraphics[width=8 cm, height=6 cm]{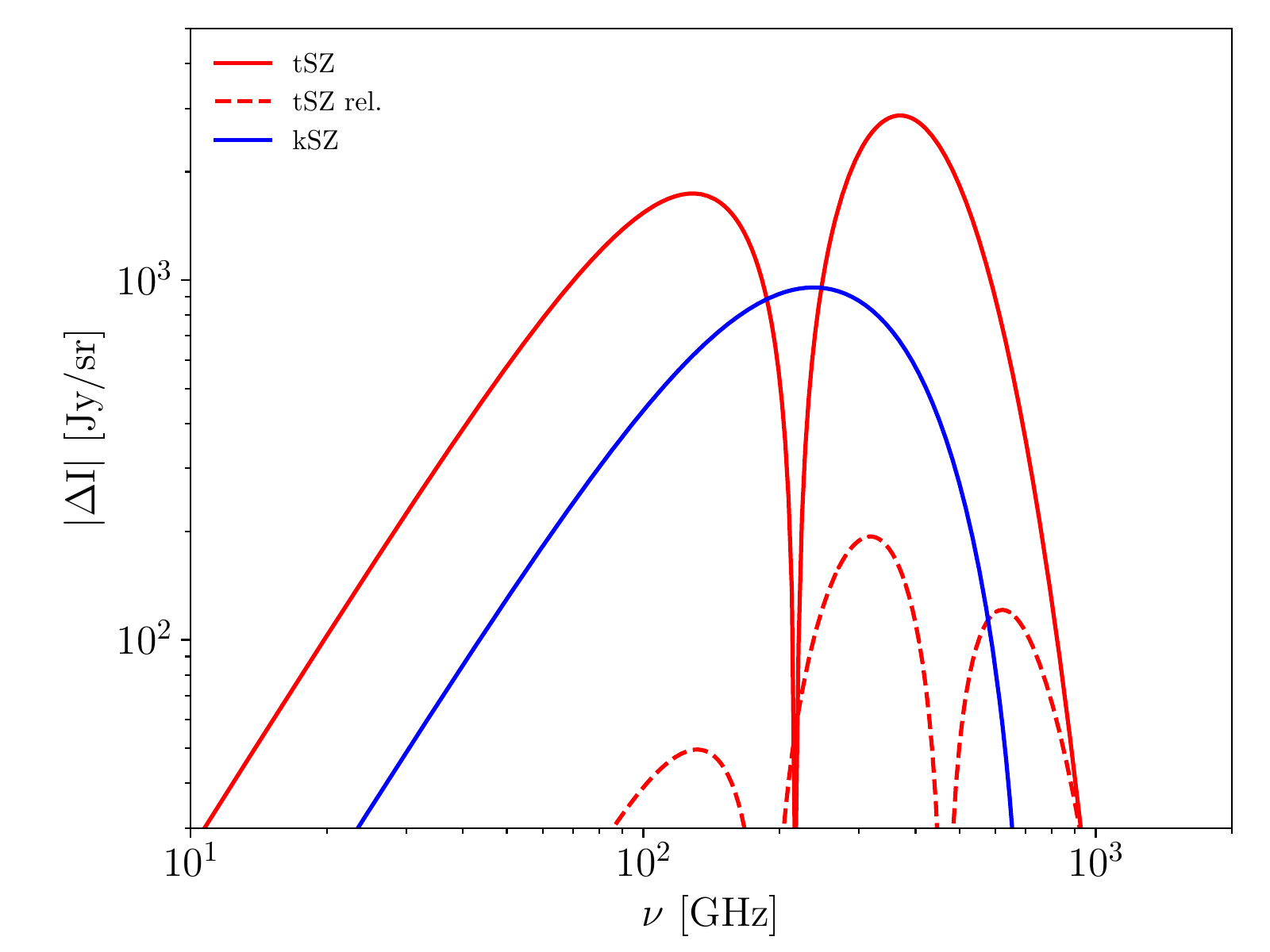}
\caption{SDs caused in the reionization epoch. The solid red curve represents the contribution from the tSZ, which is the dominant one, and the dashed red line shows the relativistic corrections. The blue curve recalls the predicted SDs caused by the kSZ effect.}
\label{fig: distortions_reio}
\end{figure}

\dnew
At this stage of the evolution of the universe the main contribution to SDs is given by the so-called Sunyaev-Zeldovich (SZ) effect \cite{zeldovich1969interaction}. It predicts that when CMB photons travel through a galaxy cluster -- or any pocket of electron gas -- they might interact with local free electrons. Since the  electrons have gained energy due to previous galactic dynamics and gravitational collapse, they are going to be much hotter than the CMB photons. In this way, an inverse \CS might occur, which transfers energy to the photons and thereby perturbs the \BB distribution. 
\dnew
Two different contributions to the SZ effect are usually distinguished: the so-called thermal SZ (tSZ) effect arises from the interaction of photons and thermally distributed electrons; while the kinematic SZ (kSZ) effect has to be accounted for due to the proper motion of the hosting galaxy cluster in the direction of the observer, additionally boosting the velocity of the electron along the line of sight (LOS). Moreover, besides the effects originating from galaxy clusters, further contributions to the SZ effect can be found more broadly in intracluster and intergalactic media (ICM and IGM). The total SDs created during the reionization epoch can thus be parametrized as
\begin{align}\label{eq:Delta_I_reio}
\Delta I_{\rm reio}=\Delta I_{\rm tSZ}+\Delta I_{\rm kSZ}~.
\end{align}
\dnewnoindent
Assuming that the electron temperature in the considered clusters does not exceed a few~keV, we can expand the tSZ signal  in powers of the dimensionless electron temperature $\theta_e \equiv T_{e}/m_{e}$ \cite{Sazonov1998, Challinor1998, Itoh98}. Since the first order term is a pure $y$ distortion, and higher order terms account for relativistic corrections, the tSZ signal can be decomposed as
\begin{align}
\Delta I_{\rm tSZ}=\tilde{y}\mathcal{Y}+\Delta Y_{\rm rel}~.
\end{align}
The amplitude of the term linear in $\theta_e$ is given by $\tilde{y} \approx 4\Delta\tau \, \theta_{e}$, where $\Delta\tau$ is the scattering optical depth along the line of sight in the cluster frame. The relativistic correction $\Delta Y_{\rm rel}$ becomes relevant for $\theta_{e} \gtrsim 10^{-2}$ and induces distortions with a more complicated shape. A possible formulation for $\Delta Y_{\rm rel}$ is given in \cite{chluba2012fast} (\final{see also \cite{stebbins1997cmbr, Itoh98, Sazonov1998, Challinor1998, chluba2013sunyaev} for similar discussions}). In our implementation we stopped at fourth order in $\theta_e$\,, thus obtaining
\begin{align}
\Delta Y_{\rm rel}= \sum_{k=1}^{2}\Delta \tau\theta_{e}^{k+1}Y_k +\mathcal{O}(\theta_{e}^{4})~,
\end{align}
with the $Y_k$ defined as
\begin{align}
Y_k=\sum_{n=1}^{2k+2}a_{n}^{(k)}x^n\partial_{x}^{n}\mathcal{B}(x)~.
\end{align}
The numerical coefficients $a_{n}^{(k)}$ are found in Table B1 of \cite{chluba2012fast}. Higher order terms can also be found in \cite{chluba2012fast}. Note that this perturbative expansion of the problem in $\theta_{e}\ll 1$ does not accurately describe high frequency distortions (for an explicit discussion, see e.g. \cite{chluba2012fast}).
\dnew
The kinematic SZ contribution depends additionally on the total peculiar velocity of the cluster $\beta$ (in natural units) and on the angle $\vartheta$ of its velocity with respect to the LOS. An expansion in the two small parameters $\theta_{e}$ and $\beta$ gives at lowest orders (\finalr{see equation (23) of \cite{chluba2012fast}}{see Equations (1) and (2a)-(2d) of \cite{chluba2013sunyaev}}):
\begin{align}\label{eq: kSZ effect definition}
\nonumber \Delta I_{\rm kSZ} = & \mathcal{N}x^3\Delta\tau\beta  \left[ \sum_{k=0}^{2} \theta_{e}^{k+1}( P_0\beta M^{\rm low}_{\final{k}} + P_1 D^{\rm low}_{\final{k}}+ P_2 \beta Q^{\rm low}_{\final{k}}) \right. \\ \nonumber & \hspace{1.7 cm} \left. \final{+\frac{1}{3}P_0\beta(Y+G)+P_1G+\frac{11}{30}P_2 \beta (Y+4G)} \right] \\ & +\mathcal{O}(\theta_{e}^{4}, \beta\theta_{e}^{3}, \beta^{2}\theta_{e}^{2})~,
\end{align}
where ${\cal N}=2(k_B T_0)^3/(hc)^2$ as before. The Legendre polynomials $P_n$ are evaluated in $\cos \vartheta$, and the distortions $M^{\rm low}_{\final{k}}(x)$, $D^{\rm low}_{\final{k}}(x)$, $Q^{\rm low}_{\final{k}}(x)$ are defined \finalr{in}{as in Appendix A of} \cite{chluba2013sunyaev}. As seen in Equation \eqref{eq: kSZ effect definition}, the contributions from the monopole and quadrupole are suppressed by a factor $\beta$, so that the leading order is given by the dipole. \final{Furthermore, as expected from a linearized Lorentz boost \cite{Sunyaev1972CoASP,Ostriker1986Generation,Rephaeli1991Peculiar}, the term at first order in $\beta$ has the shape of a temperature shift $G(x)$ multiplied by the observer's tangential velocity $\beta P_1 = \beta \cos \vartheta$ and the optical depth $\Delta \tau$, with additional relativistic corrections that are of the order $\mathcal{O}(\theta_e)$.}
\dnew
Note that, as pointed out in \cite{chluba2012fast}, different conventions for the optical depth $\Delta \tau$ are present in the literature. This does not affect the expression of the tSZ  signal, but it changes the splitting between first order and higher order terms in the expression of the kSZ signal. Thus, for the sake of completeness and generality, in our {\sc class} implementation we also include the conventions of \cite{nozawa2005improved} as an option, where the authors do not define the optical depth in the cluster rest frame.
\dnew
The observable SZ effect depends on each direction in the sky. In order to compute the average SZ signal, we can use the previous results with some effective average values of the free parameters $\Delta \tau$, $T_{e}$, $\beta$ and $\cos\vartheta$. For a simple estimate of the average tSZ contribution, we follow  \cite{hill2015taking} and take $T_{e} = 4$ keV, $\theta_e\approx0.01$, and $\Delta \tau=2\times10^{-4}$, which yields $y = \tilde{y}/4 \approx 1.6\times 10^{-6}$. For the kSZ effect we further fix $\beta = 0.01$ and $\vartheta=0$ as in \cite{chluba2012fast}. These numbers are the default values in our numerical 
implementation\footnote{Of course, these numbers are just rough estimates and should be taken with a grain of salt. Several works have discussed the uncertainty on these parameters \cite{nozawa2005improved, chluba2012fast, chluba2013sunyaev}. For instance, the authors of \cite{hill2015taking} find the $y$-weighted temperature $T_{e} = 1.3$ keV and  $ \Delta \tau\approx 3.89\times10^{-3}$ within the standard SZ halo-model \citep{Komatsu1999}, which yields a value $y=1.77\times10^{-6}$ very similar to that in our baseline model. The average relativistic temperature is indeed dominated by low-mass halos $\simeq \text{few}\times 10^{13}\,M_{\odot}$ rather than those with $T_e \simeq 5$ keV that contribute most to the power spectrum, see e.g., \cite{Remazeilles2019} for a recent discussion.}.
The resulting SDs are displayed in Figure~\ref{fig: distortions_reio}, where the solid red curve represents the leading tSZ contribution, the dashed red line shows the relativistic tSZ corrections, and the blue curve the kSZ contribution. The tSZ effect is the dominant contribution and its maximum is well above the detection threshold of PIXIE-like detectors. The mapping of the thermal and kinematic SZ effects across the sky is a very active field in observational cosmology. In the last decade, several collaborations have been able to infer the matter distribution in the galactic neighborhood from this effect~\cite{staniszewski2009galaxy, hasselfield2013atacama, reichardt2013galaxy, story2013measurement, gralla2014measurement, ade2014planck, bleem2015galaxy}.
\dnew
Finally, we should mention the existence of alternative treatments of the contribution to \SD{s} from the reionization era. The authors of  \cite{Titarchuk:2019tvt} based themselves on known solutions of the thermal Comptonization problem in a finite medium \cite{sunyaev1980comptonization, pozdnyakov1983comptonization, titarchuk1995power} and studied the effect of the presence of a bounded spherical plasma cloud on the CMB spectrum. 
\dnew
However, accounting for the spatial structure of the medium is only significant if second order contributions in $\Delta \tau$ are relevant, i.e., if multiple scatterings are possible. However, as shown in \cite{Chluba:2013ada, Chluba:2013paa}, such second order corrections are negligible, so that we can safely neglect the spatial extension of the medium.
\subsubsection{Heating mechanisms in exotic scenarios \label{ssec:heatingExotic}}
In addition to the the heating rates predicted within the standard cosmological model, many other effects can be found that predict different kinds of energy injection or extraction. The most famous and frequently studied ones depend on the presence of annihilating, decaying, or interacting DM, but also Primordial Black Hole (PBH) accretion or evaporation, and early dark energy scenarios that may influence the heating history of the photon bath. In the following paragraphs we are going to describe a few examples.

\begin{figure}[t]
\centering
\begin{minipage}{\textwidth}
\includegraphics[width=7.5 cm, height=5.5 cm]{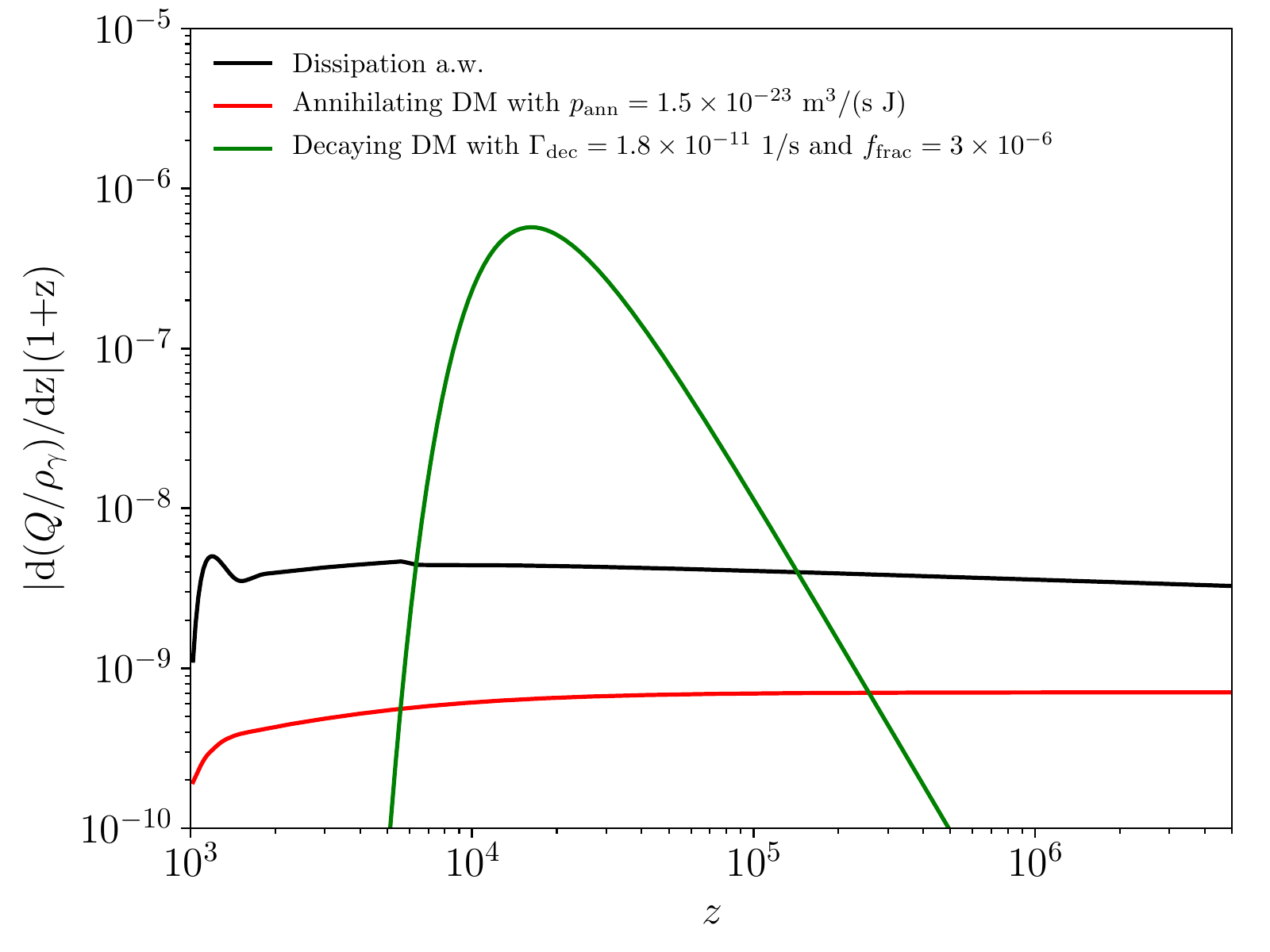}
\includegraphics[width=7.5 cm, height=5.5 cm]{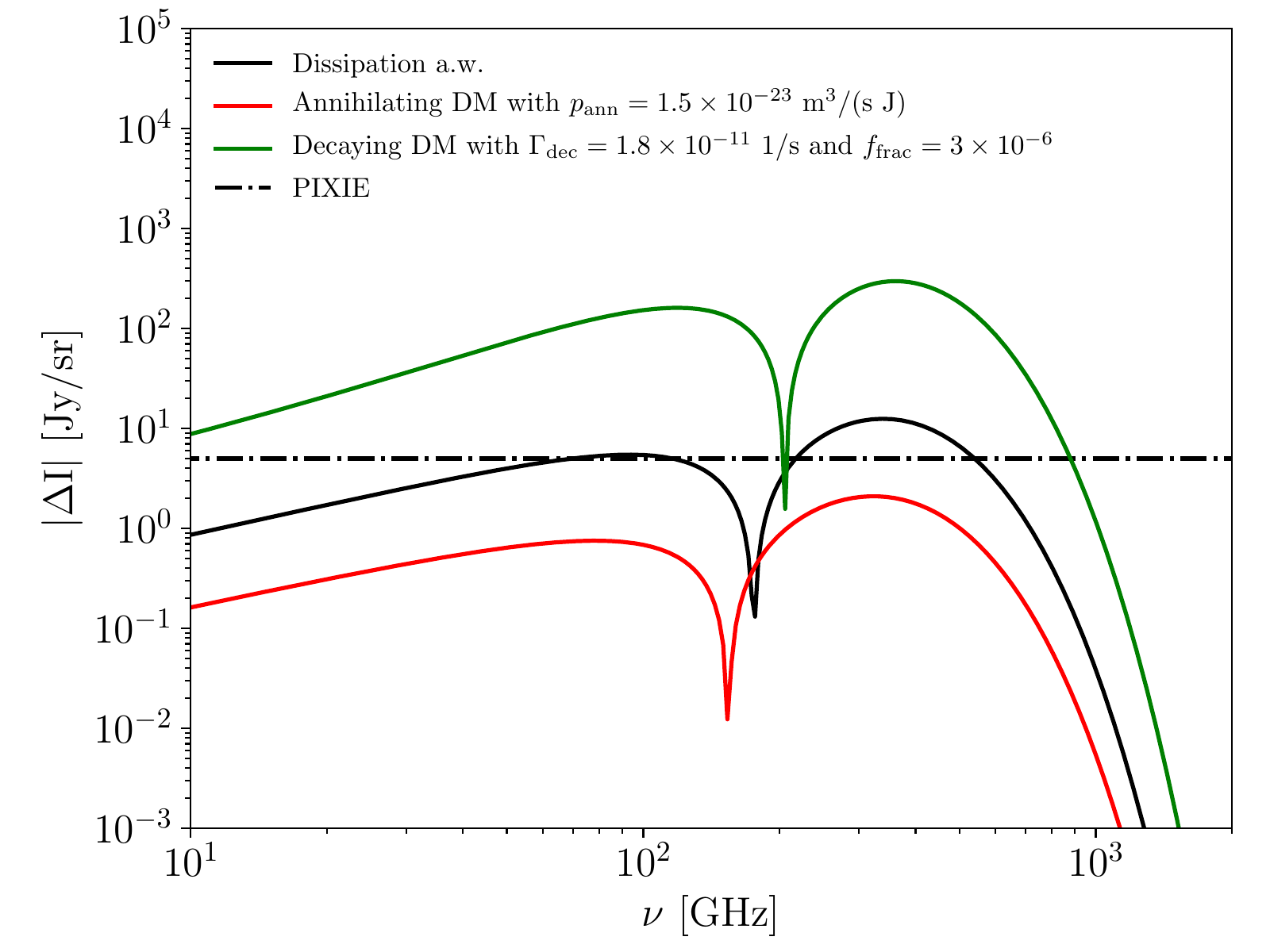}
\end{minipage}
\caption{Heating rate (left panel) and SDs (right panel) caused by DM annihilation (red line) and decay (green line). The heating rate caused by the dissipation of acoustic waves (black line) is given as a reference. In the right panel the dot-dashed line represents once more the predicted PIXIE sensitivity.}
\label{fig: heating_ann_dec}
\end{figure}

\ssec{Dark matter annihilation}
In the case of annihilating DM, the energy injection rate can be written as
\begin{align}\label{eq: heating rate annihilating DM 1}
\dot{\mathcal{Q}}= \rho_{\rm cdm}^{2}f_{\rm frac}f_{\rm eff}\frac{\langle \sigma v \rangle}{\text{M}_\chi} \equiv \rho_{\rm cdm}^{2}p_{\rm ann}~,
\end{align}
where $f_{\rm frac}$ represents the fraction of annihilating DM with respect to the total DM content, $\langle \sigma v \rangle$ is the annihilation cross section, and $\text{M}_\chi$ refers to the mass of DM particle. Since the free parameters $f_\mathrm{eff}$, $f_{\rm frac}$, $\langle \sigma v \rangle$ and $\text{M}_\chi$ are degenerate, they are usually grouped under a single quantity $p_{\rm ann}$ called annihilation efficiency (e.g.~\cite{galli2009cmb,galli2013systematic,Poulin:2015pna,Aghanim:2018eyx}).
The red line in the left panel of Figure~\ref{fig: heating_ann_dec} shows the evolution of the heating rate  $\chi_h \, \dot{\mathcal{Q}}$ for a given value of $p_{\rm ann}$ and assuming maximum deposition efficiency, $f_\mathrm{eff}(z) = 1$ (we recall that we use the GSVI2013 model \cite{galli2013systematic} for the $\chi_h$). The right panel displays the corresponding SD.
\dnew
Note that \fulleqref{eq: heating rate annihilating DM 1} is true only for the case of s-wave annihilation. If we wanted to consider an annihilating DM with p-wave annihilation cross-section $\langle \sigma v\rangle \propto (1+z)$ we would have to introduce \finalr{an additional factor $(1+z)$}{additional powers of $(1+z)$ (for more in-depth discussions regarding the origin of this factor see e.g., \cite{PhysRevD.63.023001, PhysRevD.87.065016, Chen_2013, chluba2013distinguishing})}. However, in this case, reference \cite{chluba2013distinguishing} has shown that BBN and light element abundances set much stronger bounds on the annihilation efficiency than SDs. Therefore, we will not discuss this class of models any further.
\dnew
Another limitation of the model is given by the clustering of DM \cite{huetsi2009constraints,Poulin:2015pna}. In fact, as also argued in  \cite{chluba2010could}, at low redshifts the averaged squared DM density $\langle \rho_{\rm cdm}^{2} \rangle$ is enhanced by a so-called clustering boost factor $B(z)$. However, this factor is negligible when investigating \SD{s}, as in our case, and we will not take it into consideration for the following discussions. The factor is, nonetheless, implemented in the code.
\dnew
Note that assuming a PIXIE detection threshold and all DM annihilating into EM particles only with maximum efficiency, i.e. assuming a constant value of $f_{\rm eff}(z)=1$, the constraint on $p_{\rm ann}$ from SDs would be on the order of $5\times 10^{-27}$ cm$^3/$(s GeV), which is still about one order of magnitude worse than the current constraint given by Planck, which is \mbox{$f_\mathrm{eff}(z=600) \, p_{\rm ann}<3.2\times10^{-28}$ cm$^3/$(s GeV)} at 95\% CL \cite{Aghanim:2018eyx}.

\ssec{Dark matter decay}
Another way to transfer energy from the dark sector to photons and baryons is through the decay of unstable dark matter relics. One can assume that some fraction of the DM decays with a given lifetime $\tau_{\rm dec}$ and a corresponding decay width $\Gamma_{\rm dec}=1/\tau_{\rm dec}$. 
\dnew
Depending on the value of the lifetime, different approaches can be considered to constrain the parameters of the model. In particular, for lifetimes larger than the time of recombination, $\tau_{\rm dec}\geq 10^{13}$ s, CMB anisotropies are by far the most constraining observation (see e.g., \cite{poulin2017cosmological}). Furthermore, for $\tau_{\rm dec}$ in the range from 0.1 s to $\approx 10^{8}$ s, deviations from BBN predictions have the largest constraining power \cite{kawasaki2005big, jedamzik2008bounds}. However, for lifetimes in the intermediate range, SDs could be the main source of information \cite{hub1993thermalization, chluba2011evolution, chluba2014teasing}.
\dnewnoindent
One can define the energy injection rate due to DM decay as
\begin{align}
\dot{\mathcal{Q}}=\rho_{\rm cdm}f_{\rm frac} f_\mathrm{eff}\Gamma_{\rm dec}e^{-\Gamma_{\rm dec} t}~.
\label{eq: heating rate decaying DM 1}
\end{align}
Note that once the age of the universe becomes much larger than the lifetime of the particle, the exponential term drives the heating to zero, ceasing to perturb the energy density of the photon bath. The green line in the left panel of Figure~\ref{fig: heating_ann_dec} shows the heating rate evolution for some arbitrarily chosen values of  $(f_{\rm frac}, \Gamma_{\rm dec})$, assuming again maximum deposition efficiency ($f_\mathrm{eff}(z) = 1$). The right panel displays the corresponding SD.

\begin{figure}[t]
	\centering
	\begin{minipage}{\textwidth}
		\includegraphics[width=7.5 cm, height=5.5 cm]{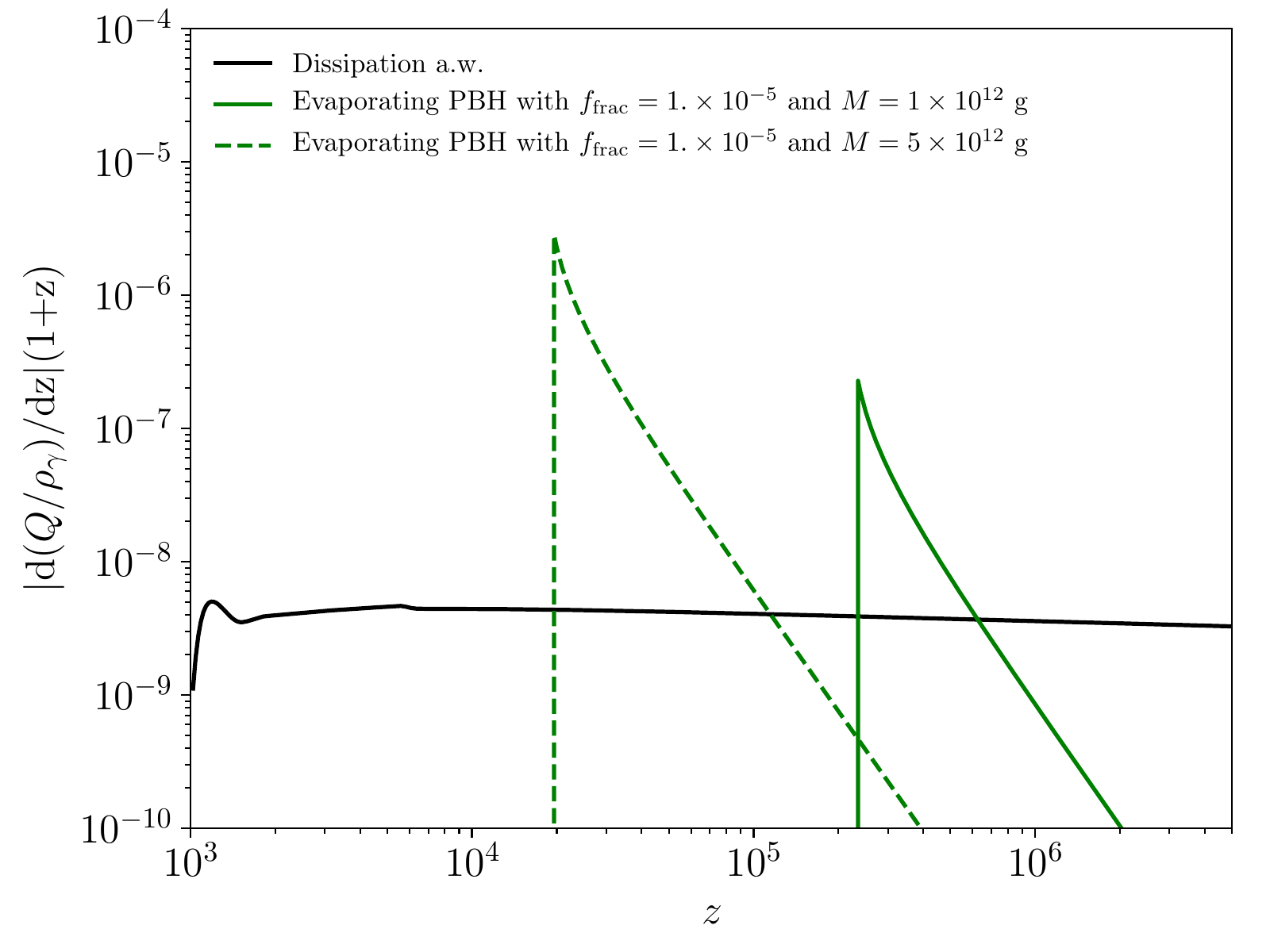}
		\includegraphics[width=7.5 cm, height=5.5 cm]{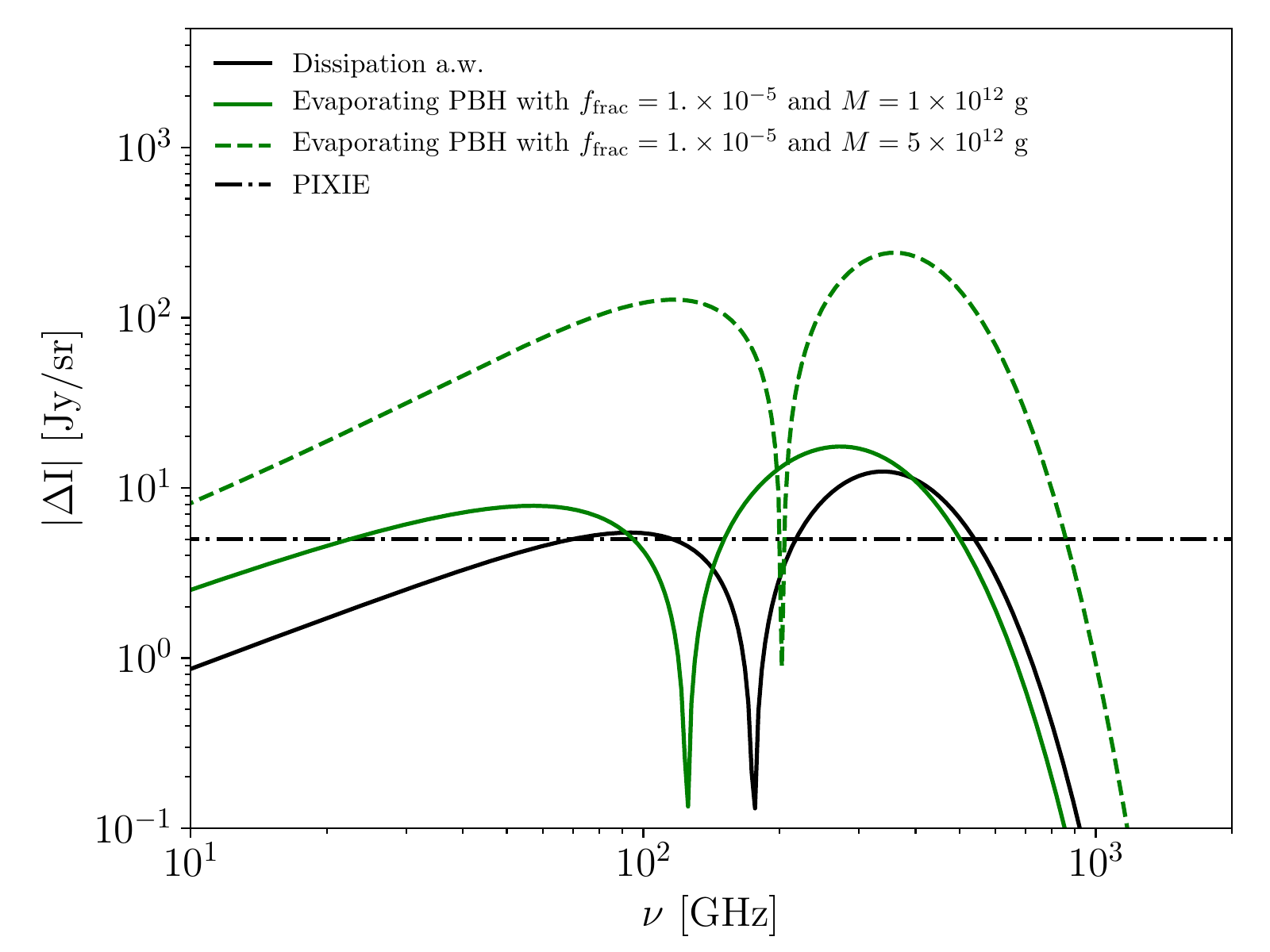}
	\end{minipage}
	\caption{Heating rate (left panel) and SDs (right panel) caused by PBH evaporation (green line). The heating rate caused by the dissipation of acoustic waves (black line) is given as a reference. Once more, the dot-dashed line in the right panel represents the predicted PIXIE sensitivity.}
	\label{fig: heating_PBH_eva}
\end{figure}

\ssec{Evaporation of Primordial Black Holes}
In the last few decades PBHs have attracted particular attention as a possible DM candidate (see e.g. \cite{carr2010new, carr2016primordial} for recent reviews, and \cite{poulin2017cosmological, stocker2018exotic} for further interesting discussions). Furthermore, according to the formation mechanism that is commonly assumed, their mass is tightly connected to the shape of the inflationary potential (see e.g. \cite{carr2016primordial} and the many references listed in Section II therein, as well as \cite{josan2009generalized, frampton2010primordial}). In particular, their abundance is believed to be intrinsically related to a possible non-Gaussianity of the density perturbations \cite{young2013primordial, bugaev2013primordial}. Moreover, it has been argued that a potential detection of a PBH might rule out several WIMP models \cite{Lacki:2010zf, Kohri2014Testing, Eroshenko:2016yve, Boucenna_2018, Adamek:2019gns, Bertone:2019vsk}.
\dnew
However, many uncertainties are involved in the modeling of PBHs, especially within the extent to which one can assume mass monochromaticity, the collapsing process at formation time, the presence of Hawking radiation, and the accretion mechanism, if present at all. Many of these open questions could be answered through observing the impact of these different assumptions on the thermal history of the universe.
\dnew
As a first example, we focus on the evaporation of PBHs. In this case, Hawking radiation \cite{hawking1974black} is expected to dominate the mass evolution of PBHs according to \cite{macgibbon1990quark, macgibbon1991quark}
\begin{align}\label{eq: mass-loss PBH}
\der{M}{t}=-5.34\times10^{25}\frac{\text{g}}{\text{s}} \times \mathcal{F}(M)\, M^{-2}~,
\end{align}
where $M$ is the mass of the PBH, while the function $\mathcal{F}(M)$ represents the effective number of species emitted by the PBH (the shape and characteristics of this function are further discussed in Appendix~\ref{ap:f_of_m}). Note also that our current numerical implementation assumes a single initial PBH mass, but would be easy to generalize to extended spectra. In the monochromatic case, and assuming that PBHs account for a fraction $f_{\rm frac}$ of DM, the energy injection rate can be calculated as \cite{poulin2017cosmological, stocker2018exotic}
\begin{align}
\dot{\mathcal{Q}}=\rho_{\rm cdm} f_{\rm frac} f_\mathrm{eff} \frac{\dot{M}}{M}~.
\end{align}
In the case of PBH evaporation, in contrast to the case of DM annihilation or decay, it is never possible to assume $f_{\rm eff}=1$, because the spectrum of emitted particles and thus the value of $f_{\rm em}$ varies greatly depending on the mass of PBHs at a given time and their related temperature. To calculate $f_{\rm eff}$, we work in the on-the-spot limit $f_{\rm eff}=f_{\rm em}$ (which is a very good approximation at least before recombination and thus for the calculation of SDs). We have devised a new approximation for $f_{\rm em}(M)$, presented in Appendix~\ref{ap:f_of_m},
which extends the range of validity of a previous scheme introduced in reference \cite{stocker2018exotic} towards much lower masses. The evolution of $\mathcal{F}(M)$ as a function of the PBH mass is displayed in the left panel of Figure~\ref{fig: PBH_f_M}, while the right panel shows the corresponding $f_{\rm em}(M)$.
\dnew
Two examples of heating rate evolution are shown in the left panel of Figure~\ref{fig: heating_PBH_eva}, with the corresponding predictions for the SDs in the right panel. Evaporating PBHs and decaying DM produce rather similar heating rate evolutions and could be difficult to distinguish through SDs. However, the PBH heating rate is more sharply peaked. This feature would be difficult to probe at the level of $\mu$ or $y$ distortions, but leaves a unique signature in the residual distortions. In addition, towards the final stages of the evaporation process, the mean energy of the particles will become very large, such that non-thermal effects are expected to become important. This can be expected to modify the shape of the \SD{s} as well as the heating efficiencies, but we leave both aspects for future analyses.
\dnew
Finally, we can integrate Equation \eqref{eq: mass-loss PBH} in order to obtain the PBH lifetime. In this way one finds that only PBHs initially lighter than $10^{13}-10^{13.5}~\mathrm{g}$ can evaporate before recombination and thus cause strong SDs, while more massive PBHs can still influence the evolution of photons and baryons through their energy release and leave a detectable signature on the CMB anisotropy spectrum~\cite{stocker2018exotic}.

\ssec{Accretion of matter into Primordial Black Holes}
PBHs could also influence the thermal history of the universe by accreting matter around them. The accreting matter could heat up, ionize, and consequently radiate high-energy photons. Up to now, no complete numerical simulation of this process over cosmological time scales has been performed. However, several approximate analytical solutions have been found (see e.g., \cite{ricotti2008effect, ali2017cosmic, poulin2017cmb}). According to these works, one of the biggest sources of uncertainty is the shape of the infalling matter distribution surrounding the PBH.
\dnew
Before recombination it is common to approximate the accretion as spherical, which provides a conservative estimate of the luminosity \cite{ali2017cosmic}. Furthermore, according to \cite{poulin2017cmb}, disk accretion becomes a natural option after~\mbox{$z\approx10^3$}. However, this is still a source of debate. The two main scenarios, i.e. disk and spherical accretion, could in principle be discriminated through their different impact on the angular power spectra of the CMB (as shown, e.g., in Figure 3 of \cite{poulin2017cmb}).
\dnew
However, SDs are mainly influenced by energy injection before recombination and we conservatively assume spherical accretion up to this point. As argued by \cite{ali2017cosmic}, this type of PBH accretion does not produce an appreciable level of SDs. Therefore, we will not discuss this case further, although it is implemented within our code for completeness.

\section{Numerical implementation \label{method}}
In the previous sections we  outlined the general picture of how the heating history of the universe could affect the shape of \SD{s} -- on top of its signature on CMB anisotropies, well described in the previous literature, see e.g.~\cite{slatyer2009cmb,galli2009cmb,slatyer2013energy,Giesen:2012rp,galli2013systematic,Slatyer2015,Poulin:2015pna,poulin2017cosmological,stocker2018exotic,liu2019darkhistory}, and not reviewed again here. Furthermore, we provided multiple examples of possible effective heating rates $\dot{\mathcal{Q}}$ and summarized their deposition properties. The corresponding numerical computation of the different heating rates, including their related injection efficiency and deposition function, as well as the calculation of the CMB anisotropies and \SD{s}, has been performed with an expanded version of {\sc class}~\cite{blas2011cosmic}, which will be released as part of {\sc class~v3.0}. 
\dnew
The new version of the code has several differences with respect to previous ones. We are going to discuss in Section \ref{class} how we handle the impact of heating rates on the thermal history of the Universe, which is crucial for both  calculations of CMB anisotropies (mainly because the ionization fraction $x_e(z)$ affects the Thomson scattering rate and the photon visibility function) and SDs (since $x_e(z)$ also affects other quantities described in section \ref{SD_causes} like the fractions $\chi_c$, the diffusion scale $k_D$ or the heat capacity $\alpha_h$). Then, in Section \ref{Greens}, we are going to describe the practical method with which \SD{s} and the corresponding residuals are calculated in our implementation.  
\dnew
As a general remark, it is important to underline the fact that many of the modifications discussed here rely on the work already performed by different groups and previous public codes. First, our implementation of the thermal history in presence of energy injection is based on a previous public branch of {\sc class } specifically designed for the precise treatment of exotic energy injections, called {\sc ExoCLASS} \cite{stocker2018exotic}, which incorporates itself some previous numerical results, in particular from \cite{Slatyer2015}. 
\dnew
However, we implemented many improvements with respect to {\sc ExoCLASS} such as a more consistent implementation of the heating rates with respect to the overall structure of the program, as well as more details discussed in Section \ref{class}. Second, for the calculation of \SD{s}, we adapted a few parts the public codes {\sc CosmoTherm} \cite{chluba2011evolution, chluba2013green} and {\sc SZpack} \cite{chluba2012fast, chluba2013sunyaev} (see \cite{chluba2011evolution, chluba2012cmb, chluba2014teasing, chluba2016spectral} and \cite{nozawa2005improved,hill2015taking} for related discussions). Also in this case, several improvements have been implemented, regarding in particular the principal component analysis (PCA) expansion and the construction of detector settings discussed in Section \ref{Greens}. 
\dnew
Finally, our implementation of likelihoods, which are used to explore the constraining power of current and future \SD{s} measurements, relies on the overall infrastructure of the {\sc MontePython} \cite{audren2013conservative, Brinckmann:2018cvx} parameter inference code and is further described in Section~\ref{likelihoods}.

\subsection{Thermal history with energy injection \label{class}}
Before computing CMB anisotropies and/or SDs, any code must compute the evolution of background thermodynamical quantitites like the free electron fraction $x_e$ or the baryon temperature $T_m$. In {\sc class} this is done by a separate \tquote{thermodynamics} module. The recombination equation can be approximated at various levels and can be calculated using three numerical packages: {\sc Recfast} \cite{Wong:2007ym}, {\sc HyRec} \cite{AliHaimoud:2010dx}, or {\sc CosmoRec} \cite{chluba2011towards}. The changes describe below are generic and compatible with these three packages.
\dnew
The presence of at least one of the many heating rates mentioned in Section \ref{SD_causes} is inevitably going to influence the thermal history of universe. In particular, the fraction of free electrons will change according to
\begin{equation}
\dot{x}_e=I_X+I_s-R_s~,
\end{equation}
where $R_s$ and $I_s$ correspond to the standard recombination and reionization rates, respectively, and $I_X$ refers to the contribution from non-standard sources. The presence of this additional factor can be attributed to the fact that effects injecting additional photons into the CMB bath increase the ionization probability of H and He atoms from the ground state, and produce additional Lyman-$\alpha$ photons that boost the $n=2$ population. 
\dnew
As a consequence, the rate of photoionization of these excited states by the CMB rises. $I_X$ can thus be split in the different contributions as
\begin{equation}\label{eq: I_X value}
	I_X=I_\mathrm{ion}+I_{X\alpha}\,,
\end{equation}
where $I_\mathrm{ion}$ and $I_{X\alpha}$ represent the ionization rate due to ionizing and Lyman-$\alpha$ photons, respectively. These quantities are proportional to the injected energy according to
\begin{equation}
I_{\mathrm{ion}/\alpha}\propto \frac{\dot{\mathcal{Q}}\chi_{\mathrm{ion}/\alpha}}{n_H E_{\mathrm{ion}/\alpha}}~,
\end{equation}
where $\chi_{\mathrm{ion}/\alpha}$ represents the deposition fraction
into the ionization or Lyman-$\alpha$ channel (these are two of the \tquote{$c$} channels already introduced in Section \ref{ssec:injdep}), $E_\mathrm{ion}$ refers to the average ionization energy per baryon, and $E_{\alpha}$ is the difference in binding energy between the $1s$ and $2p$ energy levels of H (for the precise definitions and a more complete discussion see, e.g., \cite{galli2013systematic}). Note that within the {\sc HyRec} \cite{AliHaimoud:2010dx} and {\sc CosmoRec} \cite{chluba2011towards} codes, the different energy levels of the H atoms are modeled, and thus the precise form of $I_X$ is in general more complicated than the form expressed in \fulleqref{eq: I_X value}.
\dnew
One particular difference to older versions of {\sc class} is the handling of the recombination coefficients within {\sc Recfast}. As argued in \cite{Chluba:2015lpa,Oldengott:2016yjc}, these should depend on the radiation temperature rather than the matter temperature. While during most of the cosmic history, this difference is negligible, it is important when strong energy injections affect the ionization history at late times after recombination.
\dnew
Another quantity that will change depending on the energy injection history is the matter temperature. Its evolution equation can be written as \final{(see e.g., \cite{chen2004particle, galli2013systematic, Poulin:2015pna}, although with different notations)}
\begin{equation}\label{eq:evolution_Tm}
\dot{T}_m=-2H T_{m} - 2 R_\gamma \frac{\mu_b}{m_e} (T_{m}-T_{z})+ \frac{\dot{\mathcal{Q}}\chi_h}{\alpha_{h}}~,
\end{equation}
where we have used the photon interaction rate $R_\gamma$\, defined as
\begin{equation}
R_\gamma = R \dot{\kappa} = \frac{4}{3} \frac{\rho_{\gamma}}{\rho_b} n_e \sigma_T~,
\end{equation}
which is related to \CS, and the mean baryon molecular weight $\mu_b$ with form
\begin{equation}
\mu_b=\frac{\rho_\mathrm{bar}}{n_\mathrm{bar}}=\frac{\rho_H+\rho_{He}+\rho_e}{n_H+n_{He}+n_e}\approx m_H \frac{n_H+4n_{He}}{n_H+n_{He}+n_e}~.
\end{equation}
In \fulleqref{eq:evolution_Tm}, the first term describes the heat loss due to the adiabatic expansion of the universe, the second term refers to the coupling of baryons to photons, and the third term to the heating due to non-standard energy injections. Finally, the function $\chi_h$ is the the deposition fraction into the heating channel, and $\alpha_h$ is the heat capacity of the IGM defined in equation (\ref{eq:heat_capacity}).
\dnew
As a technical remark, note that in the new {\sc class} implementation, we do not follow the evolution of the matter temperature $T_m$, but rather of the difference $\Delta T_{mz}$ between $T_{m}$ and $T_{z}$. The reason for this is that for most part of the evolution of the universe, the two temperatures have extremely similar values, and their difference can be below the level of numerical precision of the differential equation solver 
(although in our new version of {\sc class} we now use the advanced differential equation solver {\tt ndf15} \cite{shampine1997matlab,blas2011cosmic} even for solving the thermal history). Without such an approach, the photon interaction term $R_\gamma (T_m - T_z)$ would be dominated by large cancellations between the terms, and thus drastically influenced \final{by} the numerical noise. The same procedure is followed in {\sc CosmoTherm}~\cite{chluba2011evolution}.
\dnew
As a final note, the realistic modeling of the matter temperature during reionization is a very challenging task. Since this is not crucial for the applications and analyses discussed in this work, we leave it for future work. Our version of {\sc class} relies on the same simplistic  treatment of $T_m$ around reionization as previous versions of the code.

\subsection{Spectral distortions from the Green's function approximation \label{Greens}}

\subsubsection{General Green's function approach}
As already discussed in Section \ref{SD_calc}, it is possible to write the total \SD as
\begin{equation}\label{eq: DI definition 2}
\Delta I_\mathrm{tot}(x) = \mathcal{G}(x)\tilde{g} + \mathcal{Y}(x) \tilde{y} + \mathcal{M}(x) \tilde{\mu} + R(x,z')~,
\end{equation}
excluding second order contributions in $\tilde{g}$. In Appendix~\ref{ap:model} we show that the calculation of the total \SD can also be performed using a Green's function approach \cite{chluba2013green}
\begin{equation}\label{eq: DI definition 3}
\Delta I_\mathrm{tot}(x,z) = \int^{\infty}_{z} \id z' G_{\rm th}(x,z') \frac{\id Q(z')/\id z'}{\rho_{\gamma}(z')}~.
\end{equation} 
The Green's function $G_{\rm th}(x,z')$ translates an energy injection/extraction at redshift $z'$ to a distortion of frequency $x$ observed at the current time. It decouples the cosmology-independent redistribution of photons, and the cosmology-dependent energy injection history. In this way, the knowledge of the heating history of the universe is enough to approximately determine the shape of the \SD{s}.
\dnew
The Green's function has been computed in \cite{chluba2013green}, using a code that follows the full evolution of the \ppsd \cite{chluba2011evolution}, and computing the response of the plasma to $\delta$-like heating terms approximated as narrow Gaussian peaks.
The results are contained in a data file published together with the current public version of {\sc CosmoTherm}\footnote{The full {\sc CosmoTherm} is not currently available, and instead the Green's function based approach is found in the {\sc Greens} code available at \cite{greensLink}.}. The same file is employed in our new version of {\sc class} as well. 
\dnew
However, once the final shape of the Green's function is known, it is interesting to try splitting this function into terms with a straightforward physical interpretation. Using \fulleqref{eq: DI definition 2} and \fulleqref{eq: SDs amplitudes definition 1}, one can expand the total Green's function as
\begin{equation}\label{eq: Greens function 1}
G_{\rm th}(x,z') = \mathcal{G}(x) \mathcal{J}_g(z') + \mathcal{Y}(x) \mathcal{J}_y(z') + \mathcal{M}(x) \mathcal{J}_\mu(z') + R(x,z')~.
\end{equation}
Several approximations have been used in the past to calculate the required branching ratios $\mathcal{J}_a(z)$ (see Appendix \ref{ap:br} for a complete discussion). Recently, \cite{chluba2014teasing} has suggested an exact calculation of the branching ratios from the full Green's function, also allowing for the determination of the residual function $R(x,z)$. Our implementation follows this method, and performs the calculation steps described below.
\subsubsection{Branching ratios from discretized Green's function}
The approach of \cite{chluba2014teasing} is based on least-squares fitting the $\mathcal{G}(x)$, $\mathcal{Y}(x)$, and $\mathcal{M}(x)$ with time-dependent coefficients to the overall $G_\mathrm{th}(x,z)$. The coefficients are then the branching ratios, while the residuals correspond to the residual distortion. In practice, it is simpler to work directly with quantities that are discretized in frequency space. Then, the least-squares fit can be replaced by Gram-Schmidt orthogonalization. We will denote the discretized version of any quantity  $A(\nu)$ as \mbox{${\mathbf{A}=A(\nu_i)}$}. 
We introduce an orthonormal basis of distortion shapes, ordered following the convention of \cite{chluba2014teasing} as $\mathbfcal{Y}$, $\mathbfcal{M}$, $\mathbfcal{G}$\footnote{The ordering does not affect the shape of the residual distortion, but leads to different relative sizes of the $y$, $\mu$, and $g$ distortions. These, however, do not affect the final constraints, as the disambiguation of $y$ and $\mu$ distortions is artificial anyway. Furthermore, some choices can lead to numerical problems.}.
\newpage
\dnewnoindent
The basis vectors are defined as
\begin{equation}
\mathbf{e}_{y}=\mathbfcal{Y}/|\mathbfcal{Y}|~, \quad \mathbf{e}_{\rm \mu}=\mathbfcal{M}_{\bot}/|\mathbfcal{M}_{\bot}|~, \quad \text{and} \quad \mathbf{e}_{g}=\mathbfcal{G}_{\bot}/|\mathbfcal{G}_{\bot}|~,
\end{equation}
where the orthogonal components are
\begin{equation}
\mathbfcal{M}_{\bot}=\mathbfcal{M}-\mathcal{M}_y\mathbf{e}_y~, \quad \text{and} \quad \mathbfcal{G}_{\bot}=\mathbfcal{G}-\mathcal{G}_y\mathbf{e}_y-\mathcal{G}_\mu\mathbf{e}_\mu~,
\end{equation} 
with $\mathcal{M}_y=\mathbfcal{M}\cdot \mathbf{e}_y$\,, $\mathcal{G}_y=\mathbfcal{G}\cdot \mathbf{e}_y$\,, and $\mathcal{G}_\mu=\mathbfcal{G}\cdot \mathbf{e}_\mu$. We can then find the branching ratios at any redshift $z$ by projecting the total Green's function along each basis vector,
\begin{align}
\mathcal{J}_g(z) & = (\mathbf{e}_g \cdot \mathbf{G}_{\rm th}(z))/|\mathbfcal{G}_{\bot}|~, \\
\mathcal{J}_\mu(z) & = (\mathbf{e}_\mu \cdot \mathbf{G}_{\rm th}(z)-\mathcal{G}_\mu\mathcal{J}_g(z))/|\mathbfcal{M}_{\bot}|~, \\
\mathcal{J}_y(z) & = (\mathbf{e}_y \cdot \mathbf{G}_{\rm th}(z)-\mathcal{M}_y\mathcal{J}_\mu(z) -\mathcal{G}_y\mathcal{J}_g(z))/|\mathbfcal{Y}|~, \\
\mathcal{J}_R(z) & = 1- \mathcal{J}_g(z) -\mathcal{J}_y(z) -\mathcal{J}_\mu(z)~.
\end{align}
The residual $\mathbf{R}(z)$ is just given by the difference between the full Green's function and the sum of the $\mathbfcal{Y}$, $\mathbfcal{M}$ and $\mathbfcal{G}$ shapes weighted by the branching ratios, such that
\begin{align}\label{eq: Greens function 2}
\mathbf{G}_{\rm th}(z') = \mathbfcal{G} J_g(z') + \mathbfcal{Y} J_y(z') + \mathbfcal{M} J_\mu(z') + \mathbf{R}(z')~.
\end{align}
The branching ratios resulting from this approach are shown in Figure \ref{fig: plot_branching_ratios} for the case of the PIXIE detector.

\subsubsection{The PCA of the residual distortion}
\begin{figure}[t]
	\centering
	\begin{minipage}{\textwidth}
		\includegraphics[width=8 cm, height=6 cm]{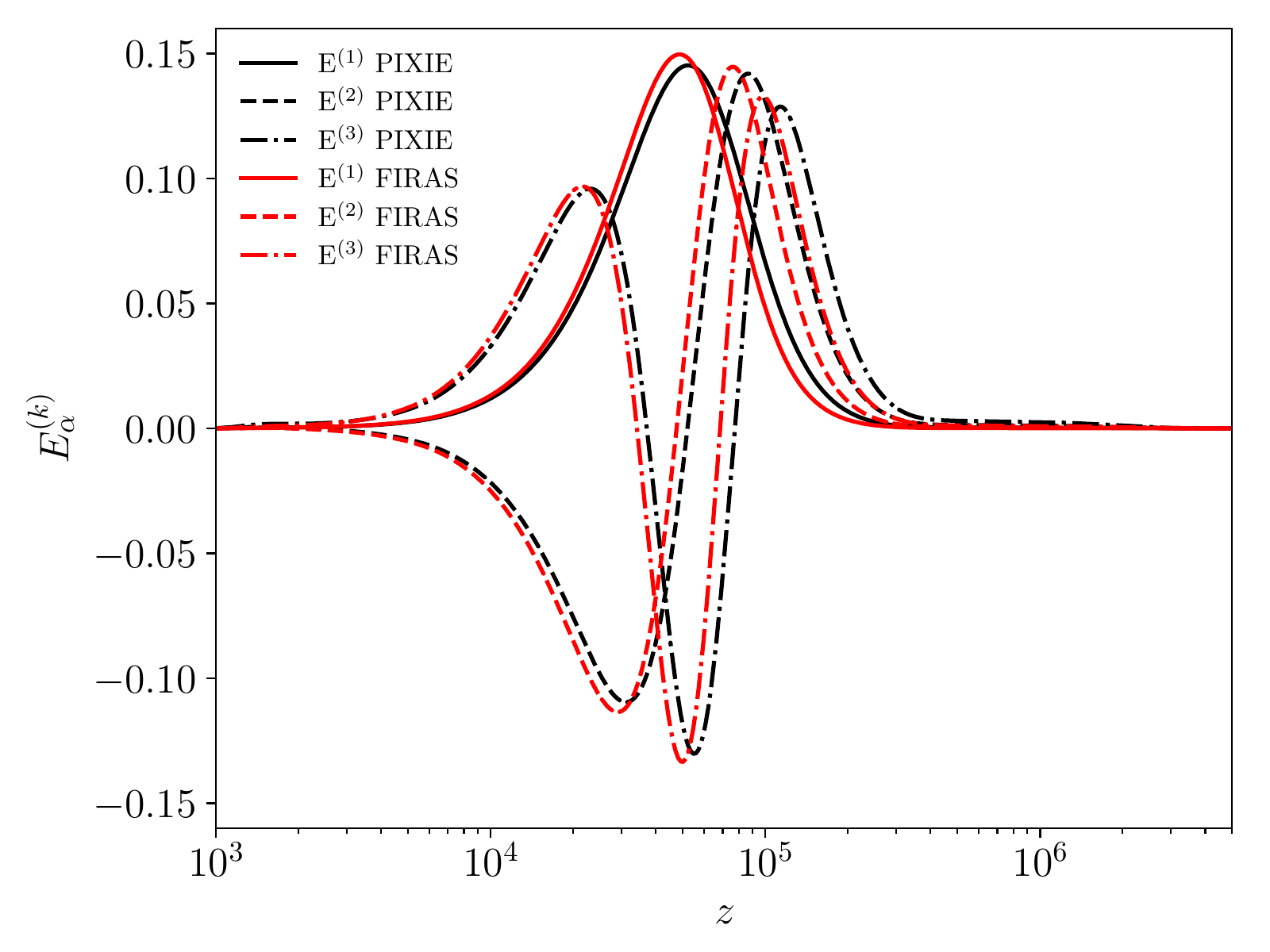}
		\includegraphics[width=8 cm, height=6 cm]{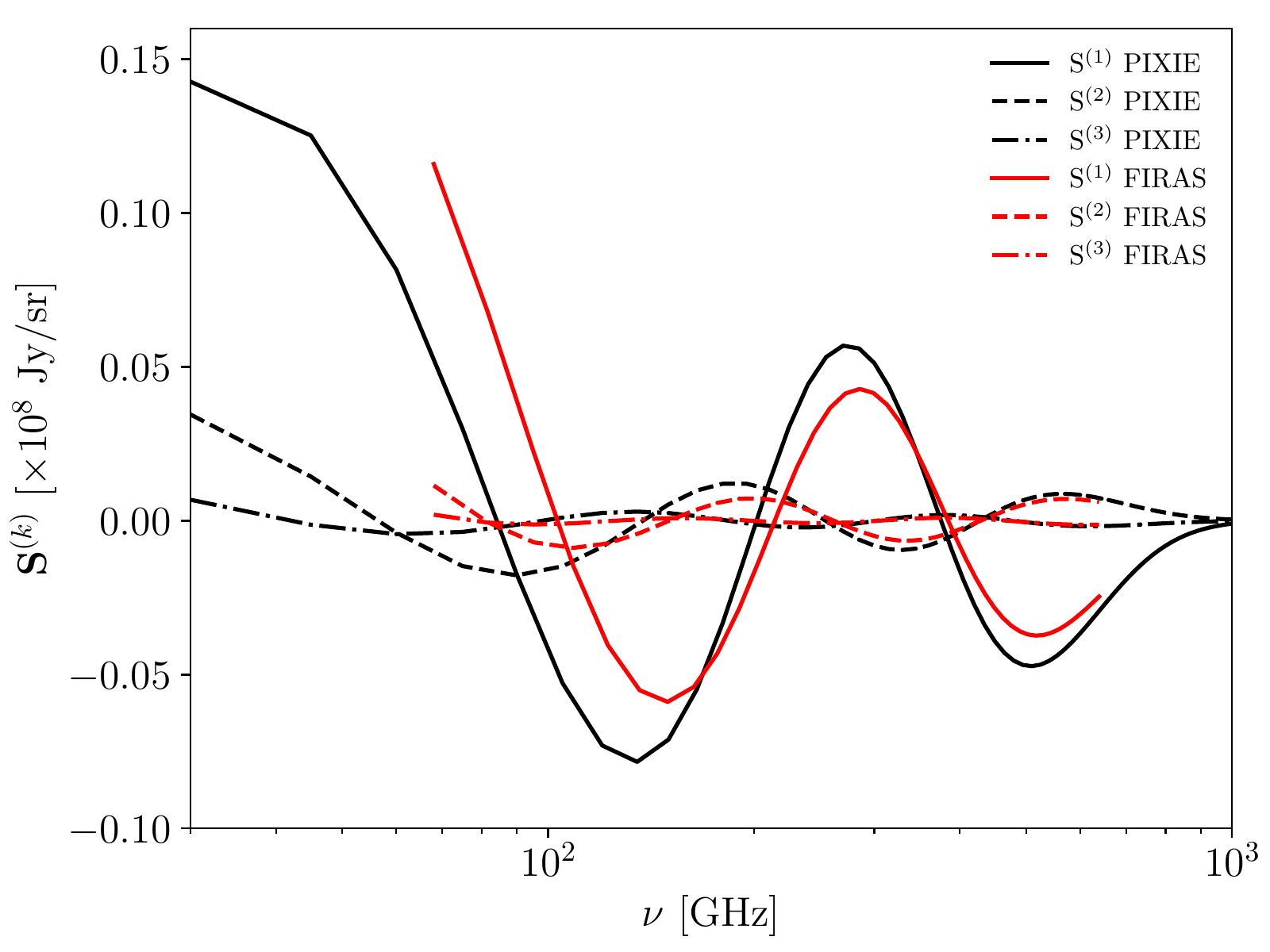}
	\end{minipage}
	\caption{Eigenvectors $E_{\alpha}^{(k)}$ of the Fisher matrix (left panel) and corresponding SD signal $\mathbf{S}^{(k)}$ (right panel) for PIXIE (black lines) and FIRAS (red lines) detector specifics. Note that in the right panel the minimum and maximum frequencies are detector dependent, as discussed in Section \ref{likelihoods}.}
	\label{fig: pca_multipoles}
\end{figure}

While this approach correctly predicts the branching ratios and the residual distortion, it gives no insight on the characteristics and on the physical origin of the the latter. For that purpose, it is beneficial to use a principal component analysis to further decompose the residual distortion into shapes $\mathbf{S}^{(k)}$ and amplitudes $\mu^{(k)}$. As argued in \cite{chluba2014teasing}, this separation should be performed on the basis of the amount of information that can be extracted by a given experiment from a given shape, which can be quantified through the Fisher information matrix. This approach maximizes the signal to noise ratio of the components $\mu^{(k)}$ of the residual distortion. 
\dnewnoindent
The residual distortion of the final spectrum can be expressed with discrete notations as
\begin{equation}\label{eq: DI residuals 1}
\Delta I^{\rm R}_{i}=\sum_\alpha \hat{R}_{i\alpha} dQ_\alpha~,
\end{equation}
where the Latin indices refer to frequencies $\nu_i$, and Greek indices to redshifts $z_\alpha$. We have defined \mbox{$\hat{R}_{i\alpha}=\hat{R}(x_i,z_\alpha)=R(x_i,z_\alpha)\,\Delta\ln z_\alpha$} and $dQ_\alpha=dQ(z_\alpha)=\left.\left[\left(\id Q/\id\ln z\right)/\rho_\gamma\right]\right|_{z_\alpha}$\, to simplify the following equations.
\dnew
The Fisher-information matrix defines the information content that can be gained by measurements for a given sensitivity $\delta I_\mathrm{noise}(\nu_i)$, and its principal components define the distortion shapes that have the highest signal to noise ratio for the considered experiment. Assuming diagonal noise covariance, we can calculate it as
\begin{equation}
\mathcal{F}_{\alpha\beta}=\sum_i \frac{1}{\delta I_{\rm noise}(\nu_i)^{2}}\frac{\partial \Delta I^{\rm R}_{i}}{\partial {dQ}_\alpha}\frac{\partial \Delta I^{\rm R}_{i}}{\partial {dQ}_\beta}=\sum_i \frac{\hat{R}_{i\alpha}\hat{R}_{i\beta}}{\delta I_{\rm noise}(\nu_i)^{2}}~.
\end{equation}
The orthonormal eigenvectors $E_{\alpha}^{(k)}$ of this symmetric matrix allow for the computation of the optimal amplitudes and shapes
\begin{equation}
\mu_k = \sum_\alpha E_{\alpha}^{(k)} {dQ}_\alpha~, \quad \qquad \mathbf{S}^{(k)} = \sum_\alpha E_{\alpha}^{(k)} \hat{\mathbf{R}}_{\alpha}~.
\end{equation}
where $\mu^{(k)}$ is the amplitude (in the sense of \fulleqref{eq: SDs amplitudes definition 1}) and $\mathbf{S}^{(k)}$ is the distortion signal of the $k^{th}$ eigenmode.
Consequently, this allows us to decompose the residual distortions as
\begin{equation}
\mathbf{\Delta I}^{\rm R}\approx\sum_k \mu^{(k)} \mathbf{S}^{(k)}~.
\label{eq: DI residuals 2}
\end{equation}
The result of this PCA is displayed in Figure \ref{fig: pca_multipoles} for the case of PIXIE (black line) and FIRAS (red line).
\dnew
Using a similar approach, one can also obtain the residual parameter $\epsilon$ introduced in \fulleqref{eq: delta rho definition 2} from
\begin{equation}
\epsilon \approx\sum_k C_S^{(k)} \mu_k~,
\end{equation}
where $C_S^{(k)} = \sum_i S^{(k)}(x_i)\Delta x_i /\sum_i \mathcal{G}(x_i)\Delta x_i$ is normalized in the same way as other contributions, see \fullmanyeqref{eq:normalization 1}-\eqref{eq:normalization 2}\footnote{Note that the ratio simplifies to $\sum_i S^{(k)}(x_i) /\sum_i \mathcal{G}(x_i)$ as in \cite{chluba2014teasing} only if the grid is equally spaced.}.
\dnew
We want to underline that our {\sc class} implementation automatically generates the orthogonal branching ratios and the PCA of the residual distortions. For this, it needs a binned frequency array together with a corresponding sensitivity, as in the case of FIRAS, or a minimal set of free parameters  \{$\nu_{\rm min}$, $\nu_{\rm max}$, $\Delta\nu_c$, $\delta I_{\rm noise}$\}, as in the case of PIXIE. For the latter case, the noise is assumed to be frequency independent (i.e. $\delta I_{\rm noise}(\nu_i)= \delta I_{\rm noise}$), and the binning to be of equal width $\Delta \nu_c$ between [$\nu_{\rm min},\nu_{\rm max}$].

\subsection{Experimental settings and likelihoods \label{likelihoods}}

In the previous sections we described how, for a chosen cosmological model, we can accurately predict the total \SD{s} 
parameterized as in \fulleqref{eq: DI definition 2}. Furthermore, we also saw how different experimental configurations of possible \SD missions can be used to determine the shape of the branching ratios through a PCA analysis, and thus the observed \SD{s}. We are now interested in seeing how well different cosmological scenarios can be constrained, given current and future \SD{s} experiments. Thus, we want to address the question: \emph{Assuming an experiment has measured a total \SD $\Delta I_{tot}$\,, which cosmological parameters can we constrain?}
\dnew
With this in mind, we have added a new family of likelihoods to the parameter extraction code {\sc MontePython}~\cite{audren2013conservative, Brinckmann:2018cvx} to deal with \emph{any} \SD mission, making use of the MCMC forecast method detailed in \cite{Perotto:2006rj}: a future (or current) experiment is encoded as a mock likelihood, providing the probability that the generated mock data is true given the model assumed at each step of the MCMC parameter exploration. In more detail, for our likelihood
\begin{enumerate}
\item We choose a fiducial model (for example $\Lambda$CDM with cosmological parameters as measured by Planck \cite{Aghanim:2018eyx}).
\item For this model, we use {\sc class} to compute the total \SD{s} in each frequency bin of the experiment. This is stored as our \emph{observed} \SD.
\item For each step in the MCMC, we update the cosmological parameters of our model, and use {\sc class} to compute the \emph{predicted} total \SD{s} in each frequency bin of the experiment for this model.
\item For each step a new $\chi^2$ is computed by comparing the \emph{predicted} model with the \emph{observed} one, taking into consideration the  sensitivity of the experiment to the signal,
\begin{equation}
\chi^2 = \sum_{\nu_i} \left(\frac{\Delta I_\mathrm{predicted}(\nu_i) - \Delta I_\mathrm{observed}(\nu_i)}{\delta I_\mathrm{noise}(\nu_i)}\right)^2~,
\end{equation}
where we have assumed the noise in different bins $i$ to be uncorrelated.
\end{enumerate}
With this family of likelihoods, in order to define a new detector, we simply need to provide the code with either the free parameters \{$\nu_{\rm min}$, $\nu_{\rm max}$, $\Delta\nu_c$, $\delta I_{\rm noise}$\} (assuming constant noise and equal binning), or the full binned frequency array $\nu_i$ together with the corresponding sensitivities $\delta I_\mathrm{noise}(\nu_i)$. 
\dnew
Our method to compute the distortions assumes that \finalr{$g$ distortions}{temperature shifts} vanish at the earliest time (or maximal redshift $z_\mathrm{max}$) included in the integral of \fulleqref{eq: SDs amplitudes definition 1}. This is implicitly equivalent to a normalization of the reference temperature $T_0$ such that $T_z=T_0(1+z)$ coincides with the true $T_\gamma(z)$ at $z=z_\mathrm{max}$. In reality, the measurement of $T_\gamma(0)$ by experiments like FIRAS aims at minimizing \finalr{$g$ distortions}{temperature shifts}, which is equivalent to fixing $T_0=T_\gamma(0)$ (up to an experimental error). Thus, in our mock likelihood, we do not need to take into account the value of $g$ coming from \fulleqref{eq: SDs amplitudes definition 1}: we can simply fix $g$ to zero. However, to be consistent, we must still take into account the experimental uncertainty on $T_\gamma(0)$. For this purpose,
we marginalize over a temperature shift distortion to second order, i.e., we marginalize over the parameter $\Delta_T$ which affects the final spectrum as
\begin{equation}
	\Delta I_\mathrm{marg}(x) = \Delta_T (1 + \Delta_T) \mathcal{G}(x) + \Delta_T^2/2 \, \mathcal{Y}(x)~.
\end{equation}
For the nuisance parameter $\Delta_T$, we choose a Gaussian prior with a standard deviation matching the current experimental resolution of the CMB \BB temperature, which has a relative uncertainty of about $\sigma(T_\gamma(0))/T_\gamma(0) = 0.00022$ \cite{fixsen1996cosmic, fixsen2009temperature}. Since experiments like PIXIE or PRISM are designed for a differential measurement of the CMB frequency spectrum, it is not clear that they will significantly reduce this error. To be conservative, we keep the same $\sigma(T_\gamma(0))$ even in our PIXIE and PRISM mock likelihoods. Together with the set of experimental errors $\delta I_\mathrm{noise}(\nu_i)$, this marginalization over $\Delta_T$ contributes to the final uncertainty on the extracted $y$ distortions. Note that the typical \finalr{$g$ distortion}{temperature shift} coming from \fulleqref{eq: SDs amplitudes definition 1} is of order $10^{-8}-10^{-9}$, thus several orders of magnitude smaller than $\sigma(T_\gamma(0))$. Therefore, in practice, setting $g=0$ in the likelihood or keeping the value inferred from \fulleqref{eq: SDs amplitudes definition 1} makes no difference.
\dnew
Similarly to what was done in \cite{chluba2014teasing}, we could also single out the contributions from reionization, and marginalize over the corresponding $y$ parameter as well as the galaxy-dependent quantities described in Section \ref{ssec:heatingLCDM}. However, here we assume that every contribution from the epoch of reionization can be modeled and calculated with better precision than present and future measurement errors, and subtracted from the data. As a consequence, we set it to zero in the fiducial and fitted models, and we do not marginalize over $y_{\rm reio}$\,. Note that neglecting all late time effects (such as the SZ effect) is an optimistic approximation since in reality, uncertainties in the modeling of reionization could leave residuals at the level~of~\mbox{$\delta y\approx 10^{-8}- 10^{-7}$}.
\dnew
With this framework we have developed likelihoods for three detectors: the existing FIRAS \cite{fixsen1996cosmic}, and the proposed future missions PIXIE \cite{kogut2011primordial} and PRISM \cite{andre2014prism}. For the latter two we assume idealized scenarios, in which the foreground removal is almost optimal within a certain range of frequencies. This frequency range is discussed and motivated in \cite{Vince2015, DeZotti:2015awh}. A more realistic treatment of the foregrounds, as in \cite{Abitbol:2017vwa}, is left for future work. Such a treatment could degrade the sensitivity to $\mu$ distortions in particular, due to degeneracies with low-frequency foregrounds \cite{Abitbol:2017vwa}.
However, modifications of future instruments and combination with external datasets can improve the foreground-removal capabilities \cite{Kogut2019}. Thus our estimates provide at least a useful benchmark.
\dnew
For FIRAS we have used the binned frequency array provided in Table 4 of~\cite{fixsen1996cosmic}. For PIXIE we have made the same assumptions as in~\cite{chluba2013distinguishing, chluba2014teasing}, which are based on the more extensive calculation conducted in \cite{kogut2011primordial}: we have assumed equidistant, independent frequency channels in the range $\left[30\,\rr{GHz} -1\,\rr{THz} \right]$, with a bin width of $\Delta \nu=15\,\rr{GHz}$. Furthermore, we have assumed that the measurement is only limited by uncorrelated instrumental noise, and all foregrounds can be removed with higher frequency channels. This gives us an overall constant noise of $\delta I_{\rm noise}(\nu_i)\approx 5\times 10^{-26}\,$W\,m$^2$\,s$^{-1}\,$Hz$^{-1}$\,sr$^{-1}$ for each frequency bin (see Appendix~\ref{ap:PIXIE_noise} for a detailed calculation). Finally, for PRISM we assumed the same frequency channel characteristics as PIXIE, but with a sensitivity improved by one order of magnitude, i.e. $\delta I_{\rm noise}(\nu_i)\approx 5\times 10^{-27}\,$W\,m$^2$\,s$^{-1}\,$Hz$^{-1}$\,sr$^{-1}$. This choice is compatible with (although slightly more optimistic than) the forecasted PRISM performance proposed in Table 2 of~\cite{andre2014prism}.
\dnew
The detector FIRAS has a corresponding sensitivity to the $\mu$ parameter of $\delta\mu = 3\times10^{-5}$, PIXIE of $\delta\mu = 9\times10^{-9}$, and PRISM of $\delta\mu = 9\times10^{-10}$ (at 2$\sigma$ level). 
These levels of precision are optimistic compared to those found respectively in references \cite{fixsen1996cosmic, kogut2011primordial, andre2014prism}, as we do not marginalize over the $y$-contribution from reionization. It is thus more instructive to compare these sensitivities to the $\mu$ distortion parameter, since the spectral sensitivity $\delta I_{\rm noise}(\nu)$ is usually affected significantly by the precise design of an experiment and the final difficulty of foreground subtraction. The given bounds for the $\mu$ parameter identify the detector sensitivity in a less experiment-dependent way. Any other experiment with a similar bound on $\mu$ can still expect similar exclusion regions, even if their $\delta I_\mathrm{noise}(\nu)$ is not the same.
\dnew
The fact that our code can accept arbitrary experimental setups as an input has a crucial advantage: it allows to analyze the influence of different choices for the frequency array (see the difference between FIRAS and PIXIE), which may affect the ability to disambiguate $\mu$ and $y$ distortions, as well as the role of the sensitivity (see the difference between PIXIE and PRISM), which determines the overall constraining power. This method can also be applied to optimize the characteristics of a planned \SD mission for a given physics-motivated target.

\section{Results \label{results}}
In the following section we are going to apply the theoretical framework described in Section~\ref{Theory} to different cosmological models, using the numerical implementation defined in Section~\ref{method}. In particular, the main goal is to show many different aspects of the synergy between CMB \SD{s} and anisotropies, extending and complementing the analysis of \cite{chluba2013distinguishing, chluba2014teasing}.
\dnew
As a first application of our setup, in Section~\ref{results_running} we will explore the constraining power provided by the combination of \SD{s} and CMB anisotropies on a minimal extension of the $\Lambda$CDM model with a running of the spectral index of the primordial power spectrum, $n_{\rm run}$\,. Then, in Sections \ref{results_annihilation}, \ref{results_decay}, and \ref{results_PBH}, we will consider extensions of $\Lambda$CDM which include exotic energy injection mechanisms like DM annihilation, DM decay, and the evaporation of PBHs respectively. For each of these models we will single out the role of new physical parameters and investigate how strongly the constraints are affected by the inclusion of \SD. 
\dnew
To cover a large spectrum of physical effects and detector sensitivities, we will base our analysis on completed missions like Planck \cite{ade2016planck} and FIRAS \cite{fixsen1996cosmic}, which we will then compare to future experiments or proposed projects with improved characteristics such as LiteBIRD~\cite{Matsumura:2013aja, Suzuki2018LiteBIRD}, CMB-S4 \cite{Abazajian:2016yjj, Abitbol2017CMB, Abazajian:2019eic}, PIXIE \cite{kogut2011primordial}, and PRISM \cite{andre2014prism} (considered both separately and/or in combination with one another). For all these experiments, we employ mock likelihoods\footnote{As long as we are only interested in the sensitivity to cosmological parameters, we can use mock likelihoods even when we account for completed experiments like FIRAS and Planck. This means that we do not include real data, but use a likelihood with the same sensitivity as the actual one, leading to the same error bars. This approach offers several technical and numerical advantages over using the actual likelihoods \cite{Brinckmann:2018owf}.}, making use of the method described in Section \ref{likelihoods}, and following the prescriptions~of~\cite{Brinckmann:2018owf}. 
\dnew
For all following examples, the fiducial models mentioned in Section \ref{likelihoods} for all likelihoods have been created assuming $\Lambda$CDM, with the parameter basis $\{h, \omega_{\rm b}, \omega_{\rm cdm}, n_s, A_s, z_{\rm reio} \}$ and fiducial values close to the Planck 2018 best-fit values \cite{Aghanim:2018eyx}.
\dnew
For the particular case of the \SD likelihoods, on top of exotic energy injection rates from DM or PBHs, we only considered the contributions to the heating rate from adiabatic cooling of electrons and baryons, and dissipation of acoustic waves. The additional $y$ parameters caused by the CMB multipoles and by reionization are implemented in our code but deactivated in this analysis, as we treat them as perfectly distinguishable contributions, so that it is always possible to subtract them from an eventual observation.
\dnew
Furthermore we also neglect the presence of galactic and extra-galactic observational foregrounds, which are expected to have a large impact at frequencies above 1 THz, as discussed e.g. in \cite{Vince2015,DeZotti:2015awh,Abitbol:2017vwa} (see Figure~2 of \cite{Vince2015} and Figure~1 of \cite{DeZotti:2015awh}). Our current approach consists in considering only frequencies $\nu \leq 1$ THz in the mock likelihoods. We leave a more accurate implementation of these foregrounds for future work.

\subsection{Running spectral index}\label{results_running}
\begin{figure}[t]
	\centering
	\includegraphics[width=8 cm, height=8 cm]{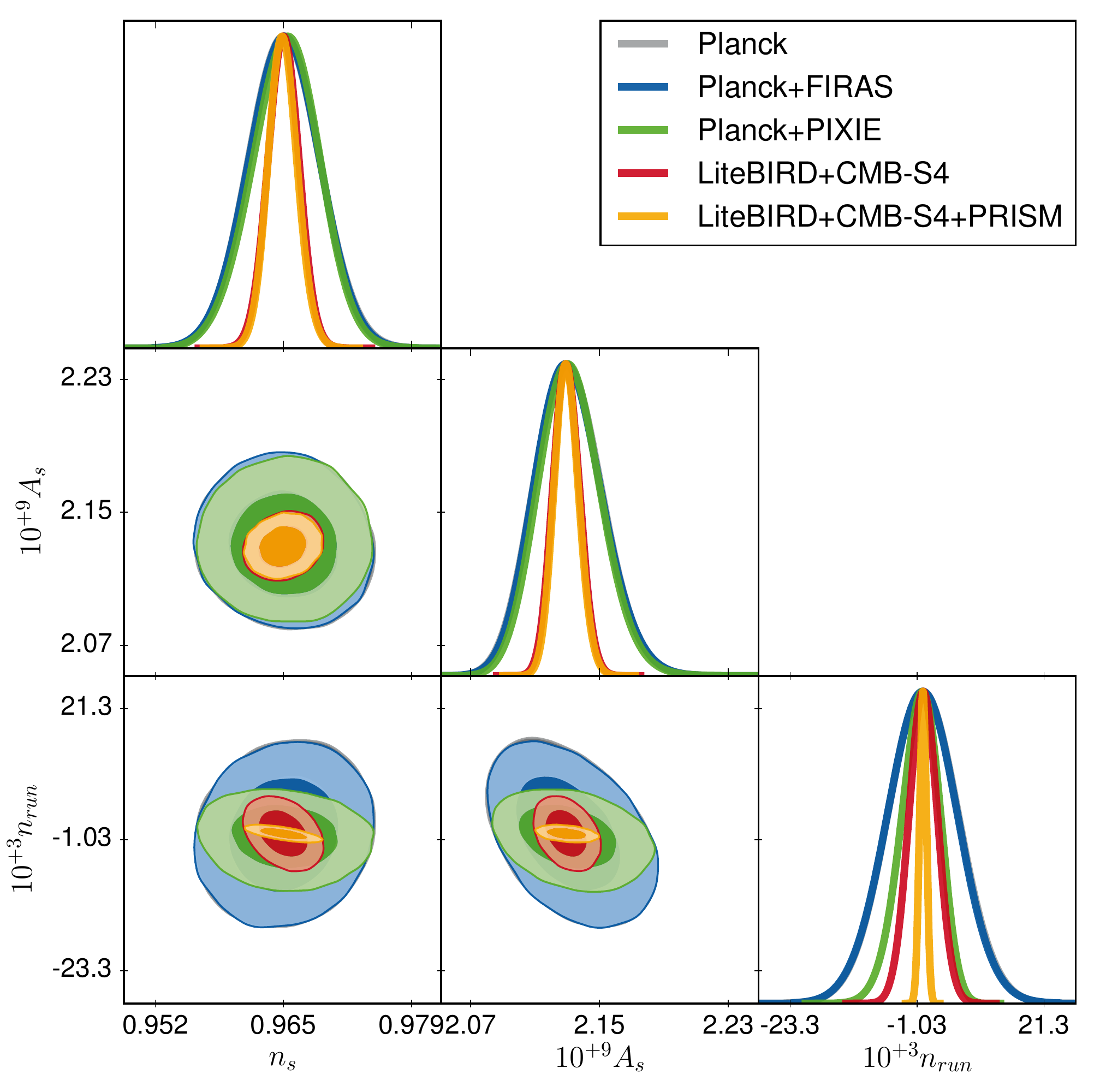}
	\includegraphics[width=6 cm, height=6 cm]{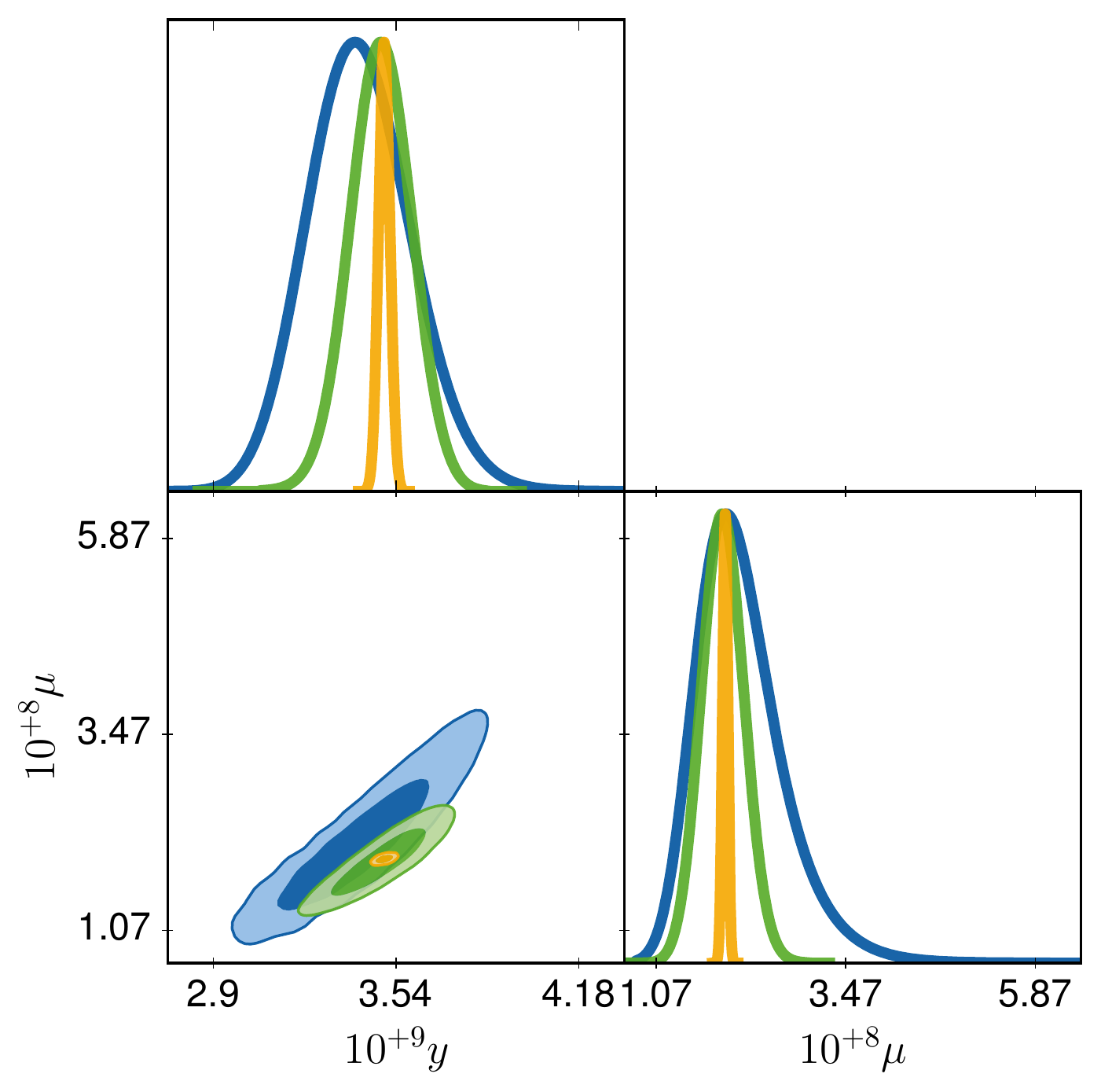}
	\caption{One-dimensional posteriors and two-dimensional contours (68\% and 95\%  CL) for the parameters mostly affected by \SD{s}, i.e. the spectral amplitude $A_s$, the spectral index $n_s$ and its running $n_{\rm run}$, and the derived \SD parameters $y$ and $\mu$.}
	\label{fig: MCMC_run}
\end{figure}

Since the main dependence of the heating rate predicted by $\Lambda$CDM is on the primordial power spectrum, the first cosmological model to which we apply our joint CMB anisotropy and SD pipeline is a minimal extension of the standard $\Lambda$CDM model including the running of the spectral index $n_\mathrm{run}$. This additional parameter is predicted to be very small in single-field slow-roll models \cite{Kosowsky1995CBR}, and constraining its value could help distinguish between several inflationary scenarios (see e.g., \cite{Silverstein2008Monodromy, Czerny2014Running, Minor2015Inflation}). 
\dnew
In this case, although the constraining power of CMB anisotropy measurements has already provided remarkably good results, the inclusion of \SD{s} could not only improve the current bounds, but also extend their validity up to much smaller scales than those considered by any previous or future CMB anisotropy mission \cite{Chluba:2014sma,Byrnes:2018txb} (see e.g. in Figure~4 of \cite{Chluba:2014sma} and Figure~9 of \cite{Byrnes:2018txb}). In fact, the power-law scale-dependence of the primordial power spectrum $P_\mathcal{R}(k)$ at Fourier modes higher than 1 Mpc$^{-1}$ is still poorly constrained, and considering PIXIE-like experiments would help us achieve tighter bounds up to scales in the order of $10^{4}$~Mpc$^{-1}$. The distortions are furthermore determined by small-scale perturbations while they still are in the linear regime. This is a different situation than with future 21cm and LSS surveys, in which the effects of a running spectral index and of non-linear structure formation could be difficult to discriminate at $k>1$~Mpc$^{-1}$ due to non-linear effects.
\dnew
We fitted the fiducial model with our mock likelihood and flat priors on the parameters $\{h, \omega_{\rm b}, \omega_{\rm cdm}, n_s, A_s, z_{\rm reio}, n_\mathrm{run} \}$. The amplitude and spectral index are defined at the pivot scale $k=0.05$ Mpc$^{-1}$. In Figure \ref{fig: MCMC_run} we show the resulting two-dimensional posterior distributions of the three primordial power spectrum parameters, as well as the derived $y$ and $\mu$ parameters for Planck, Planck$+$FIRAS, Planck$+$PIXIE, LiteBIRD$+$CMB-S4, and LiteBIRD$+$CMB-S4$+$PRISM. A summary of the bounds obtained is given in Table \ref{tab:best fits lcdm}.
\begin{table}
	\centering
	\scriptsize
	\label{tab:best fits lcdm}
	\caption{Expected $1\sigma$ sensitivity to the parameters of the $\Lambda$CDM model extended with the running of the spectral index, of different combinations of CMB anisotropy and SD experiments. We also report the 95\% CL upper bounds on the derived $y$ and $\mu$ parameters.}
	\begin{tabular}{|c|c|c|c|c|c|}
		\hline
		Parameter & Planck & Planck+FIRAS & Planck+PIXIE & LiteBIRD+CMB-S4 & \thead{\scriptsize LiteBIRD+CMB-S4\\ \scriptsize +PRISM} \\
		\hline
		$\sigma\left(10^{2}\omega_{\rm b}\right)$ & $0.016$ & $0.017$ & $0.016$ & $0.0038$ & $0.0031$\rule{0pt}{3.0ex} \\[0.1 cm] 
		$\sigma\left(\omega_{\rm cdm}\right)$ & $0.0013$ & $0.0013$ & $0.0012$ & $0.00029$ & $0.00027$\\[0.1 cm]
		$\sigma\left(h\right)$ & $0.0058$ & $0.0057$ & $0.0057$ & $0.0012$ & $0.0011$ \\[0.1 cm]
		$\sigma\left(z_{\rm reio}\right)$ & $0.46$ & $0.46$ & $0.45$ & $0.20$ & $0.20$ \\[0.1 cm]
		$\sigma\left(10^{9}A_s\right)$ & $0.021$ & $0.021$ & $0.020$ & $0.0083$ & $0.0079$ \\[0.1 cm]
		$\sigma\left(n_s\right)$ & $0.0038$ & $0.0037$ & $0.0036$ & $0.0017$ & $0.0016$ \\[0.1 cm]
		$\sigma\left(10^{3}n_{\rm run}\right)$ & $6.4$ & $6.3$ & $3.5$ & $2.6$ & $0.60$ \\[0.1 cm]
		\hline
		$10^{9}y$ & - & $<3.79$ & $<3.70$ & - & $<3.54$\rule{0pt}{3.0ex} \\[0.1 cm]
		$10^{8}\mu$ & - & $<3.36$ & $<2.46$ & - & $<2.01$ \\
		\hline
	\end{tabular}
\end{table}
\dnew
As expected, the addition of FIRAS to the Planck likelihood does not provide any additional information. With the addition of PIXIE, while the bounds on $A_s$ and $n_s$ do not change with respect to the Planck ones, a substantial improvement on $n_{\rm run}$ bounds is present, which shows that \SD{s} have the potential to improve the current CMB bounds on the running of the spectral index \cite{chluba2012cmb, Khatri2013forecast, chluba2014teasing, Cabass2016}. This improvement of the error bars, driven by the presence of the PIXIE, also suggests that \SD missions would indeed be able to extend the validity of the bounds up to $10^{4}$ Mpc$^{-1}$, as mentioned before. Similarly, the addition of PRISM to the LiteBIRD+CMB-S4 case significantly improves the bounds by a factor 4.5.
It is interesting to note that in the particular case of the running of the spectral index, the addition of PIXIE to the Planck likelihood provides nearly the same constraints as the combination of LiteBIRD and CMB-S4.
\dnew
Note that in Figure~\ref{fig: MCMC_run}, there is a vertical offset between the best-fit values of the $y$ and $\mu$ parameters in the different forecasts, while one would expect all contours to be centered around some fixed fiducial values. Still, this is consistent. Indeed, Section~\ref{Greens} show that the choice of the frequency array for a given detector strongly influences the Gram-Schmidt decomposition into $y$ and $\mu$ distortions. This also affects the shape of the branching ratios. As a consequence, the position of the $y$ and $\mu$ contour in Figure~\ref{fig: MCMC_run} is affected by the frequency array. This can be interpreted as the detector disambiguating $\mu$ and $y$ distortions differently. Our bounds on other parameters are derived from the full distortion shape, and are not affected by this disambiguation effect. On the other side, the width of the contour is set by the sensitivity of the mission, as clear from the comparison of the PIXIE and PRISM cases.
\dnew
Finally, note that there are several other modifications of the standard primordial power spectrum that could be tested with \SD{s}. Some particularly interesting examples are, for instance, given in Section~5 of \cite{Chluba2012inflaton}, \cite{chluba2013distinguishing}, and references therein. 
\dnew
These include the specific cases of steps in the amplitude or slope of the primordial power spectrum. Additional extensions of the standard power spectrum could include the running of the running or even higher derivative parameters, to stay within this framework of slow-roll inflation (see e.g. \cite{Cabass2016SRparams}). In order to constrain this class of models, \SD{s} might be particularly well suited. In fact, to avoid the tight CMB anisotropy constraints, many models predict deviations from the the standard primordial power spectrum only at scales larger than 1 Mpc$^{-1}$. As already mentioned before, in this region of parameter space, \SD{s} are a more powerful constraining tool than CMB anisotropies.

\subsection{Dark matter annihilation}\label{results_annihilation}
We performed a similar analysis for a $\Lambda$CDM model including s-wave DM annihilation, parametrized with the annihilation efficiency parameter $p_\mathrm{ann}$ presented in \ref{ssec:heatingExotic} (under the simplified and optimistic assumption $f_\mathrm{eff}(z)=1$). We do not show our results for that case, because they simply confirm erlier works \cite{chluba2013distinguishing} in that \SD experiments will not improve bounds on $p_\mathrm{ann}$ with respect to CMB anisotropy observations.
\dnew
We should note, however, that if DM annihilation was detected, the combination of CMB anisotropy and \SD{} data would be very useful for discriminating between different models, since the spectral distortion spectrum and the angular polarization spectrum would be sensitive to different assumptions on the branching ratios and on the energy spectrum of the annihilation products.

\subsection{Dark matter decay}\label{results_decay}
In the third case we are going to focus on, which is the decay of DM into electromagnetic particles, the importance of \SD{s} is particularly striking. Indeed, in this scenario, the energy injection history directly depends on the lifetime and abundance of the decaying DM particle \cite{chluba2011evolution, chluba2013distinguishing, chluba2014teasing}. Consequently, constraints from CMB anisotropies and \SD{s} are complementary and apply to two separate regions of the parameter space. On the one hand, DM particles with lifetimes between $10^5$ s and $10^{12}$ s mainly affect the thermal history of the universe before recombination, thus almost exclusively causing \SD{s}. On the other hand, lifetimes between $10^{12}$ s and $10^{25}$ s are strongly constrained by CMB observations (see e.g., Figure~11 of \cite{chluba2014teasing} and Figure~5 of \cite{poulin2017cosmological} for a graphical representation).
\dnew
In this analysis we consider a simplified model with $f_{\rm eff}=1$, i.e., we adopt the on-the-spot approximation (wich is normally very accurate for energy injection prior to recombination) and we assume that all decay products interact electromagnetically.
\dnew
In order to derive bounds on the fraction $f_{\rm frac}$ of DM that decays and its lifetime $\tau_{\rm dec}=1/\Gamma_{\rm dec}$, we consider a 6+2 extension of the standard $\Lambda$CDM model with flat priors on $\{h, \omega_{\rm b}, \omega_{\rm cdm}, n_s, A_s, z_{\rm reio} \} + \{\log_{10} f_{\rm frac}, \log_{10} \tau_{\rm dec} \}$, and we scan the parameter space with an MCMC sampler. However, because of the non-convex topology of the parameter space spanned by $\log_{10} f_{\rm frac}$ and $\log_{10} \tau_{\rm dec}$, a single MCMC would be extremely slow to converge.
\begin{figure}[t]
	\centering
	\includegraphics[width=12 cm, height=9 cm]{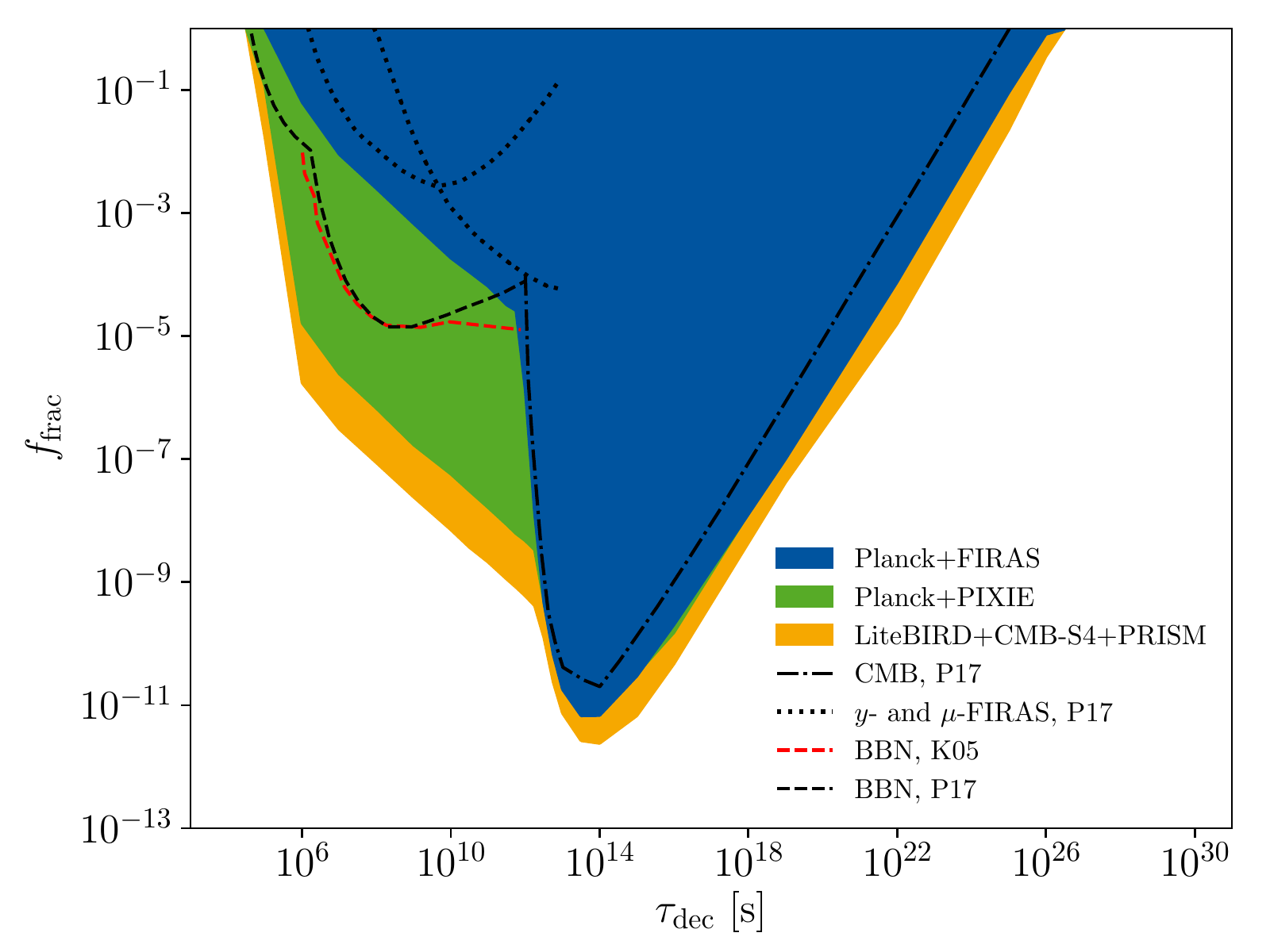}
	\caption{95\% CL bounds on the decaying DM fraction as function of the particle lifetime. The black lines correspond to the bounds displayed in Figure~5 of \cite{poulin2017cosmological} (labeled P17 for brevity), while the red dashed line corresponds to the BBN constraints from \cite{kawasaki2005big} (adapted from \cite{chluba2014teasing} and labeled K05).}
	\label{fig: decay_exclusion_region}
\end{figure}
\dnew
In order to avoid this issue, we slice the parameter space along $\tau_{\rm dec}$ at several values in the range between $10^4$ s and $10^{27}$ s, and we sample the remaining 6+1 dimensional parameter space with separate MCMC runs. As a result, we obtain an array of values for $f_{\rm frac}$ that we can interpolate along the $\tau_{\rm dec}$ slices. For each slice we adopt a top-hat prior on $\log_{10} f_{\rm frac}$ in the range from $-13$ to 0. 
\dnew
The results are displayed in Figure~\ref{fig: decay_exclusion_region}. The colored regions represent regions of parameter space \textit{excluded} by the condition $\chi^2 > \chi^2_\mathrm{min}+4$. We also show the contours calculated in \cite{poulin2017cosmological} (dotted line for \SD{s} and dashed-dotted line for CMB anisotropy constraints). These are to be compared with the case of Planck+FIRAS (blue region) and several details are worth noticing. First of all, for the exclusion bounds in the range $10^5 \mathrm{s} <\tau_{\rm dec} < 10^{13} \mathrm{s}$, which are dominated by SDs, the authors of \cite{poulin2017cosmological} derived some separate bounds from $y$ and $\mu$ distortions based on \cite{fixsen1996cosmic}. Thus our contours are smoother and in principle more reliable, especially in the region where $y$ and $\mu$ distortions are produced in similar proportions.
\dnew
When moving to higher values of the DM lifetime, for which the bounds are dominated by CMB anisotropies, we also notice that we have slightly tighter bounds than in \cite{poulin2017cosmological}. This can be attributed mainly to different assumptions on $f_\mathrm{eff}$ (we take $f_\mathrm{eff}=1$, while  \cite{poulin2017cosmological} assumes two specific values $f_\mathrm{eff}<1$ motivated by particular decay channels), as well as minor differences in the MCMC analysis:
the choice of lower prior boundary for $\ln f_\mathrm{frac}$ and the slightly different likelihood for Planck -- full Planck 2015 likelihood in \cite{poulin2017cosmological}, mock likelihood approximating Planck 2018 in this work.
\dnew
Finally, in Figure~\ref{fig: decay_exclusion_region}, we compare the limits on the decaying DM fraction set by SDs with the BBN bounds derived in \cite{kawasaki2005big} and \cite{Poulin:2015opa}. At the moment, BBN bounds are nearly two orders of magnitude better than FIRAS bounds. However this situation is likely to change in the future. On the one hand, we only expect marginal improvements on the BBN side (and the comparison of \cite{kawasaki2005big} and \cite{Poulin:2015opa} only shows a very small improvement over one decade). On the other hand, a PIXIE-like mission would improve over FIRAS bounds by up to two orders of magnitude, and PRISM by almost three orders, leading to much stronger bounds than BBN.
\vspace*{-1\baselineskip}

\subsection{Primordial Black Hole evaporation}\label{results_PBH}
\begin{figure}[t]
	\centering
	\includegraphics[width=12 cm, height=9 cm]{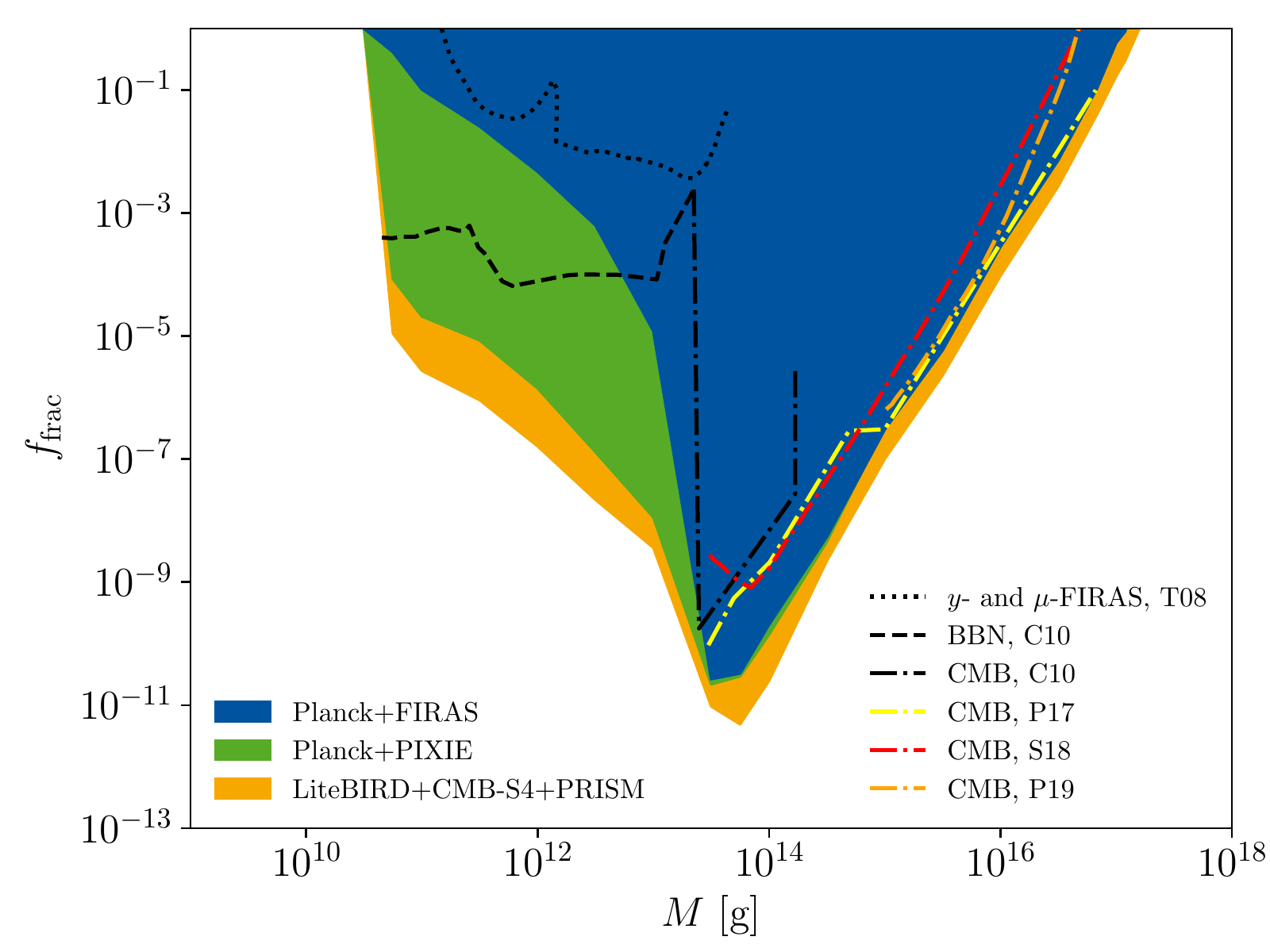}
	\caption{95\% CL bounds on the PBHs fraction as function of their mass. The black dashed-dotted line corresponds to the bounds from \cite{carr2010new} (here labeled C10 for brevity), the red one to the bounds from \cite{poulin2017cosmological} (P17), the orange one to the bounds from \cite{stocker2018exotic} (S18) and yellow one to the bounds from \cite{Poulter:2019ooo} (P19). The dotted black line represents the \SD{s} constraints and is taken from \cite{tashiro2008constraints}, while the BBN constraints are the same as in \cite{carr2010new}.}
	\label{fig: PBH_exclusion_region}
\end{figure}
As already seen in Section \ref{ssec:heatingExotic}, the heating rate caused by the evaporation of PBHs has very similar characteristics to that of decaying DM. However, PBHs are generated through primordial overdensities that are directly related to the shape of the inflationary curvature perturbations, and can thus be used to constrain the inflationary period itself, in addition to the many other effects already mentioned in Section \ref{ssec:heatingExotic} (and references therein). Additionally, PBHs allow for a possible explanation of DM which does not involve modifications of the standard model of particle physics.
\dnew
It is thus very important to test this model at all scales. As in the case of DM decay, \SD{} constraints lay in a region of the parameter space where bounds from CMB anisotropies are absent, and can, therefore, play a crucial and complementary role. For this purpose, the authors of \cite{tashiro2008constraints} already computed semi-analytical \SD bounds on the PBH mass using FIRAS, while CMB anisotropy bounds were estimated by \cite{carr2010new, poulin2017cosmological, stocker2018exotic, Poulter:2019ooo}. Moreover, in \cite{carr2010new}, the FIRAS bounds have been compared to BBN constraints.
\dnew
In comparison, the current analysis brings several improvements, such as the calculation of the full \SD{} shape, a joint fit to current and future CMB anisotropy and \SD{} data, and the development of a new approximation to calculate the heating rate caused by PBHs at early times, especially when QCD effects are important. The latter can be found in Appendix \ref{ap:f_of_m}.
\newpage
\dnew
We consider an extension of $\Lambda$CDM featuring the fraction of DM in the form of PBHs, $f_\mathrm{frac}$, and the individual black hole mass $M$.
For each value of $M$ in the interval $[10^{10}-10^{18}]$ g, we scan the remaining 6+1 dimensional parameter space with an MCMC sampler, assuming a top-hat prior on $-15 < \log_{10} f_\mathrm{frac} \leq 0$.
\dnew
Our results are shown in Figure \ref{fig: PBH_exclusion_region}. The blue region corresponds to the Planck$+$ FIRAS case, and can be compared with the dashed-dotted lines for previous CMB anisotropy bounds (black for \cite{carr2010new}, red for \cite{poulin2017cosmological}, orange for \cite{stocker2018exotic}, and yellow for \cite{Poulter:2019ooo}), with the dotted line for previous \SD{s} constrains, and finally with dahsed lines for previous BBN constraints.

\dnew
For PBH masses bigger than approximately $10^{13.5}$ g, the constraints are dominated by CMB anisotropies. Comparing with previous bounds, we observe that for PBH masses larger than $10^{15}$ g the agreement is very good, in particular with respect to the more recent calculations performed in \cite{Poulter:2019ooo}. However, in the mass range between $10^{13.5}$ g and $10^{15}$ g our results disagree with other references by up to one order of magnitude. This is mostly due to our treatment of the QCD-confinement scales. Indeed, all references but \cite{stocker2018exotic} do not account for the effects of QCD confinement. However, the public version of {\sc ExoCLASS} used in \cite{stocker2018exotic} appears to have a small typo in the calculation of the factor $Q_{\rm QCD}$, which can be identified by comparing with \fulleqref{eq:QCD coefficient}. As a consequence, in all previous references, the value of $\mathcal{F}(M)$ was larger than in the current analysis (to appreciate the impact of the $Q_{\rm QCD}$ factor, one can compare Figure 4 and Figure 5 of \cite{stocker2018exotic}).
With our updated modeling of $\mathcal{F}(M)$, the energy injection rate is smaller for each given value of the mass $M$, but evaporation and energy injection take place over a considerably longer span of time, leading to stronger CMB anisotropy bounds.
\dnew 
For PBH masses smaller than approximately $10^{13.5}$ g, the constraints from CMB \SD{s} become dominant, and our results can be compared with those of \cite{tashiro2008constraints}. However, this pioneering analysis is more than a decade old, and has several limitations. First, it relies on WMAP best-fits values, with differences up to 10\% with respect to our fiducial values. Second, their approximation of the branching ratios is relatively strong (see the second approximation mentioned in Appendix \ref{ap:br}), and neglects the gradual transition between the $\mu$ and $y$ eras. Third, the computation of $\mathcal{F}(M)$ in  \cite{tashiro2008constraints} omits several of the particles listed in Table \ref{tab: PBH mass evolution parameters}, as well as QCD confinement effects, which could explain the relatively strong deviation from our predictions at mass scales between $10^{12}$ g and $10^{13.5}$ g. Finally, as already discussed in the case of decaying DM, employing the FIRAS bounds on the $y$ and $\mu$ parameters computed in \cite{fixsen1996cosmic} can introduce a difference by a factor 2 to 10 with respect to our 95\% CL limits, that are derived from the full distortion $\Delta I_\mathrm{tot}(\nu)$ (our bounds are however slightly optimistic since we neglect foreground contamination).
\dnew
Finally, we find that PIXIE-like missions would significantly improve the bounds for PBH masses between $10^{10}$ g and $10^{13.5}$ g. Like in the case of decaying DM, the gain with respect to BBN bounds is between 1 and 3 orders of magnitude for PIXIE, while the limits from PRISM show an even larger margin for improvement.
\dnew
\final{Note that all the aforementioned bounds originate only from the thermal effects that the PBH evaporation has on the CMB photons. This means that, for instance, we currently neglect the additional constraints caused by the transfer of energy density from DM (in form of PBHs) to baryonic matter or radiation. These additional contributions, depending on the injection time, would further affect BBN and the different CMB power spectra, thus increasing the deviations from $\Lambda$CDM. In this respect, our bounds can be considered conservative. It is, however, important to underline that such effects would only affect the results presented in this section for PBH fractions lower than roughly $10^{-3}-10^{-2}$, which corresponds to the current observational precision of the different energy densities. In our case, this would correspond to PBH masses below $10^{11}-10^{12}$ g. Since a precise implementation of the additional constraints presents several challenges such as a proper modeling of the secondary cascades, the inclusion of these additional contributions is left for future work.}

\section{Conclusions\label{conclusions}}
Within this work we have presented a detailed and pedagogical introduction to the topics of \SD{s} and energy injection throughout the thermal history of the universe. We have described the origin of \SD{s}, shown how to calculate them, and discussed their significance to our understanding of the universe. We have also touched upon the complexity of the various energy injection mechanisms in the early and late universe.
\dnew
We have subsequently presented the numerical implementation of \SD{s} and several energy injection mechanisms in the Boltzmann solver {\sc class}. Although the actual content of our extensions are largely based on some modules of two previously published codes, {\sc CosmoTherm} and {\sc ExoCLASS}, we have developed a unified numerical framework with improved structure, homogeneous conventions, and several new features. For example, it is now possible to compute \SD{s} for arbitrary heating histories and any detector setup, specifying the precise detector sensitivity per frequency bin.
\dnew
The inclusion of \SD{s} in the more general framework of {\sc class} allows for a parallel computation of CMB \SD{s}, CMB anisotropies, the matter power spectrum, and many other cosmological quantities. This synergy is particularly well suited for models such as decaying DM and evaporating PBHs, whose free parameters can span several orders of magnitude, influencing the whole thermal history of the universe at all scales. We have shown the substantial benefit that \SD{s} can bring to the other more conventional probes, either providing complementary bounds, such as in the case of CMB anisotropies, or even surpassing current constraints by 2-3 orders of magnitude, as in the case of BBN.
\dnew
Although the current implementation already considers a large variety of physical effects and can be applied to several models, there is still much work to be done. First of all, our calculation of \SD{s} in the $\Lambda$CDM model does not yet include CRR. This could be achieved implementing the work already done in the framework of {\sc CosmoSpec} into our new version of {\sc class}, which would then include all effects of $\Lambda$CDM. Furthermore, an investigation of the frequency dependence of energy injection mechanisms and its observable effects is left for future work, following the lines of \cite{Chluba:2015hma}. Additionally, to make our forecast more realistic, we should incorporate a precise treatment of observational foregrounds in the \SD{} likelihoods.
\dnew
Aside from these limitations, with one single code exploiting the synergy between CMB \SD{s} and anisotropies, it is now possible to constrain many cosmological quantities over a range of values spanning up to 25 orders of magnitude, as we have seen in the case of decaying DM. This opens the door to the analysis of several other interesting models, including for instance non-standard inflationary scenarios, or models with interactions between particles in the visible and dark sectors. In this work we have shown the far-reaching possibilities of combining CMB anisotropies and \SD{s}, and with the framework we have developed, we are now prepared to fully make use of this synergy.
\vspace*{-2\baselineskip}

\section*{Acknowledgements}
We thank Patrick St\"{o}cker, Yacine Ali-Ha\"{i}moud, and Chris Byrnes for very useful exchanges. We especially want to thank the referee for the constructive suggestions that have led to more clarity in our paper. ML is supported by the \tquote{Probing dark  matter with neutrinos} ULB-ARC convention. NS acknowledges support from the DFG grant~\mbox{LE~3742/4-1}. JL acknowledges support for this project from the DFG. DH was supported by the DFG grant~\mbox{LE~3742/3-1}, and during the last stages of this work, by the FNRS research grant number~\mbox{F.4520.19}. This work has received funding from the European Research Council (ERC) under the European Union's Horizon 2020 research and innovation program (grant agreement~No~725456, CMBSPEC) as well as the \mbox{Royal~Society} (grant~UF130435). Simulations for this work were performed with computing resources granted by JARA-HPC from RWTH Aachen University under the project jara0184.
\clearpage

\appendix
{\small
\addtocontents{toc}{\protect\setcounter{tocdepth}{1}} 

\section{Further details on the photon Boltzmann equation \label{ap:photon_Boltzmann}}

\subsection{A note on partial derivatives and the photon Boltzmann equation \label{ap:parder}}
It is a well known fact that a partial derivative depends on the chosen coordinate system, in such a sense that it depends on which quantities are held constant while taking the derivative. This especially applies to the choice of holding $x$ or $p$ constant while taking the partial time derivative of the photon phase-space distribution.
\dnew
To appreciate this fact, note that we have explicitly defined $x=p/T_z \propto a \cdot p$ so that $\id x/ \id t = 0$. This implies that the derivative holding $p$ constant has to keep track of the the phase-space distribution redshifting through the decreasing photon momentum, while the term holding $x$ constant already accounts for that fact.
\dnew
Let us denote the quantity we hold constant during the partial derivative as an index to the brackets of the derivative, i.e. $(\partial y/\partial x)_a$ will hold the quantity $a$ constant while taking the partial derivative of $y$ with respect to $x$.
\dnewnoindent
We then find explicitly for the general photon Boltzmann equation
\begin{equation}\label{eq:photonboltzmann}
\begin{split}
C[f] = \der{f(q^\mu,p^\mu)}{t} & = \parderconst{f}{q^\mu}{{p^\mu}}\der{q^\mu}{t} + \parderconst{f}{p^\mu}{{q^\mu}}\der{p^\mu}{t} ~.
\end{split}
\end{equation}
We now split the dependencies into time and space components as $q^\mu=(t,\mathbf{q})$ and $p^\mu =(E,p \mathbf{n})$. Note that since non-virtual particles obey $E^2 = p^2 + m^2$ with constant mass $m$, the dependence on $E$ is equivalent to the dependence on $p$. We could choose either variable to obtain
\begin{equation}\label{eq:photonboltzmann2}
\begin{split}
C[f] = \der{f(t,p)}{t} & = \parderconst{f}{t}{p}\der{t}{t} + \parderconst{f}{\mathbf{q}}{p}\der{\mathbf{q}}{t} + \parderconst{f}{\mathbf{n}}{t}\der{\mathbf{n}}{t} + \parderconst{f}{p}{t}\der{p}{t}~.
\end{split}
\end{equation}
Using the fact, that $\partial f /\partial \mathbf{q}=0$ and $\partial f /\partial \mathbf{n}=0$ for the homogeneous background solution, we find additionally that
\begin{equation}\label{eq:photonboltzmann3}
\begin{split}
C[f] = \der{f(t,p)}{t} & = \parderconst{f}{t}{p} + \parderconst{f}{p}{t}\der{p}{t}~.
\end{split}
\end{equation}
Finally, using $\id p/\id t = -H p $ from \fulleqref{eq:p prop to a}, we find
\begin{equation}\label{eq:photonboltzmann4}
\begin{split}
C[f] = \der{f(t,p)}{t} & = \parderconst{f}{t}{p} - Hp \parderconst{f}{p}{t}~.
\end{split}
\end{equation}
Substituting now $p \to x(t,p)$, we obtain instead also
\begin{equation}
C[f] = \der{f(t,x)}{t} = \parderconst{f}{t}{x} + \der{x}{t}\parderconst{f}{x}{t} = \parderconst{f}{t}{x}~,
\end{equation}
which is equivalent to saying
\begin{equation}
	\parderconst{f}{t}{p} - Hp \parderconst{f}{p}{t} = \parderconst{f}{t}{x}~.
\end{equation}
Holding $x=p/T_z $ constant while calculating the partial derivative with respect to time is thus fundamentally different to holding the momentum $p$ constant. The difference between these two approaches is exactly subtracting the momentum shift $p \propto a^{-1} \propto T_z$\,.
\dnew
Note that the conclusion of \fulleqref{eq:Boltzmann 1} can be similarly obtained using \fulleqref{eq:photonboltzmann4}. For a non-interacting ($C[f]=0$) photon bath, one easily finds (using $\id a = a H \id t$)
\begin{equation}
\parderconst{f}{\ln a}{p} =  \parderconst{f}{\ln p}{a}~.
\end{equation}
This is fulfilled for any solution of the kind $f(a\cdot p)$. Since, however, $p \propto 1/a$ for any individual photon, their total distribution is conserved independently of its precise shape.

\subsection{Evolution of the chemical potential $\mu$ \label{ap:mu}}
The evolution of the effective chemical potential $\mu$ of the photons is a difficult problem, which has been extensively discussed in the literature (e.g., \cite{kompaneets1957establishment, Sunyaev1970, danese1982double, hu1993thermalization, chluba2005spectral, Khatri2012BB, chluba2014science, chluba2014science}). Here, we simply want to recap the most important results.
\dnew
The general solution of the photon Boltzmann equation including \CS, \DC, and \BS permits a $\mu$ distortion, which is a function of both frequency and time.\footnote{In Section~\ref{SD_what} we described it as a constant. Note that this was true at any single time $t$, allowing for different $C$ at different times, i.e., a function $C(t)$. The $\mu$ distortion is only large and independent of frequency for $x \gg x_c$, while it is suppressed due to \DC and \BS for $x \ll x_c$ as argued in \cite{hu1993thermalization}.} Its dependence can usually be written~as
\begin{equation}
	\mu(x,t) \approx \mu_0(t) e^{-x_c(t)/x}~,
\end{equation}
where $x_c(t)$ is the critical frequency of the \DC and \BS processes (see \cite{burigana1991formation,danese1982double,hu1993thermalization,chluba2014science}), and $\mu_0(t)$ obeys the differential equation
\begin{equation}
\der{\mu_0}{\tau} = C_\mu \frac{\dot{Q}}{\rho_{\gamma}} - \gamma_N \frac{T_\gamma x_c \mu_0}{m_e}~,
\end{equation}
with $\dot{Q}$ being the effective heating rate, and $\gamma_N = (4/3) (C_\mu/ G_2)$\,, where $C_\mu$ and $G_2$ are the same as in \fullmanyeqref{eq:normalization 3} and \eqref{eq:number conservation mu-dist}, respectively. This represents an exponentially decaying effective chemical potential with decay time of approximately 
\begin{equation}
	\tau_\mu(z) \approx \gamma_N \int \frac{T_\gamma x_c}{m_e} d\tau~.
\end{equation}
This defines then the distortion visibility function as $\exp(-\tau_\mu(z))$, which can subsequently be approximated as $\exp(-(z/z_\mathrm{th})^{5/2})$, where $z_\mathrm{th}$ is defined as in \fulleqref{eq:z_th}.

\section{Further details on the spectral distortions \label{ap:specdist}}

\subsection{Second order temperature shift \texorpdfstring{\finalr{g distortion}{}}{}}\label{ap:secondorder_g}
The second order contribution to the temperature shift \finalr{distortion}{} is very easy to calculate (assuming $\epsilon=\Delta T/T_z \ll 1$):
\begin{equation}
	B\left(\frac{x}{1+\epsilon}\right) \approx B\left(\frac{x}{1+\epsilon}\right)\biggr\rvert_{\epsilon=0} + (\partial_\epsilon B)\left(\frac{x}{1+\epsilon}\right)\biggr\rvert_{\epsilon=0} \epsilon + (\partial_\epsilon^2 B)\left(\frac{x}{1+\epsilon}\right)\biggr\rvert_{\epsilon=0} \epsilon^2/2 + \mathcal{O}(\epsilon^3)~.
\end{equation}
Using now $y=x/(1+\epsilon)$, $\partial_\epsilon = -x/(1+\epsilon)^2 \partial_y$\, and $B'(x) \equiv \partial_x B(x)$, we find immediately
\begin{equation}
\begin{aligned}
B\left(y\right) 
 & \approx  B(x) - \frac{xB'(y)}{(1+\epsilon)^2} \biggr\rvert_{\epsilon=0} \epsilon + \left(\partial_\epsilon\left(-\frac{xB'(y)}{(1+\epsilon)^2}\right)\right)\biggr\rvert_{\epsilon=0} \epsilon^2/2 + \mathcal{O}(\epsilon^3) \\
 & =B(x) - x B'(x) \epsilon + \left(\left(2\frac{xB'(y)}{(1+\epsilon)^3}\right) + \frac{x^2}{(1+\epsilon)^4}\partial_y B'(y) \right)\biggr\rvert_{\epsilon=0} \epsilon^2/2 + \mathcal{O}(\epsilon^3)\\
 & =B(x) - x B'(x) \epsilon + \left(2 xB'(x) + x^2 B''(x) \right) \epsilon^2/2 + \mathcal{O}(\epsilon^3)~.\\
\end{aligned}
\end{equation}
By the definitions of $G(x) = -xB'(x)$ and $Y(x) = -1/x^2 \partial_x(x^3 G(x)) = 4x B'(x) + x^2 B''(x)$, we then obtain
\begin{equation}
\begin{aligned}
B\left(y\right) \approx B(x) + G(x) \epsilon + (2G(x) + Y(x)) \epsilon^2/2 + \mathcal{O}(\epsilon^3)~.
\end{aligned}
\end{equation}
Converting $\epsilon \to g$, we find the created \finalr{distortion}{temperature shift} as \cite{Zeldovich1972Sup, chluba2004superposition}
\begin{equation}
\begin{aligned}
\Delta f(x) \approx G(x) g(1+g) + Y(x) g^2/2 + \mathcal{O}(g^3)~.
\end{aligned}
\end{equation}
Adding now the normalization, and keeping only terms up to second order directly translates into $\tilde g(1+\tilde g/4) \mathcal{G}(x) + \tilde g^2/8 \mathcal{Y}(x)$, as stated in Section \ref{SD_calc}. 
\dnew
The reason to include the second order in $g$ is to avoid misattributing the $\tilde g^2/8$ term to the $y$ distortion. Since $g$ can vary within the error bars of the $T_0$ determination, which is up to around $2 \cdot 10^{-4}$, it adds a term of up to around $2 \cdot 10^{-8}$ to the $y$ distortion. Thus, the error made by neglecting the term is small, but possibly measurable. We chose to include it in accordance with \cite{chluba2014teasing}. Terms of order $g^3$ are of order $10^{-12}$ \cite{chluba2004superposition}, and are thus far below sensitivity and also less important than better modeling of other effects. We will neglect the third order effects here.

\subsection{Heating and Distortions \label{ap:drho_rho}}
The idea of this short derivation is to point out explicitly that the redshifting term of the \ppsd directly translates into a redshifting term for the energy density of the photons. This has to be taken into account when calculating the total \SD{s}, since it gives a slightly surprising relation in the end.
\dnewnoindent
Integrating the photon Boltzmann equation \eqref{eq:photonboltzmann} gives 
\begin{equation}
\begin{aligned}
\dot{Q} &= \int C[f] E \id^3p= \int \der{f}{t} E \id^3p = \int \der{f}{t} 4\pi p^3 \id p\\ &=\int \left[\parderconst{f}{t}{p} - H p \parderconst{f}{p}{t}\right] 4\pi p^3 \id p = \parder{\rho}{t} - 4\pi H \int p^4 \parderconst{f}{p}{t} \id p \\&= \parder{\rho}{t}  - 4\pi H \left[p^4f\right]\Big |_{0}^{\infty}+ 4 \cdot 4\pi H \int p^3 f \id p  = \parder{\rho}{t} + 4 H \rho~,
\end{aligned}
\end{equation}
where we have \textit{defined} the heating rate as the change in energy density due to the collision operator $C[f]$. We simply find
\begin{equation}\label{eq: diff eq rho}
	\parder{\rho_\gamma}{t} + 4 H \rho_\gamma = \dot{Q} \qquad \Longrightarrow \qquad \parder{(a^4\rho_{\gamma})}{t} = a^4 \dot{Q}~.
\end{equation}
The solution to this differential equation can be written as $\rho_\gamma(t) = \rho_z(t) + \Delta \rho(t)$ with homogeneous solution $\rho_z(t)$ and particular solution $\Delta \rho(t)$. 
\dnew
Since the homogeneous solution is the one where the right hand side vanishes, it corresponds to the background solution without any heating, thus evolving exactly as $\rho_z(t) \propto a^{-4}$\,. On the other hand, the particular solution can be found by simple integration, giving us
\begin{equation}\label{eq:def Delta rho}
	\Delta \rho(t) = \frac{1}{a^4} \int_0^{t} a^4 \dot{Q} \id t' = \rho_z(t) \int_0^{t} \frac{\dot{Q}}{\rho_z(t')} \id t'~. 
\end{equation}
We thus find for the total change in energy density
\begin{equation}
	\frac{\Delta \rho_{\gamma}}{\rho_{\gamma}}\biggr\rvert_\mathrm{tot} = \frac{\Delta \rho(t_0)}{\rho_z(t_0)} = \int_0^{\infty} \frac{\dot{Q}}{(1+z)H\rho_z} \id z = \int_0^{\infty} \frac{dQ/dz}{\rho_z} \id z~.
\end{equation}
Here we have used $\id z = -(1+z) H \id t$\,.
\dnew
Note furthermore that in several work present in the literature \cite{chluba2013distinguishing, chluba2013green, chluba2014teasing, chluba2016spectral}, a different definition of the heating has been adopted with respect to the one used within this work. There, instead of defining $Q$ in relation to the collision operator, the heating function has been defined as the total change in the photon energy density, i.e. $Q_{\Delta\rho}=\Delta \rho(t)$. One can easily show using \fulleqref{eq: diff eq rho} that 
\begin{align}\label{eq: 2 definitions of Q}
\frac{\dot{Q}}{\rho_{z}}=\parder{(\rho_{\gamma}/\rho_z)}{t}=\parder{(1 + \Delta \rho(t)/\rho_z)}{t} = \parder{(\Delta \rho(t)/\rho_z)}{t}~,
\end{align}
where we used again the fact that $\rho_z(t) \propto a^{-4}$ and $\rho_\gamma(t) = \rho_z(t) + \Delta \rho(t)$.
\newpage \noindent
From \fulleqref{eq: 2 definitions of Q}  it then follows that
\begin{align}\label{eq: 2 def of Q}
\parder{(\Delta \rho(t)/\rho_{z})}{z}= \der{(Q_{\Delta\rho}/\rho_{z})}{z}= \frac{1}{\rho_{z}}\der{Q}{z}~,
\end{align}
where, for the sake of clarity, we referred to $\dot{Q}$ as the heating rate following directly from the collision term and to $\dot{Q}_{\Delta\rho}$ as the one defined according to the other convention occurring in the literature, proving the equivalence of the two definitions.\footnote{Note that it is still important not to use the two conventions interchangeably. One has to consistently stick with either definition and pay attention to the corresponding factor of $\rho_z$ inside or outside of the derivative.}

\subsection{Branching ratios \label{ap:br}}
In order to determine the shape of the branching ratios $\mathcal{J}_a(z)$, there are different approaches and, correspondingly, different levels of approximation. A possible classification has been proposed in \cite{chluba2016spectral} and in the next paragraphs we are going to review the main characteristics.
\dnew
In a first very crude simplification, one could assume that the transitions between BB and $\mu$ eras and between $\mu$ and $y$ eras occur sharply at a given redshift. Historically, this values are assumed to be $z_{\rm th}\approx 2\times 10^6$ and $z_{\mu y} \approx 5 \times 10^4$ \cite{burigana1991formation,hu1993thermalization}, approximately. Furthermore, in this first simplification, we are going to neglect the presence of residuals. In this case we have $\mathcal{J}_T(z)=1$ for $ z \geq z_{\rm th}$, $\mathcal{J}_\mu(z)=1$ for $ z_{\mu y} \leq z \leq z_{\rm th}$ and $\mathcal{J}_y(z)=1$ for $ z \leq z_{\rm \mu y}$. Otherwise, all functions are equal to zero. 
\dnew
This approximation can be relaxed by introducing a so-called visibility function for spectral distortions, i.e. a quantity determining how efficient the thermalization of SDs is. As first defined in \cite{Sunyaev1970} and resulting from Appendix \ref{ap:mu}, this function takes form
\begin{align}
f(z)\approx e^{-(z/z_{\rm th})^{5/2}}~,
\label{eq: visibility function}
\end{align}
where $z_{\rm th} \approx 1.98 \times 10^6$ defines the "surface of the blackbody photosphere". In this approximation, the effect of DC emission was included \cite{danese1982double}. Using the SDs visibility function, it is possible to see that even at redshift larger than $z_{\rm th}$ small $\mu$ distortion can be created, while most of the released energy produces a change of the \BB temperature. This leads to $\mathcal{J}_\mu(z)=f(z)$ and $\mathcal{J}_T(z)=1-f(z)$ for $ z \geq z_{\mu y}$ and leaves the other conditions unchanged.
\dnew
A further improvement of the previous approximation can be achieved by considering that also the transition between $\mu$ and $y$ distortions does not take place exactly at $z \approx z_{\mu y}$ but is rather gradual. It is therefore necessary to introduce a region where a sum of $y$ and $\mu$ distortions is present. In this case, the shape of these weighting functions has been studied in \cite{chluba2013green} with the resulting form
\begin{align}
& \mathcal{J}_y(z)=\left[1+\left(\frac{1+z}{6.0\times10^4}\right)^{2.58}\right]^{-1} ~,\\
& \mathcal{J}_\mu(z)=\left\{1-\exp\left[-\left(\frac{1+z}{5.8\times10^4}\right)^{1.88}\right]\right\}f(z) ~,\\
& \mathcal{J}_{T}(z)=1-f(z)~.
\end{align}
Note however that, neglecting the presence of residuals as done in \cite{chluba2013green}, the branching ratios do not add up to unity and, consequently, energy is not exactly conserved. To solve this problem, it is possible to redefine $\mathcal{J}_\mu(z)$ as $[1-\mathcal{J}_y(z)]f(z)$, as suggested in \cite{chluba2016spectral}.
\dnew
Finally, an exact solution to the branching ratios problem exists and has been found by \cite{chluba2014teasing}. The details are discussed in depth in Section \ref{Greens}. The calculation of \SD{s} with any of the branching ratio approximations mentioned above is possible using the new {\sc class} implementation. A direct comparison of the resulting amplitudes of the spectral distortions is performed in Table~1~of~\cite{chluba2016spectral}.
\section{Further details on the heating mechanisms \label{ap:heating}}

\subsection{Possible approximations for injection efficiency and deposition function}\label{ap:f_eff_chi}
In the following paragraphs we are going to review three examples of how to model the dependence of the deposition fraction of each channel on redshift and/or free electron fraction, all of which are implemented in {\sc class}. However, the one used throughout this work is the GSVI2013 \cite{galli2013systematic} approximation summarized below.

\ssec{CK2004}
Firstly, we are going to discuss the pioneering Monte Carlo simulations of \cite{shull1985x} and the more recent approximation proposed by \cite{chen2004particle}, both of which provide excellent leading order approximations. Regarding the interaction between primary electrons and the photon bath, the computation performed in the former work considered only two main effects: Coulomb collisions with thermal electrons and collisional ionization or excitation of HI, HeI, and HeII. Nearly 20 years later, \cite{chen2004particle} proposed then an analytical approximation of their results to split the fractional energy deposition as
\begin{equation}\label{eq:dep_func}
\chi_\mathrm{ion}=\frac{1-x_e}{3}~, \quad \chi_\alpha=\frac{1-x_e}{3}~, \quad \text{and} \quad \chi_h=\frac{1+2x_e}{3}~
\end{equation}
into the various channels. Here $\chi_{\mathrm{ion}/\alpha/h}$ represents the deposition fraction into the ionization, Lyman-$\alpha$ excitation, or heating channels. Note this set of equations is also consistent with the expectation that in a fully ionized medium, all the injected energy is ultimately deposited in the form of heat. On the other side, when the medium is neutral, the energy budget is equally divided between the three channels. However, as $x_e>1$ the above approximation breaks down, and we have to set $\chi_h=1$, while the other channels remain at $\chi_\mathrm{ion}=\chi_\alpha=0$. 

\ssec{PF2005}
Several different extensions of the CK2004 approximation have been proposed in order to include helium recombination. Most notably, the approximation introduced in \cite{padmanabhan2005detecting} and used within the {\sc CosmoTherm} code has expanded the equations as
\begin{equation}\label{eq:dep_func2}
\chi_\mathrm{ion}=\frac{1}{3}-\frac{x_e}{3(1+f_\mathrm{He})}~, \quad \chi_\alpha=\frac{1}{3}-\frac{x_e}{3(1+f_\mathrm{He})}~, \quad \text{and} \quad \chi_h=\frac{2}{3}+\frac{x_e}{3(1+f_\mathrm{He})}~.
\end{equation}
Here, the approximation begins to break down only at $x_e > 1+f_\mathrm{He}$, which is the ionization fraction including singly-ionized helium. The second helium recombination is then taken into account. However, the approximations remain linear and based on the original Monte Carlo Simulations of~\cite{shull1985x}. Furthermore, helium is still not treated with a separate branching channel.

\ssec{GSVI2013}
An additional step towards a more realistic description of the deposition fractions carried out by \cite{valdes2008energy} and the updated version described in \cite{galli2013systematic}. In this case, again extensive Monte Carlo simulations have been performed. However, many differences with respect to \cite{shull1985x} are present, like e.g., a more advanced treatment of the cross-section calculations involving collisional ionization and excitation from electron-photon, electron-electron, and free-free interactions. Another improvement consists in a more accurate prediction of the radiation produced by the excitation processes, influencing the amount of energy deposited into low-energy photons not interacting with the gas, and the amount deposited to the Lyman-$\alpha$ background. In addition to that, H and He are also treated separately, splitting therefore $\chi_\mathrm{ion}$ into $\chi_{\mathrm{ion}, \rm H}$ and $\chi_{\mathrm{ion}, \rm He}$. Note furthermore that \cite{galli2013systematic}  also introduced an extended analytical approximation taking into account helium (see their Equation~(11)). The final results are summarized in Table V of \cite{galli2013systematic}, where the deposition fraction of the different channels is again given in dependence on the free electron fraction. Furthermore, a very illustrative comparison between the CK2005 and GSVI2013 approximations mentioned above is displayed in Figure 10 of \cite{galli2013systematic}.

\ssec{Others}
Other important works on the topic are \cite{slatyer2009cmb, huetsi2009constraints, chluba2010could, slatyer2013energy}, all with varying levels of refinement and details in the physics. Their adaptation into the code is left for future work.

\subsection{PBH statistical factors}\label{ap:f_of_m}
The PBH has a temperature given by the Hawking temperature,
\begin{equation}
	T_\mathrm{BH} = \frac{1}{8 \pi G M}~,
\end{equation}
which predominantly guides the particles that can be emitted from the PBH. In \fulleqref{eq: mass-loss PBH} describing the evolution of the PBH mass, the function $\mathcal{F}(M)$ \cite{stocker2018exotic,macgibbon1990quark,macgibbon1991quark},
\begin{align}\label{eq: f PBH}
\mathcal{F}(M)=\sum_{\text{species }i} \,g_i \,f_{s,q} \,e^{-M/(\beta_s\tilde{M}_i)}~,
\end{align}
represents the effective number of species emitted by the PBH and it is normalized to unity at $M \gg 10^{17}\mathrm{g}$. At those PBH masses only massless particles (photons and neutrinos) are emitted due to the low PBH temperature. Here $g_i$ is the numbers of internal degrees of freedom of the particle species $i$, $f_{s,q}$ is a statistical factor depending on particle spins $s$ and charges $q$, $\tilde{M}_i$ is the BH mass that would satisfy $(8 \pi G \tilde{M}_i)^{-1}=m_i$\,, and $\beta_s$ is a spin dependent quantity which accounts for shifts between the peak of the \BB distribution (which should be centered at the particle's mass) and the mean PBH temperature given through $\tilde{M}_i$ \cite{macgibbon1991quark}. The parameters of this approximation are given in Table \ref{tab: PBH mass evolution parameters}.
\begin{table}[t]
	\begin{center}
    \centering
    \begin{tabular}{|c|ccc|cc|}
    \hline
        Particles & $g_i$ & $f_{s,q}$ & $\beta_s$ & $m_i$ [GeV]           & $\beta_i\tilde{M}_i$ [g] \\
    \hline
        $\gamma$  & 2       & 0.060     & 6.04      & 0.00                  & $\infty$                 \\
        $\nu$     & 6       & 0.147     & 4.53      & 0.00                  & $\infty$                 \\  \hline \Tstrut{2.6}      
        $e$       & 4       & 0.142     & 4.53      & 5.11$\times10^{-4}$   & 9.3969$\times10^{16}$    \\       
        $\mu$     & 4       & 0.142     & 4.53      & 0.104                 & 4.5429$\times10^{14}$    \\       
        $\tau$    & 4       & 0.142     & 4.53      & 1.77                  & 2.7022$\times10^{13}$    \\ \hline \Tstrut{2.6}      
        $u$       & 12      & 0.142     & 4.53      & 2.2$\times10^{-3}$    & 2.1826$\times10^{16}$    \\       
        $d$       & 12      & 0.142     & 4.53      & 4.7$\times10^{-3}$    & 1.0217$\times10^{16}$    \\       
        $s$       & 12      & 0.142     & 4.53      & 9.6$\times10^{-3}$    & 5.0019$\times10^{14}$    \\       
        $c$       & 12      & 0.142     & 4.53      & 1.28                  & 3.7514$\times10^{13}$    \\       
        $b$       & 12      & 0.142     & 4.53      & 4.18                  & 1.1488$\times10^{13}$    
        \\ 
        $t$       & 12      & 0.142     & 4.53      & 173.1                 & 2.7740$\times10^{11}$    \\  \hline \Tstrut{2.6}      
        $g$       & 16      & 0.060     & 6.04      & 0.6                   & 1.0671$\times10^{14}$    \\       
        $\pi^0$   & 1       & 0.267     & 2.66      & 0.1350                & 2.0886$\times10^{14}$    \\       
        $\pi^\pm$   & 2       & 0.267     & 2.66      & 0.1396                & 2.0198$\times10^{14}$    \\  \hline \Tstrut{2.6}      
        $W$       & 6       & 0.060     & 6.04      & 80.39                 & 7.9642$\times10^{11}$    \\       
        $Z$       & 3       & 0.060     & 6.04      & 91.19                 & 7.0209$\times10^{11}$    \\       
        $h$       & 1       & 0.267     & 2.66      & 125.1                 & 2.2541$\times10^{11}$    \\       
    \hline
    \end{tabular}
    \caption{Constants for the definition of $\mathcal{F}(M)$ given in \fulleqref{eq: f PBH}. Note that several values for $f_{s,q}$ has been updated compared to Table 1 of \cite{stocker2018exotic}.}
    \label{tab: PBH mass evolution parameters}
\end{center}
\end{table}
\dnew
The current form of \fulleqref{eq: f PBH} assumes that a BH only emits fundamental particles kinematically available at the BH temperature scale. However, for a temperature below the QCD-confinement scale, $T_{\rm QCD}\approx300$ MeV, quarks and gluons are not emitted singularly, but are instead released in form of pions, which will then subsequently decay into electrons and photons. 
\newpage
On the other hand, above $T_{\rm QCD}$ quarks and gluons can be radiated directly. This effect suppresses the quark production for PBHs with masses larger than $\approx10^{13.5}\mathrm{g}$, and replaces it with pion emission. Mathematically, this can be accounted for introducing the weighting factor
\begin{align}\label{eq:QCD coefficient}
Q_{\rm QCD}(T_{\rm BH})= \left[1+\exp\left(-\frac{\log_{10}(T_{\rm BH}/T_{\rm QCD})}{\sigma}\right)\right]^{-1}~,
\end{align}
where $\sigma$ defines the width of the exponential cut-off and has been set to $0.1$. Then \fulleqref{eq: f PBH} can be decomposed as
\begin{align}\label{eq: f PBH 2}
\mathcal{F}=\mathcal{F}_{\rm EM}\final{+\mathcal{F}_{\mu,\tau}}+\mathcal{F}_{\rm \nu}+\mathcal{F}_{\rm q,g}Q_{\rm QCD}+\mathcal{F}_{\pi}(1-Q_{\rm QCD})+\mathcal{F}_{\rm bosons}\,,
\end{align}
where $\mathcal{F}_{\rm EM}$ represents the contribution from \finalr{leptons}{electrons} and photons\final{, $\mathcal{F}_{\mu,\tau}$ from muons and taus}, $\mathcal{F}_{\rm \nu}$ from neutrinos, $\mathcal{F}_{\rm q,g}$ from quarks and gluons, $\mathcal{F}_{\pi}$ from pions, and $\mathcal{F}_{\rm bosons}$ from $W$, $Z$ and $h$. The different contributions are displayed in the left panel of Figure \ref{fig: PBH_f_M}. The introduction of $Q_{\rm QCD}$ also allow us to underline the fact that protons are never directly emitted by the PBH. In fact, below the QCD threshold they are too heavy to be produced, and above the same threshold only the individual quark content is emitted.
\begin{figure}[t]
	\centering
	\begin{minipage}{\textwidth}
		\includegraphics[width=7.5cm, height=5.5cm]{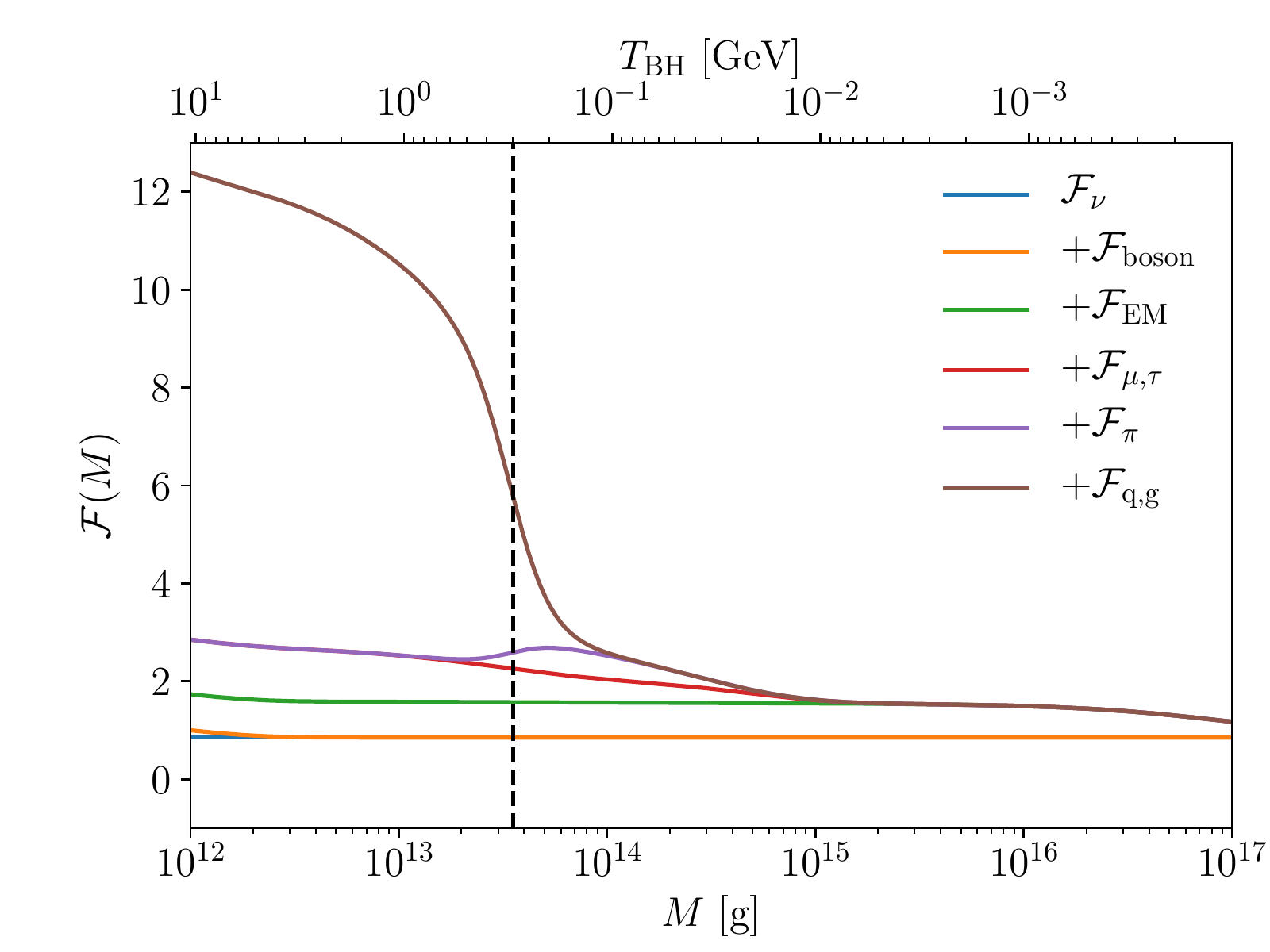}
		\includegraphics[width=7.5cm, height=5.5cm]{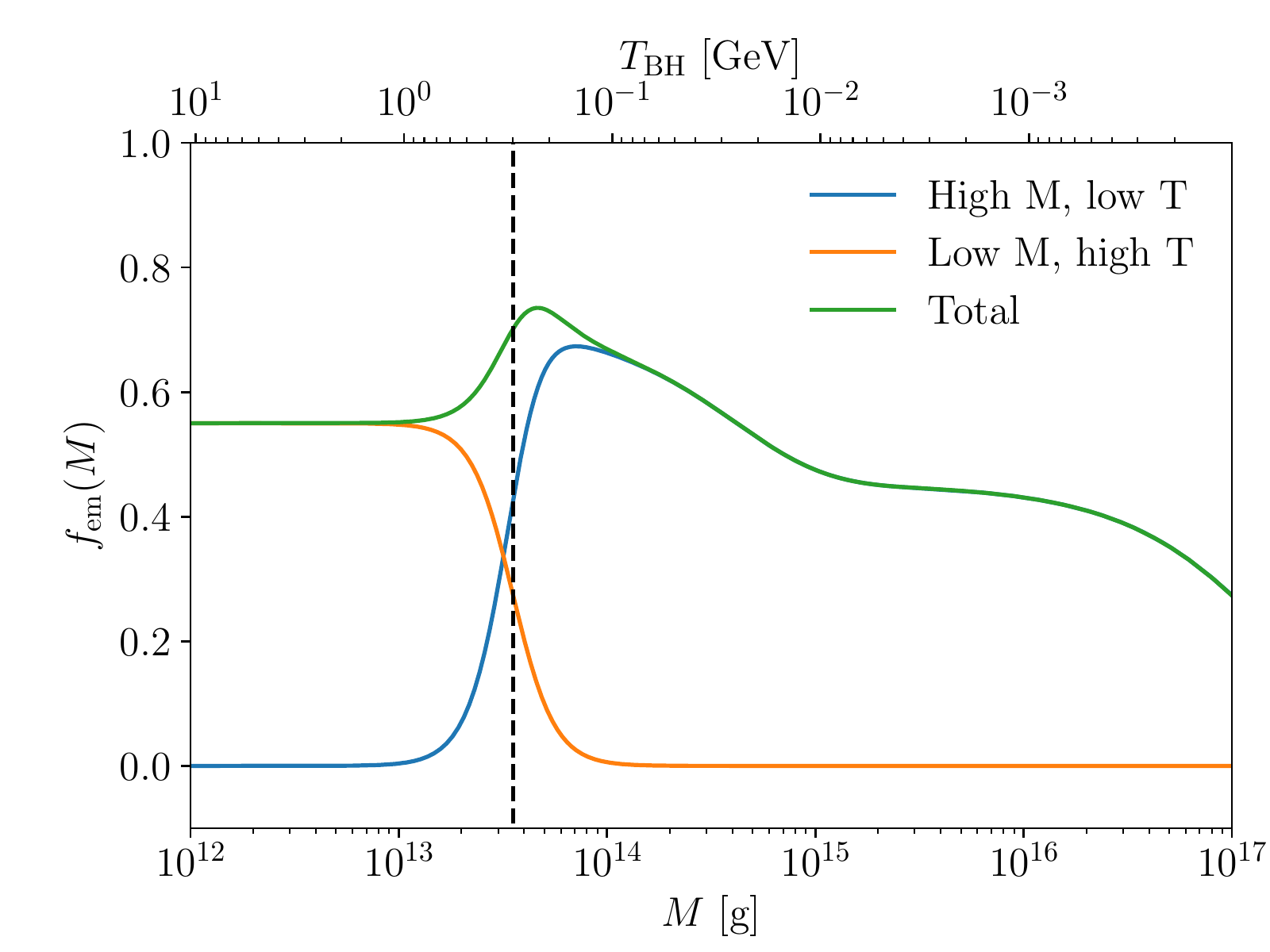}
	\end{minipage}
	\caption{\textbf{Left panel}: Different contributions to the number of species emitted by the PBH. The quarks-gluon and pions contribution are multiplied with the exponential cut-offs as defined in \fulleqref{eq: f PBH 2}. \textbf{Right panel}: Fraction of energy going into EM contribution.}
	\label{fig: PBH_f_M}
\end{figure}
\dnew
As seen in \fulleqref{eq: f PBH 2} and the left panel of Figure \ref{fig: PBH_f_M}, it is possible for a PBH to produce \finalr{non-electromagnetic}{} particles \final{not interacting electromagnetically}, in particular for low PBH masses. These would then not contribute to the total deposited energy. For instance, for PBHs with temperatures above $T_\mathrm{QCD}$, the contribution from particles like, e.g., quarks and pions, becomes dominant. The number of primary ejection species is described by \fulleqref{eq: f PBH}, while the secondary abundances can be computed only by taking into account the decay branching ratios of all non-stable particles. In the end, the secondary emission will only consist of protons (only created in secondary processes), electrons, photons and neutrinos.
\dnew
However, as already mentioned at the beginning of Section \ref{SD_causes}, we are only interested in the electromagnetic contribution, and thus need to additionally isolate and remove the contribution from neutrinos. To accomplish this task, we rely on the calculations carried by \cite{macgibbon1990quark} (in particular on the results summarized in their Table III). There, the authors show that, for BH temperatures above $T_{\rm QCD}$ up to $100$ GeV, around 45\% of the final emitted power in all secondaries is carried by neutrinos and would therefore not contribute to the deposited energy. This value of 45\% is relatively constant, and thus allows us to simply introduce the fraction of energy contributing electromagnetically for relatively low PBH masses as
\begin{align}\label{eq: PBH f_em 2}
f_{\rm em}^\text{low M, high T}=(1-0.45)~.
\end{align}
On the other hand, for large PBH masses, the total contribution consists mostly of EM particles and neutrinos. To subtract these, we use the fraction of energy contributing electromagnetically for relatively high PBH masses as
\begin{align}\label{eq: PBH f_em 1}
f_{\rm em}^\text{high M, low T}=\frac{\mathcal{F}-\mathcal{F}_{\rm \nu}}{\mathcal{F}}~,
\end{align}
which removes \final{most of} the neutrino contribution\finalr{, as also done in \cite{stocker2018exotic}}{}.
\dnew
Using the $Q_\mathrm{QCD}$ for a smooth transition from above to below the QCD confinement scale, the \fullmanyeqref{eq: PBH f_em 1} and \eqref{eq: PBH f_em 2} become
\begin{align}
f_{\rm em}=0.55Q_{\rm QCD}+(1-Q_{\rm QCD})\frac{\mathcal{F}-\mathcal{F}_{\rm \nu}}{\mathcal{F}}~.
\label{eq: PBH f_em 3}
\end{align}
The shape of the three functions is shown in the right panel of Figure \ref{fig: PBH_f_M}.
\dnew
\final{Note that this approximation neglects the production of neutrinos from the decay of pions and unstable leptons. This approximation is justified considering that -- as shown in the left panel of Figure \ref{fig: PBH_f_M} -- the evaporation of the PBH into these particles is only allowed below masses of approximately $10^{15}$ g and is subdominant with respect to $\mathcal{F}_{\rm EM}$ and $\mathcal{F}_{\nu}$ down to masses in the order of $10^{14}$ g, where this high-mass approximation for $f_{\rm em}$ anyway begins to be suppressed (see the right panel of the same figure). Moreover, only a fraction of the already subdominant contribution from pions and unstable leptons is going to decay into neutrinos, which further lowers the impact of our approximation. A more accurate analysis will only possible with the help of tools like \textsc{DarkAges} and \textsc{DarkHistory}, which will be subject of future work.}

\section{Further details on the Green's function approximation \label{ap:greens}}

\subsection{Model independent Green's functions \label{ap:model}}
For linear differential equations of the type
\begin{equation}\label{eq:ap:lindgl}
\mathcal{D} y(x) = f(x)~,
\end{equation}
with differential operator $\mathcal{D}$, one can always find solutions of the type 
\begin{equation}\label{eq:ap:lindglsol}
y(x) = \int G(x,x') f(x')\id x'~,
\end{equation}
where the Green's function $G(x,x')$ is a solution to
\begin{equation}\label{eq:ap:lindglgreens}
\mathcal{D} G(x,x') = \delta(x-x')~.
\end{equation}
When inserting \fulleqref{eq:ap:lindglsol} into \fulleqref{eq:ap:lindgl} and using \fulleqref{eq:ap:lindglgreens}, we find
\begin{equation}
\mathcal{D} y(x) = \mathcal{D} \int G(x,x')f(x')\id x' = \int \delta(x-x') f(x')\id x' = f(x)~.
\end{equation}
The advantage of using the Green's function to solve the system is that the differential equation can be solved for an arbitrary right-hand side $f(x)$. Note, that the Green's function can be found according to \fulleqref{eq:ap:lindglgreens} without knowing any particular $f(x)$.
\dnew
For the case of \SD{s}, this means that the problem can be solved for arbitrary heating histories\footnote{Assuming that the distortion problem is a linear differential equation, which is valid at first order in the distortions \finalr{and at sufficiently high photon energy}{of the \ppsd}.}. All of the information about photon redistribution will be captured within the Green's function, which can be found independently of the heating history. The heating term can then simply be convolved with the Green's function to immediately give the total distortion. One would write similarly to \fulleqref{eq:ap:lindglsol} that
\begin{equation}
\Delta I(x,z) = \int \id z' G_{\rm th}(x,z,z') \frac{\id Q(z')/\id z'}{\rho_{\gamma}(z')}~.
\end{equation}
where one has to find 
\begin{equation}\label{eq:ap:findgreens}
\mathcal{D} G_{\rm th}(x,z,z') = \delta(z-z')~,
\end{equation}
with the differential operator $\mathcal{D}$ here representing all interactions, including \CS, \DC, and \BS. Comparing to equation \fulleqref{eq: SDs amplitudes definition 1}, we find the representation of the Green's function using the branching ratios and distortion shapes as 
\begin{equation}
G_{\rm th}(x,0,z') = \mathcal{G}(x)J_g(z') + \mathcal{Y}(x) J_y(z') + \mathcal{M}(x) J_\mu(z') + R(x,z')~.
\end{equation}
Solving the full \fulleqref{eq:ap:findgreens} thus allows us to find the residual term $R(x,z)$ as well. Note in conclusion that this approximation is only possible as long as the distortions are very small, as we know to be the case.

\section{Further details on the likelihood \label{ap:likelihood}}
\subsection{PIXIE detector sensitivity}\label{ap:PIXIE_noise}
According to \cite{kogut2011primordial} (and references therein), the PIXIE noise equivalent power (NEP) is defined as
\begin{equation}
{\rm NEP}^{2}=2A\Omega T^5\int \alpha\epsilon f \frac{x^4}{e^x-1}\left(1+\frac{\alpha\epsilon f}{e^x-1}\right)\id x~,
\end{equation}
where $A$, $\Omega$ and $\alpha$ are the detector area, solid angle and absorptivity, respectively, while $T$ and $\epsilon$ are the temperature and emissivity of the source. The parameter $f$ corresponds to the power transmission through the optics. Thus, the noise can be defined as
\begin{equation}
\delta P =\frac{{\rm NEP}}{\sqrt{\tau/2}}~,
\end{equation}
where $\tau$ is the integration time and the factor of 2 accounts for the conversion between the frequency and time domains. 
\dnewnoindent
The detector noise then reads approximately
\begin{equation}\label{eq:PIXIE sens 1}
\delta I_{\rm noise} =2 \frac{\delta P}{A\Omega \Delta\nu} \frac{1}{\alpha\epsilon f}~,
\end{equation}
where $\Delta \nu$ is the bandwidth. The presence of the factor 2 is due to the splitting of the sky signal among 4 detectors, which increases the noise by a factor 4, and, at the same time, the averaging of the signal over the 4 detectors lowers the noise by a factor $1/\sqrt{4}$. Inserting thus the optical parameters for the deployed calibrator ($A\Omega=4$ cm$^2$ sr, $\alpha=0.54$, $f=0.82$, and $\Delta \nu=15$ GHz as given in Table 1 of \cite{kogut2011primordial}, and NEP$=2.3$ W/Hz$^{1/2}$) into \fulleqref{eq:PIXIE sens 1} and assuming $\epsilon=1$, one obtains
\begin{equation}\label{eq:PIXIE sens 2}
\delta I_{\rm noise} \approx \frac{2.4\times 10^{-22}}{\sqrt{\tau/1s}}\frac{{\rm W}}{{\rm m}^2\, {\rm Hz}\, {\rm sr}}~.
\end{equation}
Considering then the time interval of 1 s, one obtains the Stokes parameter $\delta I^{I}_{\nu}$ given in Equation~(3.4) of \cite{kogut2011primordial}. Additionally, as argued in Appendix B of \cite{kogut2011primordial}, by multiplying $\delta I^{I}_{\nu}$ by a factor $\sqrt{2}$ accounting for the presence of polarization, one ends up with the second Stokes parameter $\delta I^{QU}_{\nu}$ given in Equation~(3.4). Finally, inserting a fiducial value of $\tau=2$ years for the total mission duration in \fulleqref{eq:PIXIE sens 2} and considering that the deployed configuration in only employed 25\,-\,35\% of the observing time, one obtains a total sensitivity of $\delta I_{\rm noise} \approx 5 \times 10^{-26}$ W/(m$^{2}$ Hz sr). Note that this value is also the one employed in \cite{chluba2014teasing} but neglects penalties due to foreground separation \cite{Abitbol:2017vwa}. Higher levels of sensitivity can be achieved by longer integration times or combining multiple copies of the telescope \cite{Kogut2019}.

}
\clearpage

{\small
\bibliography{biblio}{}
\bibliographystyle{JHEP}
}

\end{document}